\documentclass[graybox,envcountchap,sectrefs]{svmono}

\usepackage{helvet}
\usepackage{amsmath}
\usepackage{courier,hyperref}
\usepackage{type1cm}         

\usepackage{imakeidx}         
\usepackage{graphicx}        
\usepackage{multicol}        
\usepackage[bottom]{footmisc}

\usepackage{array}
\newcolumntype{x}[1]{>{\centering\arraybackslash\hspace{0pt}}p{#1}}

\newcommand{\bom}{{\mbox{$\vec{\omega}$}}}
\newcommand{\bell}{{\mbox{\boldmath $\ell$}}}
\newcommand{\ehat}{\widehat{\mbox{\boldmath $e$}}}

\def\gappeq{\mathrel{ \rlap{\raise.5ex\hbox{$>$}}
                      {\lower.5ex\hbox{$\sim$}}  } }

\def\lappeq{\mathrel{ \rlap{\raise.5ex\hbox{$<$}}
                      {\lower.5ex\hbox{$\sim$}}  } }
                      
               \renewcommand{\refname}{}

\makeindex             

\begin{document}

\author{Carlo F. Barenghi and Nick G. Parker}
\title{A Primer on Quantum Fluids}
\subtitle{}
\maketitle

\frontmatter

\preface
\vspace{-2cm}

This book introduces the theoretical description and 
properties of quantum fluids.  The focus is on gaseous atomic Bose-Einstein 
condensates and, to a minor extent, superfluid helium, 
but the underlying concepts are relevant to other forms of quantum fluids 
such as polariton and photonic condensates.  The book is pitched at the 
level of advanced undergraduates and early post-graduate students, aiming to provide the reader with the knowledge and skills to 
develop their own research project on quantum fluids.  
Indeed, the content for this book grew from introductory notes 
provided to our own research students.  It is assumed that the reader 
has prior knowledge of undergraduate mathematics and/or physics; otherwise, 
the concepts are introduced from scratch, often with references for 
directed further reading.

After an overview of the history of quantum fluids and the motivations for studying them (Chapter 1),  we introduce the simplest model of a 
quantum fluid provided by the ideal Bose gas, following the 
seminal works of Bose and Einstein (Chapter 2).  
The Gross-Pitaevskii equation, an accurate description of weakly-interacting Bose gases at low temperatures, 
is presented, and its typical time-independent solutions
examined (Chapter 3).  We then progress to solitons and waves (Chapter 4) and vortices  (Chapter 5) in quantum fluids.  
For important aspects which fall outside the scope of this book, e.g. 
modelling of Bose gases at finite temperatures, we list appropriate 
reading material.  Each chapter ends with key exercises to deepen the understanding.  Detailed solutions can be made available to instructors upon request to the authors. 

We thank Nick Proukakis and Em Rickinson for helpful comments on this work.

\vspace{\baselineskip}
\begin{flushleft}\noindent
{\it Carlo Barenghi\footnote{\texttt{carlo.barenghi@newcastle.ac.uk}}, Nick Parker\footnote{\texttt{nick.parker@newcastle.ac.uk}}}\hfill \\
Joint Quantum Centre (JQC) Durham-Newcastle, \hfill \\ 
School of Mathematics and Statistics, \hfill \\ 
Newcastle University \hfill \\
April 2016 \hfill
\end{flushleft}

\tableofcontents

%
%

\extrachap{Acronyms}


{\it List of acronyms}
\begin{description}[CABR]
\item[1D]{One-dimensional}
\item[2D]{Two-dimensional}
\item[3D]{Three-dimensional}
\item[BEC]{Bose-Einstein condensate}
\item[GPE]{Gross-Pitaevskii equation}
\item[LIA]{Local induction approximation}
\end{description}

\noindent {\it List of symbols}
\begin{description}[CABR]
\item[$A$]{Wavefunction amplitude}
\item[${\bf A}$]{Vector potential }
\item[$\alpha_j$]{Scaling solution velocity coefficients, $j=x, y, z$ }
\item[$a_0$]{Vortex core radius }
\item[$a_{\rm s}$]{{\it s}-wave scattering length }
\item[$b_j$]{Scaling-solution variables, $j=x, y, z$ or $j=r, z$ }
\item[$B$]{Dark soliton coefficient $B=\sqrt{1-u^2/c^2}$} 
\item[$\beta$]{Irrotational flow amplitude}
\item[$c$]{Speed of sound }
\item[$C_{\rm \mathcal{V}}$]{Heat capacity at constant volume }
\item[$d$]{Average inter-particle distance }
\item[$D$]{System size} 
\item[$\mathcal{D}$]{Number of dimensions}
\item[$\ehat_j$]{Unit vector. $j=x, y, z$ for Cartesian coordinates or $j=r, z, \theta$ for cylindrical polar coordinates}
\item[$\epsilon$]{Small parameter}
\item[$E$]{Energy}
\item[$E'$]{Energy per unit mass}
\item[$\eta$]{Flow angle}
\item[$f_j$]{Distribution function, with $j={\rm B}$, ${\rm BE}$ or ${\rm FD}$ for the Boltzmann, Bose-Einstein or Fermi-Dirac distributions }
\item[$F$]{Free energy }
\item[$g_i$]{Degeneracy of $i$'th energy level }
\item[$g$]{Density of states, $g(E)$ or $g(p)$ }
\item[$g$]{GPE nonlinear coefficient }
\item[$\Gamma(x)$]{The Gamma function, $\displaystyle \Gamma=\int_0^\infty t^{x-1}e^{-t}~{\rm d}t$ }
\item[$H_0$]{Cylinder height}
\item[$\hbar$]{Planck's constant, $\hbar=6.63 \times 10^{-34}~{\rm m}^2~{\rm kg}/{\rm s}$}
\item[$k$]{Wavenumber}
\item[$k_{\rm B}$]{Boltzmann's constant, $k_{\rm B}=1.38 \times 10^{-23}~{\rm m}^2~{\rm kg}~{\rm s}^{-2}~{\rm K}^{-1}$ }
\item[$\kappa$]{Quantum of circulation}
\item[$\lambda$]{Trap ratio, $\omega_z/\omega_r$, of a cylindrically-symmetric harmonic trap}
\item[$L$]{Vortex line density }
\item[$L_z$]{Angular momentum about $z$ }
\item[$\ell$]{Wavepacket size (Chapter 4)}
\item[$\ell$]{Average inter-vortex distance (Chapter 5)}
\item[$\ell_j$]{Harmonic oscillator length $\ell_j=\sqrt{\hbar/m \omega_j}$ in $j$th dimension  } 
\item[$\lambda$]{Wavelength, including de Broglie wavelength $\lambda_{\rm dB}$ }
\item[$m$]{Mass}
\item[$\mu$]{Chemical potential}
\item[$n$]{Number density}
\item[$N$]{Number of particles, including critical number of particles $N_{\rm c}$, and number of particles in $i$'th level, $N_i$}
\item[$\mathcal{N}_{\rm ps}$]{Number of phase space cells}
\item[$\omega$]{Angular frequency, e.g. of wave or trap}
\item[$\bom$]{Vorticity }
\item[$\Omega$]{Rotation frequency }
\item[$\Omega(z)$]{Complex potential }
\item[$p$]{Momentum (vector {\bf p}, magnitude $p$)}
\item[$\mathcal{P}$]{Condensate momentum}
\item[$P$]{Pressure, including quantum pressure $P'$}
\item[${\rm Pr}$]{Probability}
\item[$\phi$]{Velocity potential} 
\item[$\psi, \Psi$]{Condensate wavefunction}
\item[$q$]{Vortex charge }
\item[$\rho$]{Mass density }
\item[$r$]{Radial coordinate, $r^2=x^2+y^2+z^2$ or $r^2=x^2+y^2$}
\item[$R_j$]{Thomas-Fermi radius in $j$th dimension}
\item[$R$]{Local radius of curvature}
\item[$R_0$]{Cylinder radius }
\item[$t$]{Time}
\item[$\sigma$]{Variational width }
\item[$S$]{Phase distribution}
\item[$S$]{Entropy of vortex configuration (Section \ref{sec:2d_turbo} only)}
\item[$t$]{Time}
\item[$T$]{Temperature, including critical temperature for BEC, $T_{\rm c}$}
\item[$u$]{Soliton speed }
\item[$U$]{Internal energy }
\item[$\mathcal{U}$]{Inter-atomic interaction potential }
\item[${\bf v}$]{Fluid velocity}
\item[$v_0$]{Frame velocity}
\item[$V$]{Trapping potential}
\item[$\mathcal{V}$]{Volume }
\item[$W$]{Number of macrostates }
\item[$\xi$]{Healing length }
\item[$\xi_{\rm s}$]{Bright soliton lengthscale}
\item[$\zeta(x)$]{The Riemann zeta function, $\zeta(x)=\sum \limits^\infty_{p=1}\frac{1}{p^x}$ }
\end{description}


\mainmatter

\chapter{Introduction}
\label{intro}

\begin{figure}[b]
\centering
		\includegraphics[width=0.99\columnwidth,clip=true]{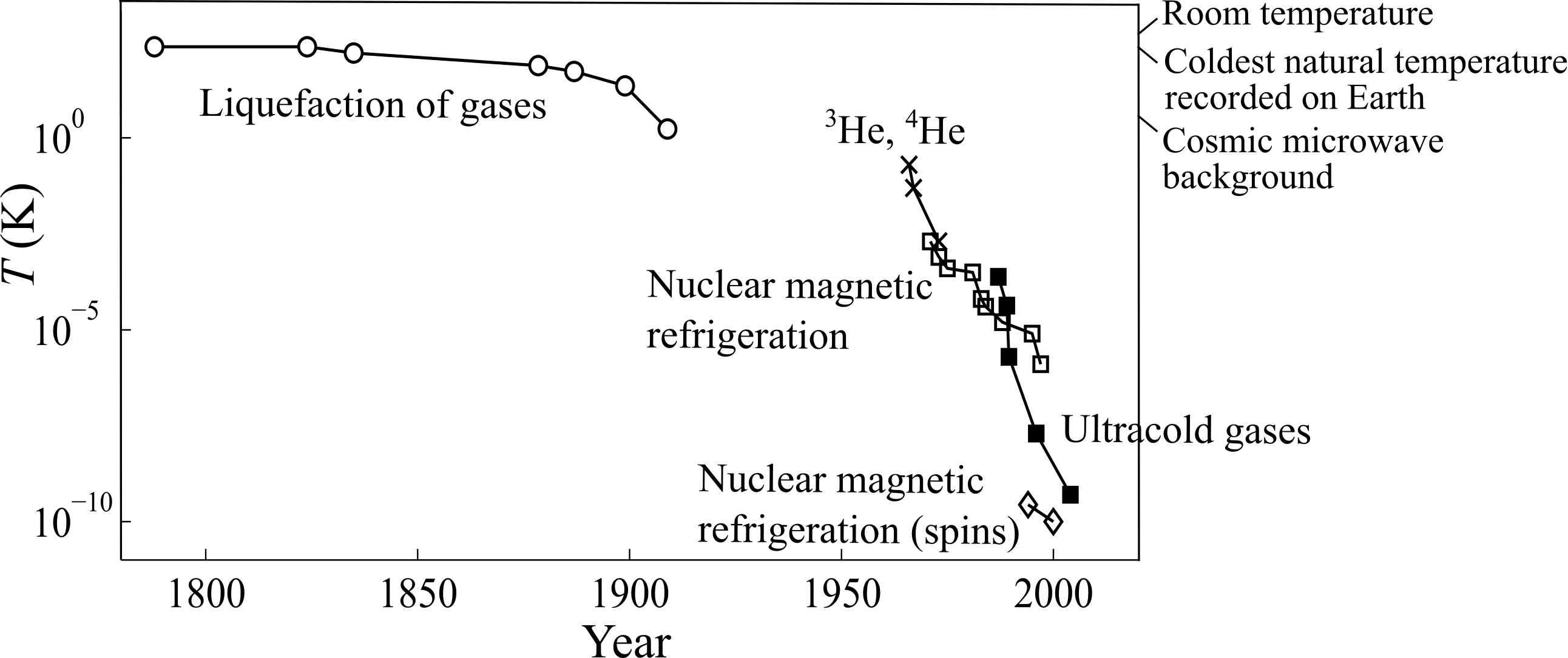}
\caption{Timeline of the coldest engineered temperatures, along with some 
reference temperatures.}
	\label{fig:temp}
\end{figure}

\abstract{Quantum fluids have emerged from scientific efforts to cool matter
to colder and colder temperatures, representing
staging posts 
towards absolute zero (Figure \ref{fig:temp}). 
They have contributed to
our understanding of the quantum world, and still
captivate and intrigue scientists with their bizarre properties.  Here we summarize the background of the two main quantum fluids to date, superfluid helium and atomic Bose-Einstein condensates. }

\section{Towards absolute zero}

The nature of cold has intrigued humankind.  Its explanation as a primordial substance, {\it primum frigidum}, prevailed from the ancient Greeks until Robert Boyle pioneered the scientific study of the cold in the mid 1600s.  Decrying the ``almost totally neglect" of the nature of cold, he set about hundreds of experiments which systematically disproved the ancient myths and seeded our modern understanding.  While working on an air-based thermometer in 1703, French physicist Guillaume Amontons observed that air pressure was proportional to temperature; extrapolating towards zero pressure led him to predict an ``absolute zero'' of approximately $-240$ $^{\rm o}$C in today's units, not far from the modern value of $-273.15$ $^{\rm o}$C (or $0$ K).  The implication was profound: the realm of the cold was much vaster than anyone had dared believe.  An entertaining account of low temperature exploration is given by Ref. \cite{shachtman_2001}.

The liquefaction of the natural gases became the staging posts as low temperature physicists, with increasingly complex apparatuses, raced to explore the undiscovered territories of the ``map of frigor".  Chlorine was liquefied at $239$~K in 1823, and oxygen and nitrogen at $T=90~\textrm K$ and $77~\textrm K$, respectively, in 1877.  In 1898 the  English physicist James Dewar liquefied what was believed to be the only remaining elementary gas, hydrogen, at 23 K, helped by his invention of the vacuum flask.  Concurrently, however, chemists discovered helium on Earth.  Although helium is the second most common element in the Universe and known to exist in the Sun, its presence on Earth is tiny.  
With helium's even lower boiling point, a new race was on.  A dramatic series of lab explosions and a lack of helium supplies meant that Dewar's main competitor, Heike Kamerlingh Onnes, pipped him to the post, liquifying helium at $4~\textrm K$ in 1908.  This momentous achievement led to Onnes being awarded the 1913 Nobel Prize in Physics.

\subsection{Discovery of superconductivity and superfluidity}
These advances enabled scientists to probe the fundamental behaviour of materials at the depths of cold.  Electricity was widely expected to grind to a halt in this limit.  Using liquid helium to cool mercury, Onnes instead observed its resistance to simply vanish below $4$ K.  {\em Superconductivity}\index{superconductivity}, the flow of electrical current without resistance, has since been observed in many materials, at up to $130$ K, and has found applications in medical MRI scanners, particle accelerators and levitating ``maglev" trains.  

\begin{figure}[t]
\centering
\includegraphics[width=0.6 \columnwidth,clip=true]{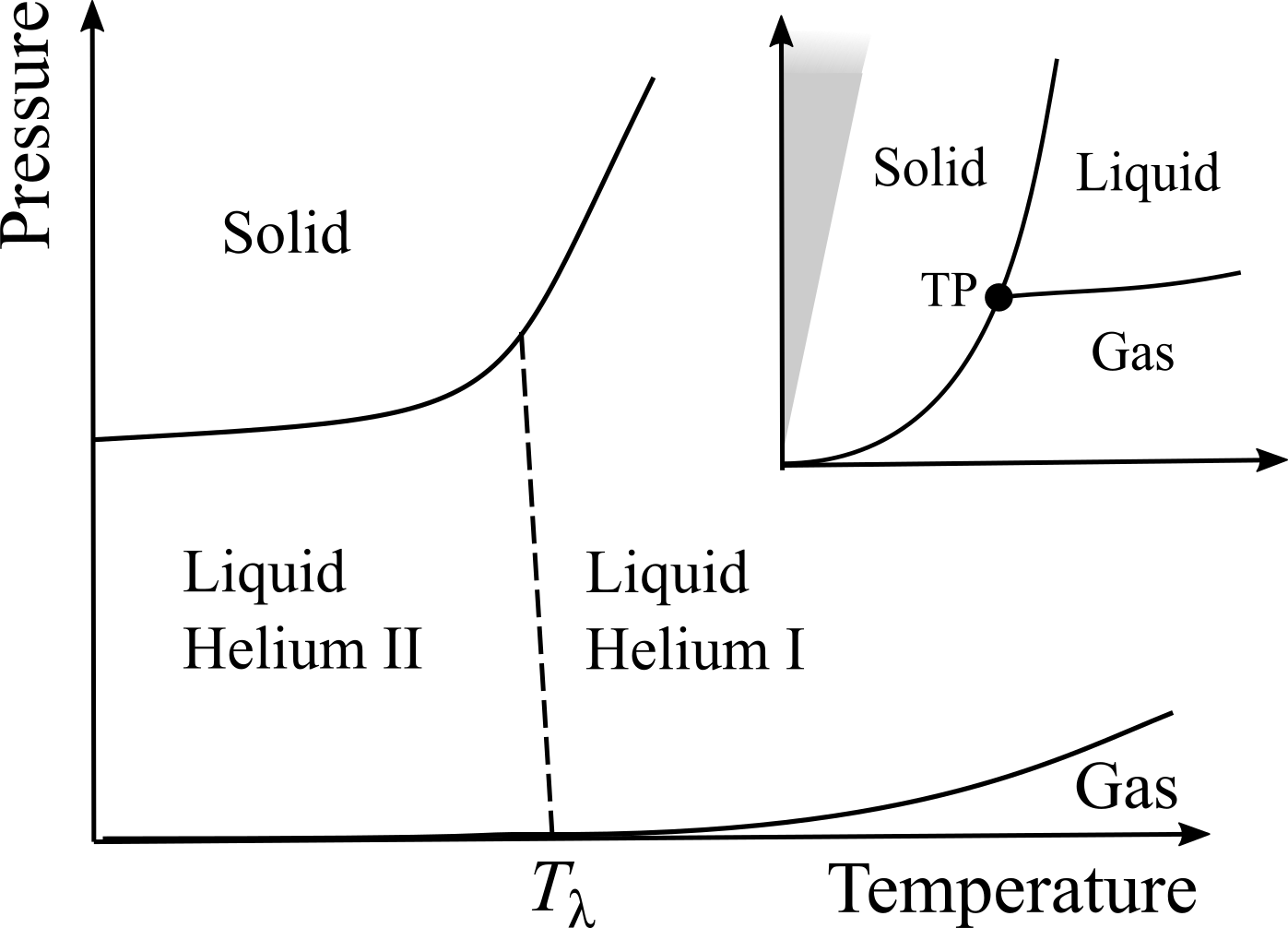}
\caption{Phase diagram of helium.  For a conventional substance (inset), there exists a {\em triple point} (TP), where solid, liquid and gas coexist. Helium lacks such a point.  The shaded region illustrates where Bose-Einstein condensation is predicted to occur for an ideal gas.  }
\label{fig:phase_diag2}
\end{figure}

Onnes and his co-workers also observed unusual behaviour in liquid helium itself.  At around $2.2$K its heat capacity undergoes
a discontinuous change, termed the ``lambda" transition due to the shape of 
the curve.  Since such behaviour is characteristic of a phase change, 
the idea developed that liquid helium existed in two phases: helium~I for 
$T>T_{\lambda}$ \index{helium!helium I} and helium~II for $T<T_{\lambda}$, \index{helium!helium II}
where $T_{\lambda}$ is the critical temperature.  \index{superfluidity} 
Later experiments revealed helium II to have unusual properties, such as it 
remaining a liquid even as absolute zero is approached, 
the ability to move through extremely tiny pores 
and the reluctance to boil. 
These two liquid phases, and the fact that helium remains liquid down 
to $T\rightarrow 0$ (at atmospheric pressure), mean that the phase diagram 
of helium (Figure~\ref{fig:phase_diag2}) is very different to 
a conventional liquid (inset).  \index{helium} \index{helium!phase diagram} In 1938, 
landmark experiments by Allen and Misener and by Kapitza revealed the most striking property of helium~II: its ability to flow without viscosity.  The amazing internal mobility of the fluid, analogous to superconductors, led Kapitza to coin the term ``superfluid".  Other strange observations followed, including ``fluid creep" (the ability of helium to creep up the walls of a vessel and over the edge) and the ``fountain effect" (generation of a persistent fountain when heat was applied to the liquid).

\subsection{Bose-Einstein condensation}

Superfluidity and superconductivity were at odds with classical physics and 
required a new way of thinking.  In 1938 London resurrected an obscure 1925 
prediction of Einstein to explain superfluidity.  Considering an ideal gas 
of quantum particles, Einstein (having developed the ideas put forward by Bose for photons)
had predicted the effect of
{\em Bose-Einstein condensation}, that at low temperatures a large proportion 
of the particles would condense into the same quantum state - the {\em condensate} - \index{Bose-Einstein condensate!ideal gas}
and the remainder of the particles would behave conventionally.  \index{Bose-Einstein condensation} This idea stalled, however, since the conditions for this gaseous phenomena lay in the solid region of the pressure-temperature diagram (shaded region in Fig. \ref{fig:phase_diag2}(inset)), making it inaccessible.
We will follow Einstein's derivation in Chapter~\ref{ideal-gases}.  
Einstein's model predicts a discontinuity in the heat capacity, suggestively similar to that observed in helium.  This, in turn, led to the development of the successful two-fluid model by Tizsa and Landau, in which helium-II is regarded as a combination of an viscosity-free superfluid and a viscous ``normal fluid". 

Bose-Einstein condensation applies to bosons (particles with integer spin,
such as photons and $^4$He atoms),\index{bosons}
but not to fermions (particles with half-integer spin, such as
protons, neutrons and electrons).\index{fermions}
The Pauli exclusion principle prevents more than one identical fermion occupying the same quantum state.  How then could Bose-Einstein condensation be responsible for the flow of electrons in superconductivity?  The answer, put forward in 1957 by Bardeen, Cooper and Schreiffer was for the electrons to form 
{\em Cooper pairs}; these composite bosons could then undergo Bose-Einstein 
condensation.  The observation of superfluidity in the fermionic 
helium isotope $^3$He in 1972 (at around 2~mK) further cemented 
this pairing mechanism.  More information on superconductivity can 
be found in Ref. \cite{annett_2004}.

Superfluid helium and superconductors are
both manifestations of Bose-Einstein condensation.  Arising from the 
macroscopic quantum state that is the condensate, they represent fluids 
governed by quantum mechanics, i.e. quantum fluids (superconductors
can be considered as fluids of charged Cooper pairs).  However, the strong 
particle interactions in liquids and solids mean that these systems 
are much more complicated that Einstein's ideal-gas paradigm, and it 
took until the 1990s for an almost ideal state to be created.  

Hallmarks of superfluidity include the capacity to flow without viscosity, the presence of a critical velocity above which superflow breaks down, the presence of quantized vortices, persistent flow, and macroscopic tunneling in the form of Josephson currents.  We will detail all of these superfluid phenomena throughout this book, with the exception of Josephson currents which can be studied elsewhere \cite{annett_2004}.

\section{Ultracold quantum gases}

\subsection{Laser cooling and magnetic trapping}
Liquids and solids have since been cooled down to milliKelvin and 
microKelvin temperatures using cryogenic refrigeration techniques 
and adiabatic demagnetization, respectively \cite{pobell_2007}, and 
the coldest recorded temperature stands at  100 picoKelvin 
for the nuclear spins in a sample
of rhodium; these achievements are shown in Fig. 1.1.  Meanwhile, the cooling of gases was advanced 
greatly by {\em laser cooling}, developed in the 1980s \cite{metcalf_2001}.   \index{Bose-Einstein condensate!atomic}
Atoms and molecules in a gas are in constant random motion with an average speed 
related to temperature, for example, around 300 m/s in room temperature air.
For a laser beam incident upon a gas of atoms (in a vacuum chamber), 
and under certain conditions, the photons in the beam can be made to impart, 
on average, momentum to atoms travelling towards the beam, thus slowing 
them down in that direction; applying laser beams in multiple directions 
then allows three-dimensional (3D) cooling.  
In 1985 this ``optical molasses" produced a 
gas at 240 $\mu$K, with average atom speeds of $\sim$ 0.5 m/s.  A few years 
later, 2 $\mu$K was achieved ($\sim1$ cm/s).  These vapours were 
extremely dilute, with typical number densities of $n \sim 10^{20}~{\rm m}^{-3}$ (c.f. $n \sim 10^{25}~{\rm m}^{-3}$ for room temperature air); this made the transition from a gas to a solid, the natural process at such cold temperatures (inset of Figure \ref{fig:phase_diag2}), so slow as to be insignificant on the experimental timescales.  In addition, magnetic fields allowed the creation of traps, bowl-like potentials to confine the atoms and keep them away from hot surfaces; with experimental advances, it is now possible to create such ultracold gases in a variety of configurations, from  toruses to periodic potentials, and manipulate them in time\index{trap!magnetic}.  The development of laser cooling and magnetic trapping techniques was recognised with the 1997 Nobel Prize in Physics \cite{nobel_1997}; further details of these techniques can be found elsewhere \cite{metcalf_2001,pethick_2008}.

\subsection{Bose-Einstein condensate  \`{a} la  Einstein}

The achievement of ultracold gases put Einstein's gaseous 
condensate within sight and a new race was on.  
Einstein's model predicted the condensate to form below a critical 
temperature $T_{\rm c} \sim 10^{-19} n^{2/3}$, but the low gas densities employed predicted $T_{\rm c} \sim 1 \mu$K, colder than achievable by laser cooling alone.  To cool even further, a stage of {\em evaporative cooling} was employed whereby the hottest atoms were selectively removed, just like how evaporation cools a cup of coffee.

\begin{figure}[t]
	\centering
		\includegraphics[width=0.6\textwidth,clip=true]{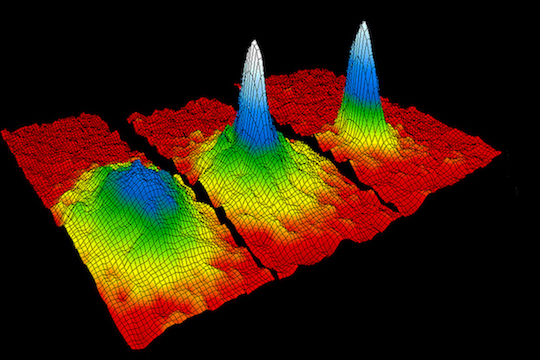}
		\caption{The first observation of a gas Bose-Einstein condensate \cite{anderson_1995}, showing the momentum distribution of a dilute ultracold gas of $^{87}$Rb atoms, confined in a harmonic trap.  As the temperature was reduced, the gas changed from a broad, energetic thermal gas (left) to a narrower distribution (right), characteristic of the condensate.  Image reproduced from the NIST Image Gallery (Reference NIST/JILA/CU-Boulder).
		}
	\label{fig:BEC2}
\end{figure}

In 1995 Cornell and Wieman cooled a gas of rubidium atoms down to $200$ nanoKelvin ($200$ billionths of a degree above absolute zero) to realize the first gaseous Bose-Einstein condensate (BEC) \cite{anderson_1995}.  Figure \ref{fig:BEC2} shows the famous experimental signature of this new state of matter.  
These images were obtained by releasing the trap which confines
the gas, thus letting the atoms fly away and the gas to expand.
Above $T_{\rm c}$ (left plot), the gas was an energetic ``thermal" gas
of atoms
characterised by a wide distribution of speed; upon opening
the trap, atoms with large speeds moved far away, hence the broad
picture in the left plot.
As the temperature was cooled through $T_{\rm c}$, a narrow distribution 
emerged from the thermal gas (middle and right plots), characteristic 
of accumulation of atoms into a state of almost zero energy and speed; these atoms are the Bose-Einstein condensate. \index{Bose-Einstein condensate!atomic}
We derive these thermal and condensate profiles in Chapter 2. 
A few months later, Ketterle independently reported a BEC of sodium 
atoms \cite{davis_1995}.  Seventy years on, Einstein's prediction 
had been realized at the depths of absolute zero.  
Cornell, Wieman and Ketterle shared the 2001 Nobel Prize for this landmark achievement \cite{nobel_2001}.  

There are now over 100 BEC experiments worldwide.  These gases are typically 10-100 micro-meters across (about the width of a human hair), exist in the temperature range $1$ to to $100$ nK, contain $10^3-10^9$ atoms, and are many times more dilute than room temperature air.   BECs are most commonly formed with rubidium ($^{87}$Rb) and sodium ($^{23}$Na) atoms, but many other atomic species, and a growing number of molecular species, have been condensed.  It is also possible to create multi-component condensates, where two or more condensates co-exist.  These gases constitute the purest and simplest quantum fluids 
available, with typically $99\%$ of the atoms lying in the condensed state.  
The last property makes condensates
amenable to first-principles modelling; the work-horse model 
is provided by the Gross-Pitaevskii equation, which will be introduced and 
analysed in Chapter 3.  Gaseous condensates
have remarkable properties, such as superfluidity, as we
see in Chapters 4 and 5. Unlike superfluid 
helium, the interaction between the atoms is very weak, which makes them
very close to Einstein's original concept of an ideal gas.

\subsection{Degenerate Fermi gases}

\begin{figure}[t]
	\centering
		\includegraphics[width=0.5\columnwidth,clip=true]{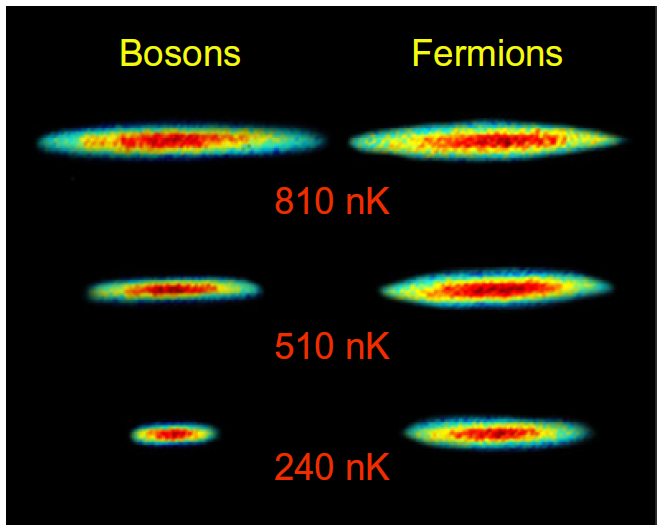}	\hspace{0.5cm}
		\caption{Change in density profile as a $^7$Li bosonic gas and a $^6$Li fermionic gas are cooled towards absolute zero.  The bosonic gas reduces to a narrow distribution corresponding to the low-energy condensate, while the fermionic gas saturates to a larger distribution due to the outwards Pauli pressure imposed by the fermions.  Reproduced from \url{apod.nas.gov} with permission from A. G. Truscott and R. G. Hulet, and corresponding to the experiment of Ref. \cite{truscott_2001}.}
	\label{fig:dfg}
\end{figure}

For a fermionic gas, cooled towards absolute zero, the particles 
(in the absence of Cooper pairing) are forbidden to enter the same quantum 
state by the Pauli exclusion principle.  Instead, they are expected fill 
up the quantum states, from the ground state upwards, each with unit 
occupancy.  This effect was observed in 1999 when a degenerate 
Fermi gas was formed by cooling potassium ($^{40}$K) atoms to below 
300 nK \cite{demarco_1999}.  \index{Fermi gas!degenerate Fermi gas}In this limit, the gas was seen to saturate 
towards a relatively wide distribution, indicating the higher average 
energy of the system, relative to a BEC.   The Pauli exclusion 
principle exerts a very strong ``pressure" \index{pressure!degeneracy pressure} against further 
contraction, an effect which is believed to stabilize neutron stars 
against collapse.  A striking experimental comparison between 
bosonic and fermionic gases as the temperature is reduced is 
shown in Figure \ref{fig:dfg}: the distribution of the
fermionic system cannot contract as the bosonic one.
More recently, experiments have examined the formation of Cooper
pairs in these systems \cite{levin_2012}. 

\section{Quantum fluids today}
We have briefly told the story of the discoveries of superfluid helium 
and atomic condensates, but what about the wider implications of these
discoveries and the current status of the field?  Here we list some examples.

\begin{description}
\item[{\it Many-body quantum systems}:] Quantum fluids embody quantum 
behaviour on a macroscopic scale of many many particles; it is this 
property that gives rise to their remarkable properties. 
As such, quantum fluids provide fundamental insight into quantum 
many-body physics. Moreover, for the case of condensates, the 
experimental capacity to engineer the system, e.g. its interactions, 
dimensionality, and the presence of disorder and periodicity, 
allows the controlled investigation of diverse many-body scenarios and 
emulation of complex condensed matter systems such as superconductors. 
\item[{\it Nonlinear systems}:] Quantum fluids represent a prototype 
fluid, free from viscosity (as we see in Chapter 3) and 
whose vorticity is constrained to take the form of
discrete, uniformly-sized mini-tornadoes.  It is interesting then 
to consider complex fluid dynamics, notably turbulence, in this 
simplified fluid; we discuss this {\em quantum turbulence} in 
Chapter 5 \index{quantum turbulence}.  Condensates also provide an 
idealized system to study 
nonlinear phenomena.  The atomic interactions in a condensate give 
rise to a well-defined nonlinearity, and experimental tricks allow 
this nonlinearity to be controlled in size and nature 
(e.g. local versus non-local nonlinearity).  Nonlinear effects such 
as solitons and four-wave mixing have been experimentally studied; 
we meet solitons in Chapter 4.  
\item[{\it Extra-terrestrial phenomena}:] Condensates are analogous 
to curved space-time and support analog black holes and Hawking radiation, 
while both condensates and helium provide analogs of the quantum vacuum 
believed to permeate the universe and be responsible for its development 
from the Big Bang.   These cosmological phenomena, not accessible on 
Earth, may thus be mimicked and explored in controlled, laboratory-based 
experiments.  
\item[{\it Cooling}:]  The excellent thermal transport property
of helium II lends to its use as a coolant; helium is therefore 
present in superconducting systems, from MRI machines in hospitals
to the Large Hadron Collider at CERN.   
\item[{\it Sensors}:]  Condensates are easily affected by
external forces, and experiments have demonstrated extreme sensitivity to 
magnetic fields, gravity and rotational forces.  Considerable efforts are currently underway to develop these ideas into next-generation sensors, 
for applications such as testing fundamental laws of physics, geological 
mapping and navigation. 
\end{description}

Since 2000, Bose-Einstein condensation has also been achieved in several new systems: magnons (magnetic quasi-particles) in magnetic insulators, polaritons (coupled light-matter quasi-particles) in semiconductor microcavities, and photons in optical microcavities.  In particular, the latter two systems have realized quantum fluids of light, with superfluid properties.

\chapter{Classical and quantum ideal gases}
\label{ideal-gases} 

\abstract{Bose and Einstein's prediction of Bose-Einstein condensation came out of their theory for how quantum particles in a gas behaved, and was built on the pioneering statistical approach of Boltzmann for classical particles. Here we follow Boltzmann, Bose and Einstein's footsteps, leading to the derivation of Bose-Einstein condensation for an ideal gas and its key properties.}

\section{Introduction}
Consider the air in the room around you.  We ascribe properties such as temperature and pressure to characterise it, motivated by our human sensitivity to these properties.  However, the gas itself has a much finer level of detail, being composed of specks of dust, molecules and atoms, all in random motion. How can we explain the macroscopic, coarse-grained appearance in terms of the fine-scale behaviour?  An exact classical approach would proceed by solving  
Newton's equation of motion for each particle, based on the forces it experiences. For a typical room (volume $\sim$50~m$^3$, air particle density $\sim 2 \times 10^{25}$~m$^{-3}$ at room temperature and pressure) this would require solving around $10^{28}$ coupled ordinary differential equations, an utterly intractable task.  Since the macroscopic properties we experience are {\em averaged} over many particles, a particle-by-particle description is unnecessarily complex.  Instead it is possible to describe the fine-scale behaviour {\it statistically} through the methodology of statistical mechanics.  By specifying rules about how the particles behave and any physical constraints (boundaries, energy, etc), the most likely macroscopic state of the system can be deduced. 

We develop these ideas for an ideal gas of $N$ identical and 
non-interacting particles, with temperature $T$ and confined to a box 
of volume $\mathcal{V}$.  The system is isolated, with no energy or 
particles entering or leaving the system\footnote{In the formalism of 
statistical mechanics, this is termed the {\em microcanonical ensemble}.}   
Our aim is to predict the equilibrium state of the gas.  
After performing this for classical (point-like) particles, we
extend it to quantum (blurry) particles.  This leads
directly to the prediction of Bose-Einstein condensation
of an ideal gas.  In doing so, we follow the seminal works 
of Boltzmann, Bose and Einstein.  Further information can be found in an introductory statistical physics textbook, e.g., \cite{mandl} or \cite{landau_lifshitz}.

\section{Classical particles}

\begin{figure}[b]
\centering
\includegraphics[width=0.6\textwidth]{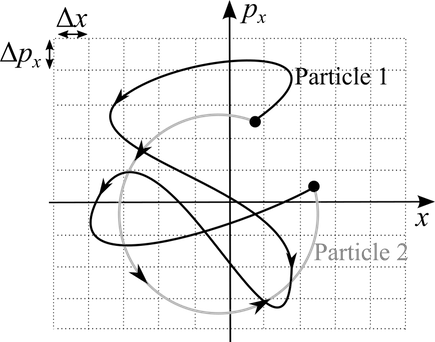}
\caption{Two different classical particle trajectories through 1D phase space ($x,p_x$), with the same initial and final states.   While classical phase space is a continuum of states, it is convenient to imagine phase space to be discretized into finite-sized cells, here with size $\Delta p_x$ and $\Delta x$.
}
\label{fig:phase_space_cells}
\end{figure}

The state of a classical particle is specified by its position ${\bf r}$ and 
momentum ${\bf p}$. 
In the 3D Cartesian world, this requires six coordinates 
$(x,y,z,p_x,p_y,p_z)$.  \index{classical particles} Picturing the world
as an abstract six dimensional {\em phase space},  \index{phase space!classical}
the instantaneous state of 
the particle is a point in this space, which traces out 
a trajectory as it evolves.   Accordingly, an $N$-particle gas is 
specified by $N$ points/trajectories in this phase space.  The accessible range 
of phase space is determined by the box (which provides a spatial constraint) 
and the energy of the gas (which determines the maximum possible 
momentum).  
Figure \ref{fig:phase_space_cells} (left) illustrates two particle trajectories in 1D phase space $(x,p_x)$.  

Classically, a particle's state (its position and momentum) can be determined to arbitrary precision. \index{state!classical} As such, classical phase space is continuous and contains an infinite number of accessible states.  This also implies that each particle can be independently tracked, that is, that the particles are {\em distinguishable} from each other.  \index{classical particles!distinguishability}

\section{Ideal classical gas}
We develop an understanding of the macroscopic behaviour of the gas from these microscopic rules (particle distinguishability, continuum of accessible states) following the pioneering work of Boltzmann in the late 1800s on the kinetic theory of gases.  Boltzmann's work caused great controversy, as its particle and statistical basis was at odds with the accepted view of matter as being continuous and deterministic.    
To overcome the practicalities of dealing with the infinity
of accessible states, we imagine phase space to be discretized into cells of finite (but otherwise arbitrary) size, as shown in Fig. \ref{fig:phase_space_cells}, and our $N$ particles to be distributed across them randomly.  \index{phase space!cells} Let there be $M$ accessible cells, each characterised by its average momentum and position.   The number of particles in the $i$th cell - its {\em occupancy number} - is denoted as $N_i$\index{occupancy number}.  The number configuration across the whole system is specified by the full set of occupancy numbers $\{N_1,N_2,...,N_M\}$.  We previously assumed that the total particle number is conserved, that is,
\begin{equation}
N=\sum_{i=1}^{i=M} N_i. \nonumber
\end{equation}
Conservation of energy provides a further constraint; for now, however, we ignore energetic considerations.

\subsection{Macrostates, microstates and the most likely state of the system}

\begin{figure}[t]
\centering
\includegraphics[width=0.7\textwidth]{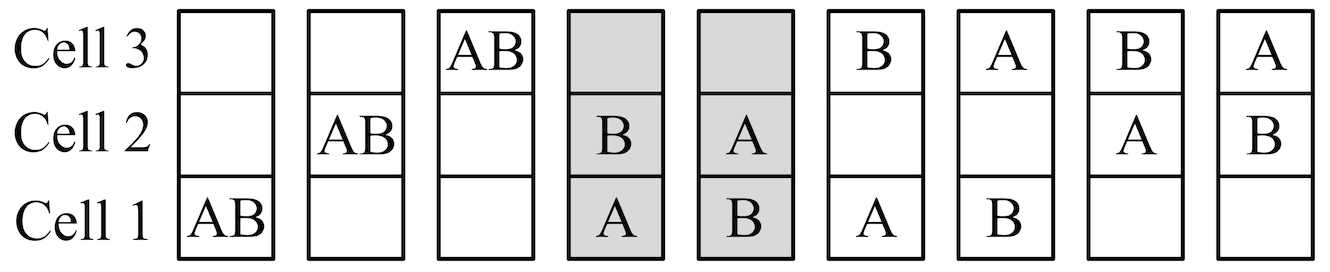}
\caption{Possible configurations of two classical particles, A and B, across three equally-accessible cells.  If we treat the energies of cells 1-3 as $0$, $1$ and $2$, respectively, and require that the total system energy is
 $1$ (in arbitrary units), then only the shaded configurations are possible.}
\label{fig:state_example}
\end{figure}

The macroscopic, equilibrium state of the gas is revealed by considering the ways in which the particles can be distributed across the cells.  In the absence of energetic constraints, each cell is
 equally likely to
be occupied.  Consider two classical particles, A and B (the distinguishability of the particles is equivalent to saying we can label them), and three such cells.  The nine possible configurations, shown in Fig. \ref{fig:state_example}, are termed {\em microstates}.  Six distinct sets of occupancy numbers are possible, $\{N_1,N_2,N_3\}=\{2,0,0\}, \{0,2,0\}, \{0,0,2\}, \{1,1,0\},\{1,0,1\}$ and $\{0,1,1\}$; these are termed {\em macrostates}.  \index{macrostates} Each macrostate may be achieved by one or more microstates.  \index{microstates}

The particles are constantly moving and interacting/colliding with each other in a random manner, such that, after a sufficiently long time, they
will have visited all available microstates, a process termed {\em ergodicity}.  It follows that each microstate is equally likely (the assumption of ``equal {\it a priori} probabilities").  Thus the most probable macrostate of the system is the one with the most microstates.  In our example, the macrostates $\{1,1,0\}$, $\{1,0,1\}$ and $\{0,1,1\}$ are most probable (having 2 microstates each).  In a physical gas, each macrostate corresponds to a particular macroscopic appearance, e.g. a certain temperature, pressure, etc.   Hence, these abstract probabilistic notions become linked to the most likely macroscopic appearances of the gas.  \index{classical particles!statistics}

For a more general macrostate $\{N_1, N_2, N_3, .., N_I\}$, the number of microstates is,  
\begin{equation}
W=\frac{N!}{\prod_i N_i!}.
\label{eqn:MB_micros}
\end{equation}
Invoking the principle of equal {\it a priori} probabilities, the probability of being in the $j$th macrostate is,\index{principle of equal {\it a priori} probabilities}
\begin{equation}
{\rm Pr}(j)=\frac{W_j}{\sum_j W_j}.
\end{equation}
$W_j$, and hence $\rm{Pr}(j)$, is maximised for the most even distribution of particles across the cells.  This is true when each cell is equally accessible; as we discuss next, energy considerations modify the most preferred distribution across cells.

\subsection{The Boltzmann distribution}

\begin{figure}[t]
~~~~~~~~~~~(a)~~~~~~~~~~~~~~~~~~~~~~~~~~~~~~~~~~~~~~~~
~~~~~(b)~~~~~~~~~~~~~~~~~~~~~~~~~~~~~~~~~~~~~~~~~~~~~~~~~~~~~~~~~~
\\
\centering
\includegraphics[width=0.75\linewidth]{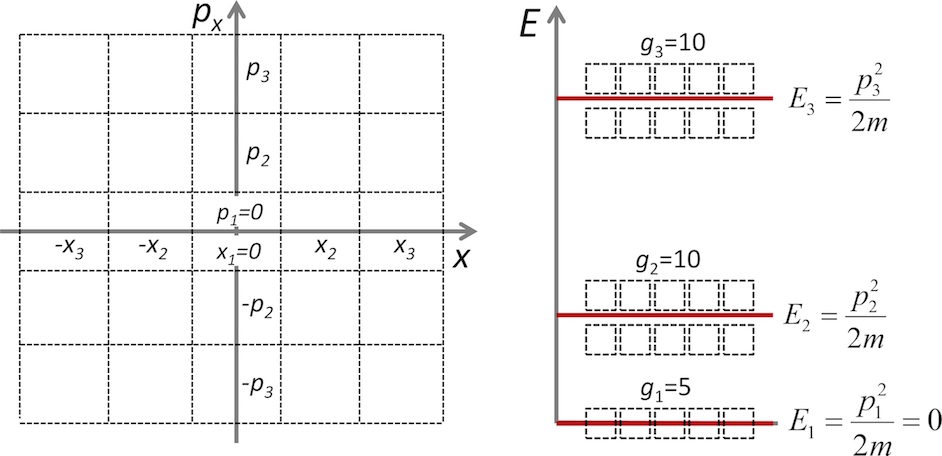}
\caption{For the phase space $(x,p_x)$ shown in (a), the discretization of phase space, coupled with the energy-momentum relation $E=p^2/2m$, leads to the formation of (b) energy levels.  The degeneracy $g$ of the levels is shown.}
\label{fig:energy_levels}
\end{figure}

In the ideal-gas-in-a-box, each particle carries only {\em kinetic energy} $p^2/2m=(p_x^2+p_y^2+p_z^2)/2m$.  Having discretizing phase space, particle energy also becomes discretized, forming the notion of energy levels (familiar from quantum mechanics)\index{energy!levels}.  This is illustrated in Fig. \ref{fig:energy_levels} for $(x,p_x)$ phase space.   Three energy levels, $E_1=0, E_2=p_1^2/2m$ and $E_3= p_2^2/2m$, are formed from the five momentum values ($p=0, \pm p_1, \pm p_2$).   In two- and three-spatial dimensions, cells of energy $E_i$ fall on circles and spherical surfaces which satisfy $p_x^2+p_y^2=2 m E_i$ and $p_x^2+p_y^2+p_z^2=2 m E_i$, respectively.  The lowest energy state $E_1$ is the {\em ground state}; \index{state!ground state}the higher energy states are {\em excited states}.  \index{state!excited state}

The total energy of the gas $U$ is,  
\begin{equation}
U=\sum_i N_i E_i,\nonumber
\end{equation}
where $E_i$ is the energy of cell $i$.  Taking $U$
to be conserved has important consequences for the microstates and macrostates.  For example, imposing some arbitrary energy values in Figure \ref{fig:state_example} restricts the allowed configurations.  Particle occupation at high energy is suppressed, skewing the distribution towards low energy.

For a system at thermal equilibrium with a large number of particles, one macrostate (or a very narrow range of macrostates) will be greatly favoured.  The preferred macrostate can be analytically predicted by maximising the number of microstates $W$ with respect to the set of occupancy numbers $\{N_1,N_2,N_3,...N_I\}$; details can be found in, e.g. \cite{mandl,landau_lifshitz}.  The result is,
\begin{equation}
N_i=f_{\rm B}(E_i),
\end{equation}
where $f_{\rm B}(E)$ is the famous Boltzmann distribution,\index{distribution function!Boltzmann}
\begin{equation}
f_{\rm B}(E) = \frac{1}{e^{(E-\mu)/k_{\rm B} T}}.
\label{eqn:MB_dist}
\end{equation}
The Boltzmann distribution
tells us the most probable spread of particle occupancy across states in an ideal gas, as a function of energy.  This is associated with the thermodynamic equilibrium state.  Here $k_{\rm B}$ is Boltzmann's constant ($1.38 \times 10^{-23}$ m$^2$ kg s$^{-2}$ K$^{-1}$) and $T$ is temperature (in Kelvin degrees, K).  On average, each particle carries kinetic energy 
$\frac{3}{2}k_{\rm B}T$ ($\frac{1}{2}k_{\rm B}T$ in each direction of motion); 
this property is referred to as the {\it equipartition theorem}.\index{equipartition theorem}

The Boltzmann distribution function $f_{\rm B}$ is normalized to the number of particles, $N$, as accommodated by the chemical potential $\mu$.\index{chemical potential}  Writing $A=e^{\mu/k_{\rm B}T}$ gives $f_{\rm B}=A/e^{E/k_{\rm B}T}$, evidencing that $A$, and thereby $\mu$, controls the amplitude of the distribution function.

\begin{figure}[t]
\centering
\includegraphics[width=0.4\linewidth]{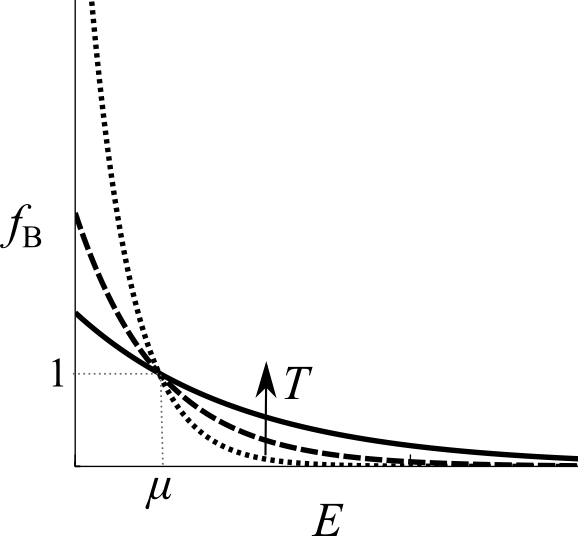}
\caption{The Boltzmann distribution function $f_{\rm B}(E)$ for 3 different temperatures (the direction of increasing temperature is indicated).}
\label{fig:MB_dist}
\end{figure}

The Boltzmann distribution function $f_{\rm B}(E)$ is plotted in Fig. \ref{fig:MB_dist}.  Low energy states (cells) are highly occupied, with diminishing occupancy of higher energy states.  As the
temperature and hence the thermal energy increases, the distribution broadens as particles can access, on average, higher energy states.  Remember, however, that this is the most {\em probable} distribution.  Boltzmann's theory allows for the possibility, for example, that the whole gas of molecules of air in
a room concentrates into a corner of the room.  Due to the strong statistical bias towards an even distribution of energy, momenta and position, such an occurrence has incredibly low probability, but it is nonetheless possible, a fact which caused great discomfort with the scientific community at the time.  

It is often convenient to work in terms of the occupancy of {\it energy levels} rather than {\it states} (phase space cells).  To relate the Boltzmann result to energy levels, we must take into account the number of states in a given energy level, termed the {\it degeneracy} and denoted $g_j$  (we reserve $i$ as the labelling of states).\index{degeneracy}  The occupation of the $j$th energy level is then,
\begin{equation}
N_j=g_j f_{\rm B}(E_j).
\label{eqn:MB_dist2}
\end{equation}


\section{Quantum particles}                

Having introduced classical particles, their statistics and the equilibrium properties of the ideal gas, now
we turn to the quantum case.  The statistics of quantum particles, developed in the 1920s, was pivotal to the development of quantum mechanics, pre-dating the well-known Schr\"{o}dinger equation and uncertainty principle.    

\subsection{A chance discovery}
Quantum physics arose from the failure of classical physics to describe the emission of radiation from a black body in the ultraviolet range (the ``ultraviolet catastrophe").  In 1900, Max Planck discovered a formula which empirically fit the data for all wavelengths and led him to propose that energy is
emitted in discrete quanta of units $hf$ ($h$ being Planck's constant 
and $f$ the radiation frequency).  Einstein extended this idea with his 1905 prediction that the light itself was quantized.

The notion of quantum particles was discovered by accident. \index{quantum particles} Around 1920, 
the Indian physicist Satyendra Bose was giving a lecture on the failure of the classical theory of light using statistical arguments; a subtle mistake led to him prove the opposite.  Indeed, he was able to derive Planck's empirical formula from first principles, based on the assumptions that a) the radiation particles are indistinguishable and b) phase space was discretized into cells of size $h^3$.   Bose struggled at first to get these results published and sought support from Nobel Laureate Einstein; Bose's paper ``Planck's law and the light quantum hypothesis" was then published in 1924 \cite{bose_1924}.  Soon after Einstein extended the idea to particles with mass in the paper ``Quantum theory of the monoatomic ideal gas" \cite{einstein_1924}.  

The division of phase space was mysterious.  Bose wrote ``Concerning the kind of subdivision of this type, nothing definitive can be said", while Einstein confided in a  colleague that Bose's ``derivation is elegant but the essence remains obscure".  It is now established as a fundamental property of particles, consistent with de Broglie's notion of wave-particle duality (that particles are smeared out, over a lengthscale given by the de Broglie wavelength $\lambda_{\rm dB}=h/p$) \index{de Broglie wavelength}and with Heisenberg's uncertainty principle (that the position and momentum of a particle have an inherent uncertainty $\Delta x \Delta y \Delta z \Delta p_x \Delta p_y \Delta p_z = h^3$). Each cell represents a distinct quantum state.  \index{state!quantum state}The indistinguishability of particles follows since it becomes impossible to distinguish two blurry particles in close proximity in phase space. \index{quantum particles!indistinguishability} 

\subsection{Bosons and fermions}
Quantum particles come in two varieties - {\it bosons} and {\it fermions}:

\begin{description}
\item[{\it Fermions}] \index{fermions} Soon after Bose and Einstein's work, Fermi and  Dirac developed {\em Fermi-Dirac statistics} for fermions.  Fermions possess half-integer spin, and include electrons, protons and neutrons.  Fermions obey the Pauli exclusion principle (Pauli, 1925), which states that two identical fermions cannot occupy the same quantum state simultaneously. 

\item[{\it Bosons}] \index{bosons} Bosons obey Bose-Einstein statistics, as developed by Bose and Einstein (above), and include photons and the Higgs boson.  Bosons have integer spin, and since spin is additive, composite bosons may be formed from equal numbers of fermions, e.g. ${^4}$He, $^{87}$Rb and $^{23}$Na.  Unlike fermions, any number of bosons can occupy the same quantum state simultaneously.
\end{description}

\begin{figure}[t]
\centering
\includegraphics[width=0.75\textwidth]{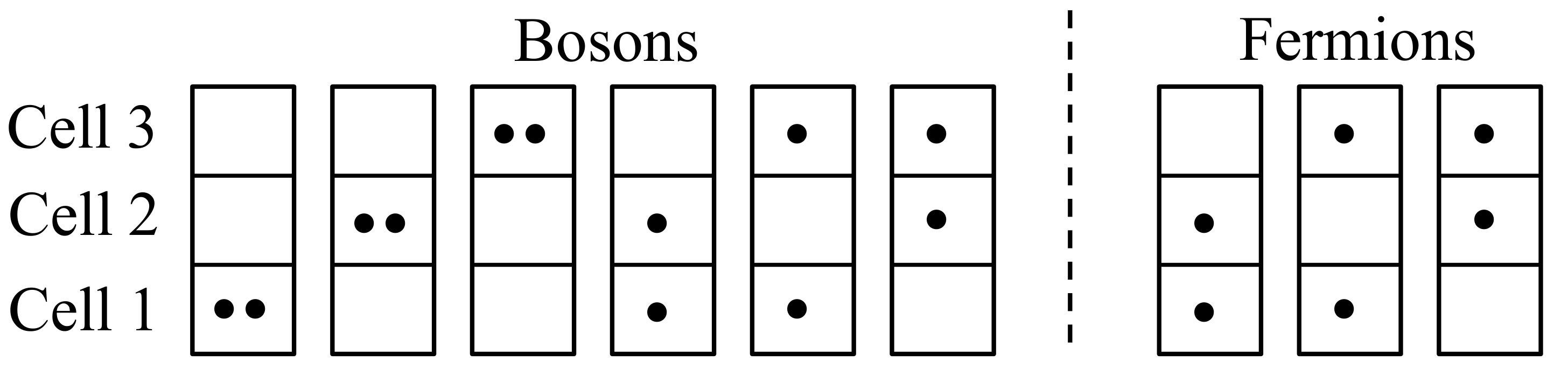}
\caption{Possible configurations of two bosons (left) and two fermions (right) across three equally-accessible cells.  The classical case was shown in Fig. \ref{fig:state_example}.}
\label{fig:bf_cells}
\end{figure}

The indistinguishability of quantum particles, and the different occupancy rules for bosons and fermions, affect their statistical behaviour.  Consider 2 quantum particles across 3 cells, as shown in Fig. \ref{fig:bf_cells}. Since the particles are indistinguishable, we can no longer label them.  For bosons there are six microstates; for fermions there are only three (compared to nine for classical particles, Fig. \ref{fig:state_example}).  The relative probability of paired states to unpaired states is $\frac{1}{3}$, $\frac{1}{2}$ and $0$ for classical particles, bosons and fermions, respectively.  Bosons are the most gregarious, having the greatest tendency to bunch up, while fermions are the most anti-social of all and completely avoid each other. \index{quantum particles!statistics}

\subsection{The Bose-Einstein and Fermi-Dirac distributions}
\label{sec:BE_dist}

\begin{figure}[t]
\centering
~~~~~(a)~~~~~~~~~~~~~~~~~~~~~~~~~~~~~~~~~~~~~~~~~~~~~~~~~~~~~~~~~~~~~(b)~~~~~~~~~~~~~~~~~~~~~~~~~~~~~~~~~~~~~~~~~~~~~~~~~~~~~~~~~~~~~~~~~
\\
\includegraphics[width=0.9\textwidth]{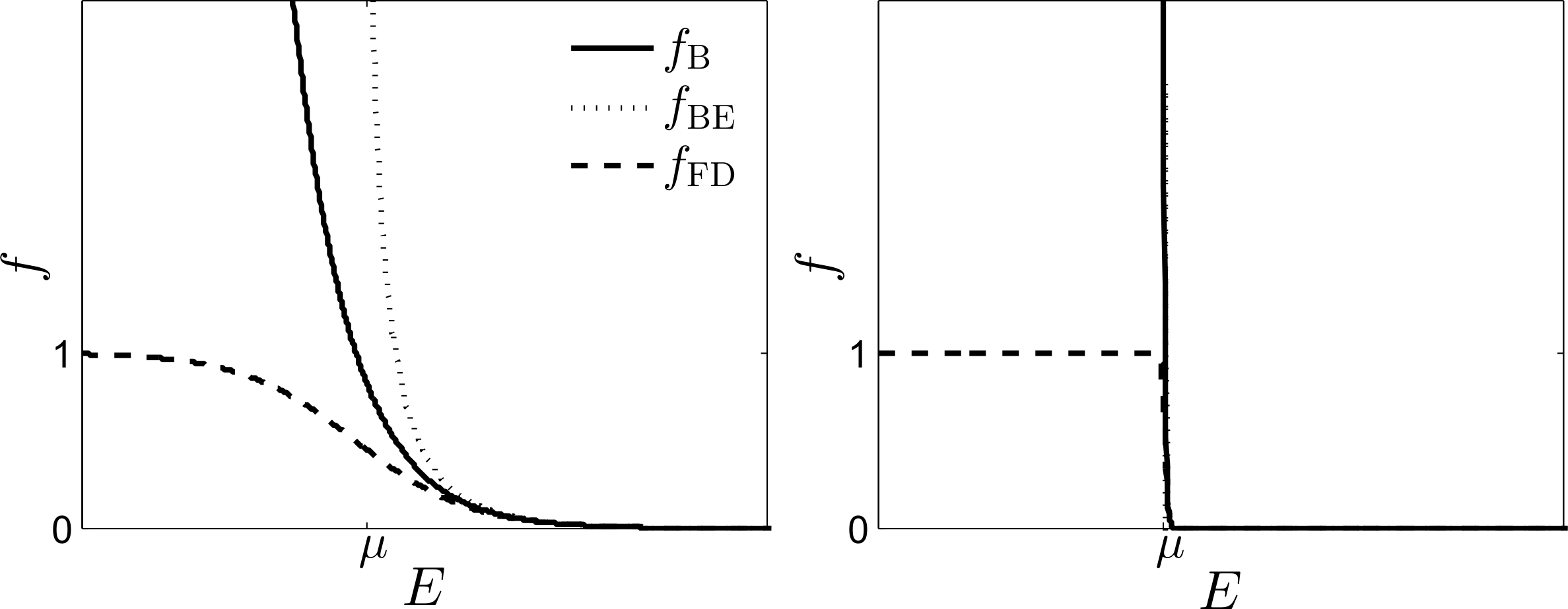}
\caption{The Boltzmann, Bose-Einstein and Fermi-Dirac distribution functions for (a) $T \gg 0$ and (b) $T \approx 0$.}
\label{fig:dist_comp}
\end{figure}

Boltzmann's mathematical trick of discretizing classical phase space becomes physical reality in the quantum world, and the same methodology can be applied to find the distribution functions for bosons and fermions (accounting for their indistinguishability and occupancy rules).  The Bose-Einstein and Fermi-Dirac particle distribution functions, which describe the mean distribution of bosons and fermions over energy $E$ in an ideal gas, are,
\begin{equation}
f_{\rm BE}(E)=\frac{1}{e^{(E-\mu)/k_{\rm B}T}-1},
\label{eqn:BE_dist}
\end{equation}\index{distribution function!Bose-Einstein}
\begin{equation}
f_{\rm FD}(E)=\frac{1}{e^{(E-\mu)/k_{\rm B}T}+1}.
\label{eqn:FD_dist}
\end{equation}\index{distribution function!Fermi-Dirac}
The rather insignificant looking $-1/+1$ terms in the denominators have profound consequences.  Figure \ref{fig:dist_comp} compares the Boltzmann, Bose-Einstein and Fermi-Dirac distributions.  

We make the following observations of the distributions functions:
\begin{itemize}\itemsep0pt
\item To be physical, the distribution functions must satisfy $f \geq 0$ (for all $E$).  This implies that $\mu \leq 0$ for the Bose-Einstein distribution.  For the Fermi-Dirac and Boltzmann distributions, $\mu$ can take any value and sign.  
\item For $(E-\mu)/k_{\rm B}T \gg 1$, the Bose-Einstein and Fermi-Dirac distributions approach the Boltzmann distribution. Here, the average state occupancy is much less than unity, such that the effects of particle indistinguishability become negligible.  Note that the classical limit condition $(E-\mu)/k_{\rm B}T \gg 1$ should not be interpreted too directly, as it seems to predict, counter-intuitively, that low temperatures favour classical behaviour; this is because $\mu$ itself has a non-trivial temperature dependence.  
\item As $E \rightarrow \mu$ from above, the Bose-Einstein distribution diverges, i.e. particles accumulate in the lowest energy states.  
\item For $E \ll \mu$, the Fermi-Dirac distribution saturates to one particle per state, as required by the Pauli exclusion principle.
\item For decreasing temperature, the distributions develop a sharper transition about $E=\mu$, approaching step-like forms for $T \rightarrow 0$.
\end{itemize}

\section{The ideal Bose gas}

A year after Einstein and Bose set forth their new particle statistics for a gas of bosons, Einstein published ``Quantum theory of the monoatomic ideal gas: a second treatise'' \cite{einstein_1925}, elaborating on this topic.  Here he predicted Bose-Einstein condensation. \index{Bose-Einstein condensation} \index{Bose-Einstein condensate!ideal gas} We now follow Einstein's derivation of this phenomena and predict some key properties of the gas. 

\normalsize

\subsection{Continuum approximation and density of states}

We consider an ideal (non-interacting) gas of bosons confined to a box, with energy level occupation according to the Bose-Einstein distribution (\ref{eqn:BE_dist}).  For mathematical convenience we approximate the discrete energy levels by a continuum, valid providing there are a large number of accessible energy levels.  Replacing the level variables with continuous quantities ($E_j \mapsto E, g_j \mapsto g(E)$ and  $N_j \mapsto N(E)$), the number of particles at energy $E$ is written,
\begin{eqnarray}
N(E)= f_{\rm BE}(E)~ g(E) = \frac{g(E)}{e^{(E-\mu)/k_{\rm B}T}-1},
\label{eqn:Nint}
\end{eqnarray}
where $g(E)$ is the {\em density of states}.\index{density of states} The total number of particles and total energy follow as the integrals,
\begin{eqnarray}
N &=& \int  N(E) ~{\rm d}E,\label{eqn:Nint2}
\\
U &=& \int E~N(E)~{\rm d}E.
\label{eqn:U}
\end{eqnarray}
These are integrated in energy upwards from the $E=0$ ($j=1$) ground state.  

\begin{figure}[t]
\centering
\includegraphics[width=0.4\columnwidth]{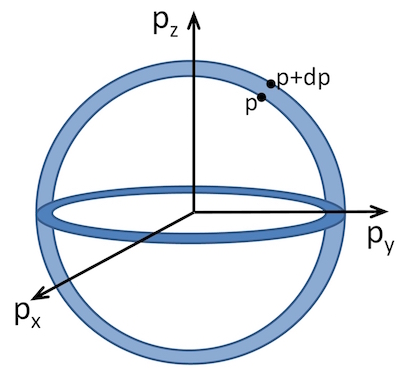}
\caption{The volume of momentum space from $p$ to $p+{\rm d}p$ is a spherical shell in 3D momentum space.}
\label{fig:dos}
\end{figure}

The density of states $g(E)$ is defined such that the total number of possible states in phase space $\mathcal{N}_{\rm ps}$ is, 
\begin{equation}
\mathcal{N}_{\rm ps}=\int g(E) {\rm d}E=\int g(p) {\rm d} p,
\end{equation}
where we have also provided the corresponding expression in terms of momentum $p$, which is more convenient to work with.  
The quantity
$g(p) {\rm d} p$ represents the number of states lying between momenta $p$ and $p+{\rm d}p$.  These states occupy a (6D) volume in phase space which is the product of their (3D) volume in position space and their (3D) volume in momentum space.  The former is the box volume, $\mathcal{V}$.  For the latter, the range $p$ to $p+{\rm d}p$ represents a spherical shell in momentum space of inner radius $p$ and thickness ${\rm d}p$, as illustrated in Fig. \ref{fig:dos}, with momentum-space volume  $4 \pi p^2 {\rm d} p$.  Hence the phase space volume is $4 \pi p^2 \mathcal{V} {\rm d} p$.  Now recall that each quantum state takes up a volume $h^3$ in phase space.  Thus the number of states between $p$ and $p+{\rm d}p$ is,
\begin{equation}
g(p) {\rm d} p=\frac{4 \pi p^2 \mathcal{V}}{h^3} {\rm d} p.
\label{eqn:p_dos}
\end{equation}
Using the momentum-energy relation $p^2=2mE$, its differential form ${\rm d}p=\sqrt{m/2E}~{\rm d}E$), and the relation $g(E)~{\rm d}E=g(p)~{\rm d}p$, Eq. (\ref{eqn:p_dos}) leads to,
\begin{equation}
g(E)=\frac{2\pi (2m)^\frac{3}{2} \mathcal{V}}{h^3}E^\frac{1}{2}.
\label{eqn:e_dos}
\end{equation}
This is the density of states for an ideal gas confined to a box of volume $\mathcal{V}$.  There are a diminishing amount of states in the limit of zero energy, and an increasing amount with larger energy.

 \begin{figure}[b]
\centering
(a)~~~~~~~~~~~~~~~~~~~~~~~~~~~~~~~~~~~~~~~~~~~~~~~~~~~~~~~~~~~~~~
(b)~~~~~~~~~~~~~~~~~~~~~~~~~~~~~~~~~~~~~~~~~~~~~~~~~~~~~~~~~~~~~~~~~~~~~~~~~~~
\\
\includegraphics[width=0.45\columnwidth]{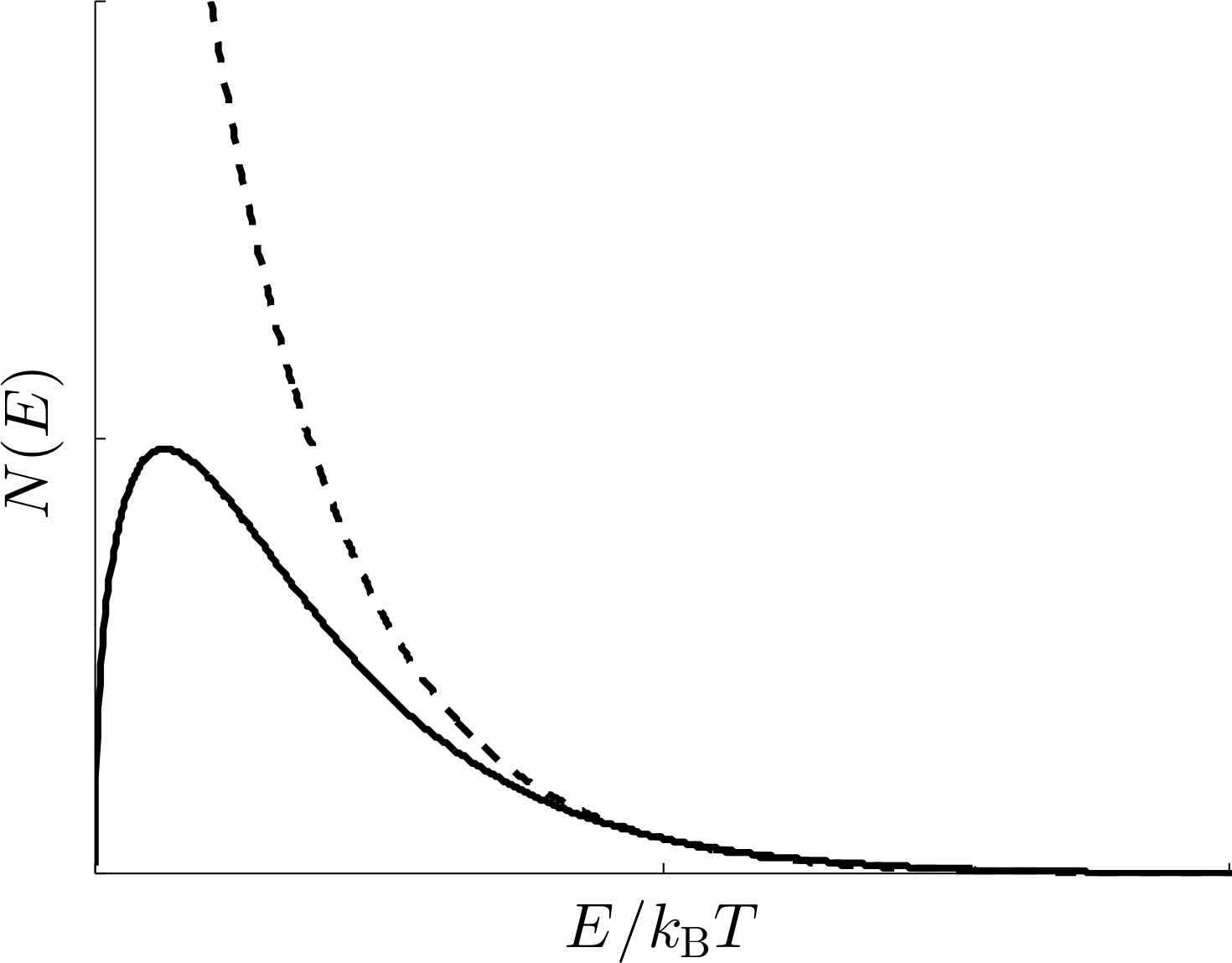}
\includegraphics[width=0.46\columnwidth]{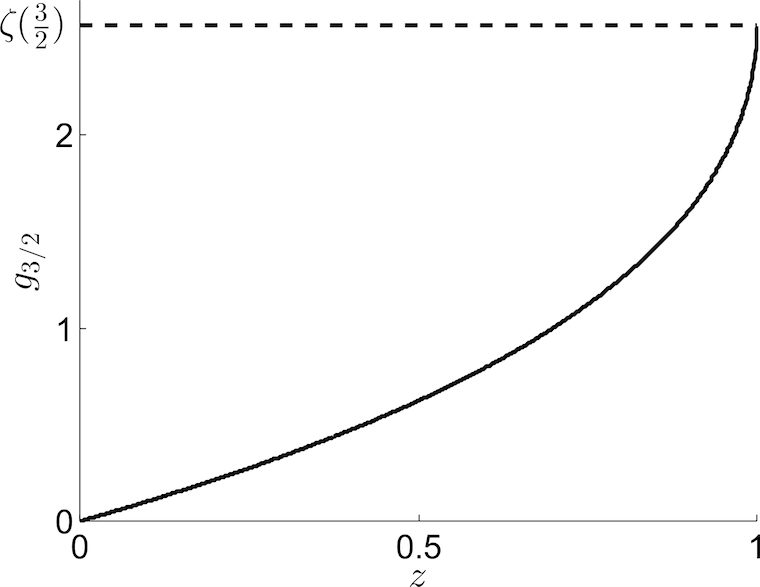}
\caption{(a)  The occupancy of energy levels $N(E)$ (solid line), compared to the Bose-Einstein distribution $f_{\rm BE}$ (dashed line).  The former vanishes as $E \rightarrow 0$ due to the diminishing density of states in this limit.  (b) The function $g_\frac{3}{2}(z)=\sum_{p=1}^\infty z^p/p^\frac{3}{2}$ over the relevant range $0 < z \leq 1$.}
\label{fig:dos_g32}
\end{figure}
While the {\em occupancy of a state} goes like $1/(e^{(E-\mu)/k_{\rm B}T}-1)$ and diverges as $E \rightarrow \mu$, the {\em occupancy of an energy level} goes like $E^\frac{1}{2}/(e^{(E-\mu)/k_{\rm B}T}-1)$ and diminishes as $E \rightarrow 0$ (due to the decreasing amount of available states in this limit).  These two distributions are compared in Fig. \ref{fig:dos_g32} (a).    

\subsection{Integrating the Bose-Einstein distribution}

Using Eqs. (\ref{eqn:Nint},\ref{eqn:e_dos}) we can write the number of particles (\ref{eqn:Nint2}) as,
\begin{eqnarray}
N=\frac{2\pi (2m)^\frac{3}{2} \mathcal{V}}{h^3}\int \limits_0^\infty \frac{E^\frac{1}{2}}{e^{(E-\mu)/k_{\rm B}T}-1} {\rm d}E.
\label{eqn:NBEC1}
\end{eqnarray}
We seek to evaluate this integral.  
To assist us, we quote the general integral\footnote{This result can be derived by introducing new variables $z=e^{\mu/k_{\rm B}T}$ and $x=E/k_{\rm B}T$ to rewrite part of integrand in the form $ze^{-x}/(1-ze^{-x})$, and then writing as a power series expansion. }, 
\begin{equation}
\int \limits_0^\infty \frac{x^{\alpha}}{e^x/z-1} {\rm d}x = \Gamma (\alpha+1) g_{\alpha+1}(z),
\label{eqn:def_integral1}
\end{equation} 
where $\displaystyle \Gamma(x)=\int_0^\infty t^{x-1} e^{-t} {\rm d}t$ is the {\it Gamma function}\footnote{Relevant values for us are $\Gamma(3/2)=\sqrt{\pi}/2$ and $\Gamma(5/2)=3\sqrt{\pi}/4$.}. \index{Gamma function} We have also defined a new function, $g_{\beta}(z)=\sum_{p=1}^\infty \frac{z^p}{p^\beta}$; an important case is when $z=1$ for which it reduces to the {\em Riemann zeta function}\footnote{Relevant values for us are $\zeta(3/2)=2.612$ and $\zeta(5/2)=1.341$}, $\displaystyle \zeta(\beta)=\sum_{p=1}^{\infty} \frac{1}{p^\beta}$.\index{Reimann zeta function}

Taking $\alpha=\frac{1}{2}$, $x=E/k_{\rm B}T$ and $z=e^{\mu/k_{\rm B} T}$ in the general result (\ref{eqn:def_integral1}), we evaluate Eq. (\ref{eqn:NBEC1}) as,
\begin{equation}
N=\frac{(2\pi m k_{\rm B} T)^\frac{3}{2}\mathcal{V}}{h^3}g_\frac{3}{2}(z),
\label{eqn:NBEC2}
\end{equation}
where we have used the result $\Gamma(3/2)=\sqrt{\pi}/2$.
Note that the relevant range of $z$ is $0<z\leq1$: the lower limit is required since $z=e^{\mu/k_{\rm B} T}>0$ while the upper limit $z \leq 1$ is required to prevent negative populations.  Note also that $\mu \leq 0$ over this range, as required for the Bose-Einstein distribution (recall Section \ref{sec:BE_dist}).  In Fig. \ref{fig:dos_g32}(b) we plot $g_\frac{3}{2}(z)$ over this range.

%

\subsection{Bose-Einstein condensation}

The prediction of Bose-Einstein condensation
in the style of
Einstein arises directly from Eq. (\ref{eqn:NBEC2}).  \index{Bose-Einstein condensation} Consider adding particles to the box, while at constant temperature.  An increase in $N$ is accommodated by an increase in the function $g_{\frac{3}{2}}(z)$.  However, $g_\frac{3}{2}(z)$ is finite, reaching a maximum value of $g_\frac{3}{2}=\zeta(\frac{3}{2})=2.612$ at $z=1$.   In other words, the system becomes {\em saturated} with particles.  This critical number of particles\index{critical number of particles}, denoted $N_{\rm c}$, follows as,
\begin{equation}
N_{\rm c}=\frac{(2\pi m k_{\rm B} T)^\frac{3}{2}\mathcal{V}}{h^3}\zeta(\frac{3}{2}).
\label{eqn:NBEC3}
\end{equation}

 Our derivation predicts a limit to how many particles the Bose-Einstein distribution can hold, but common sense tells us that it should always be possible to add more particles to the box.  In fact, we made a subtle mistake.  In calculating $N$ we replaced the summation over discrete energy levels (from the $i=1$ ground state upwards) by an integral over a continuum of energies (from $E=0$ upwards).  However, this continuum approximation does not properly account for the population of the ground state, since the density of states, $g(E) \propto E^\frac{1}{2}$, incorrectly predicts zero population in the ground state.  What we have predicted is the {\em saturation of the excited states}; any additional particles added to the system enter the ground state (which comes at no energetic cost).  For $N\gg N_{\rm c}$, the ground state acquires an anomalously large population. 

As Einstein put it \cite{einstein_1925}, ``a number of atoms which always grows with total density makes a transition to the ground quantum state, whereas the remaining atoms distribute themselves... A separation occurs; a part condenses, the rest remains a saturated ideal gas.''  This effect is {\em Bose-Einstein condensation}, and the collection of particles in the ground state is the {\em Bose-Einstein condensate}. \index{Bose-Einstein condensate!ideal gas} The effect is a condensation in momentum space, referring to the occupation of the zero momentum state.  In practice, when the system is confined by a potential, a condensation in real space also takes place, towards the region of lowest potential.  Bose-Einstein condensation is a {\em phase transition}, but whereas conventional phase transitions (e.g. transformation from gas to liquid or liquid to solid) are driven by particle interactions, Bose-Einstein condensation is driven by the particle statistics.

Based on the above hindsight, we note that the total atom number $N$ appearing in Eqs. (\ref{eqn:Nint2}), (\ref{eqn:NBEC1}) and  (\ref{eqn:NBEC2})  should be replaced by the number in excited states, $N_{\rm ex}$.
  
\subsection{Critical temperature for condensation}
If, instead, the particle number and volume are fixed, then there exists a critical temperature $T_{\rm c}$ below which condensation occurs.  \index{temperature!critical} The population of excited particles at a given temperature is given by Eq. (\ref{eqn:NBEC2}.  For $T > T_{\rm c}$, this is sufficient to accommodate all of the particles, and the gas is in the normal phase.  As temperature is lowered, however, the excited state capacity also decreases.  At the point where the 
excited states no longer accommodate all the particles, Bose-Einstein 
condensation occurs.   The critical temperature is obtained by setting $z=1$ in Eq. (\ref{eqn:NBEC2}) and rearranging for $T$,
\begin{equation}
T_{\rm c}=\frac{h^2}{2\pi m k_{\rm B}}\left(\frac{N}{\zeta(\frac{3}{2}) \mathcal{V}} \right)^\frac{2}{3}.
\label{eqn:TBEC}
\end{equation}
For further decreases in temperature, $N_{\rm ex}$ decreases and so more and more particles must enter the ground state.  In the limit $T \rightarrow 0$, excited states can carry no particles and all particles enter the condensate.  

\subsection{Condensate fraction}
\begin{figure}[t]
\centering
\includegraphics[width=0.7\columnwidth]{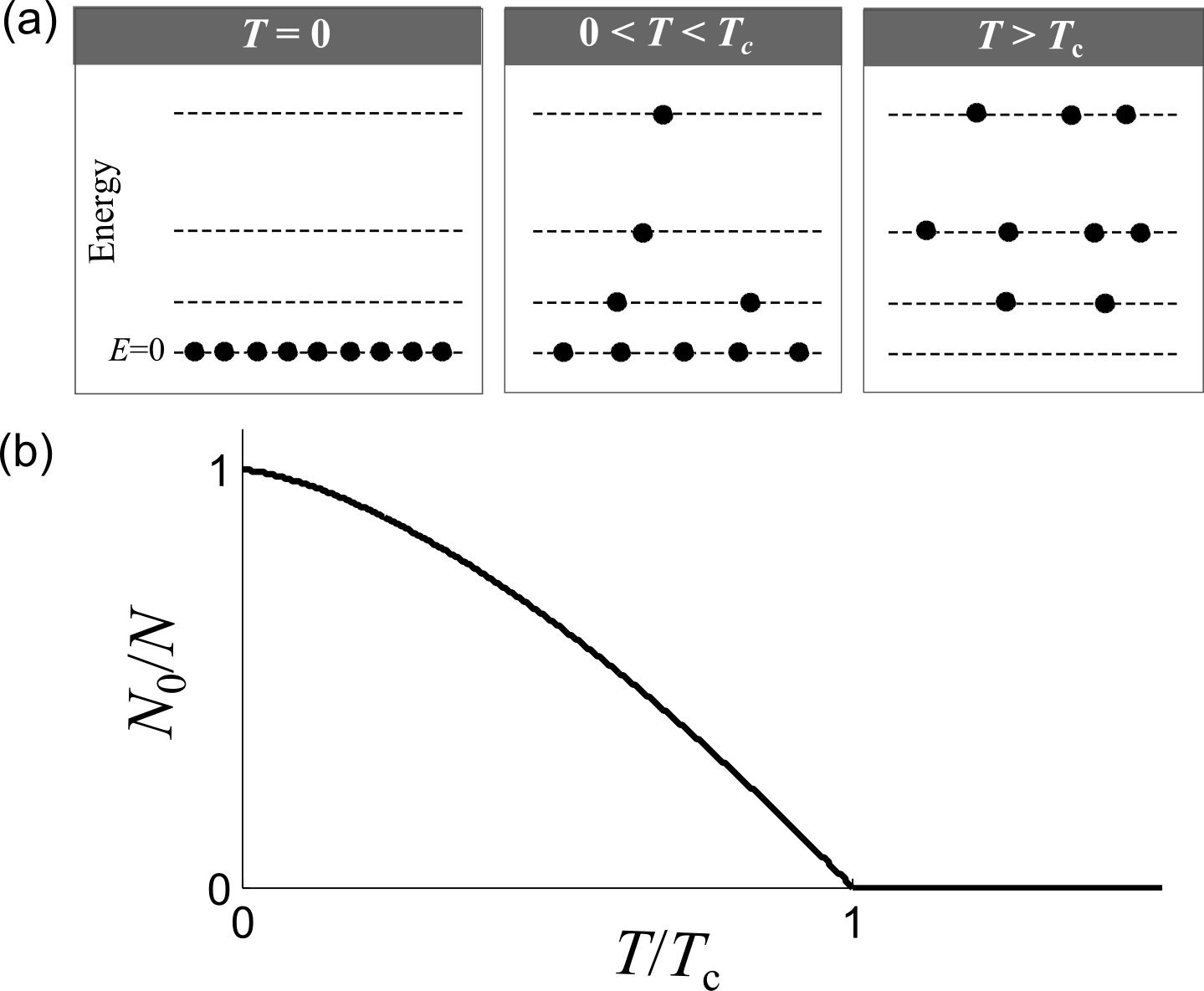}
\caption{(a) Illustration of energy level occupations in the boxed ideal Bose gas.   At $T=0$ all particles lie in the ground state.  For $0 < T < T_{\rm c}$, some particles are in excited levels but there is still macroscopic occupation of the ground state.  For $T>T_{\rm c}$, there is negligible occupation of the ground state.  (b)  Variation of condensate fraction, $N_0/N$, with temperature, as per Eq. (\ref{eqn:BEC_fraction}). 
}
\label{fig:BEC_fraction}
\end{figure}

A useful quantity for characterising the gas is the {\em condensate fraction}, that is, the proportion of particles which reside in the condensate, $N_0/N$.  \index{condensate fraction}  Let us consider its variation with temperature. Writing $N=N_0+N_{\rm ex}$ leads to,
\begin{equation}
\frac{N_0}{N}=1-\frac{N_{\rm ex}}{N}.
\end{equation}
For $T \leq T_{\rm c}$, the excited population $N_{\rm ex}$ is given by Eq.~(\ref{eqn:NBEC2}) with $z=1$, and the total population is given by Eq.~(\ref{eqn:NBEC1}) with $z=1$ and $T=T_{\rm c}$.  Substituting both into the above gives,
\begin{equation}
\frac{N_0}{N}=1-\left(\frac{T}{T_{\rm c}} \right)^{3/2}.
\label{eqn:BEC_fraction}
\end{equation}
For $T > T_{\rm c}$, we expect $N_0/N \approx 0$. This behaviour is shown in Fig. \ref{fig:BEC_fraction}.

\subsection{Particle-wave overlap}
\label{sec:overlap}

\begin{figure}[t]
\centering
\includegraphics[width=0.8\columnwidth,clip=true]{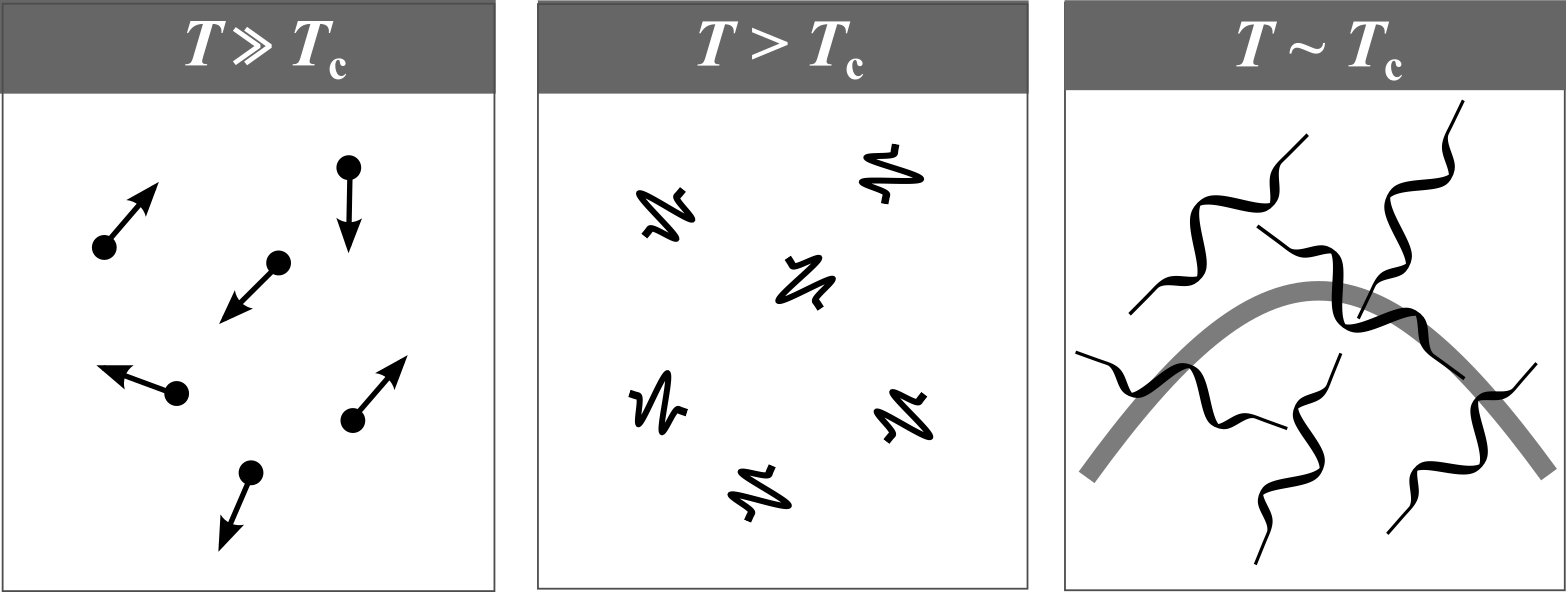}
\caption{Schematic of the transition between a classical gas and a Bose-Einstein condensate. At high temperatures ($T \gg T_{\rm c}$) the gas is a thermal gas of point-like particles. \index{thermal gas}At low temperatures (but still exceeding $T_{\rm c}$) the de Broglie wavelength $\lambda_{\rm dB}$ becomes significant, yet smaller than the average spacing $d$. At $T_{\rm c}$, the matter waves overlap ($\lambda_{\rm dB} \sim d$), marking the onset of Bose-Einstein condensation.} 
\label{fig:bec_schem2}
\end{figure}

Bose-Einstein condensation occurs when $N>N_{\rm c}$, with $N_{\rm c}$ given by Eq. (\ref{eqn:NBEC3}).  It is equivalent to write this criterion in terms of the number density of particles, $n=N/\mathcal{V}$, as,
\begin{equation}
n>\zeta \left(\frac{3}{2}\right) \displaystyle \frac{(2\pi m k_{\rm B} T)^\frac{3}{2}}{h^3}.
\label{eqn:n_criteria}
\end{equation}
According to de Broglie, particles behave like waves, with a wavelength $\lambda_{\rm dB}=h/p$.  For a thermally-excited gas, the particle wavelength is $\displaystyle \lambda_{\rm dB}=\frac{h}{\sqrt{2\pi m k_{\rm B} T}}$.  Employing this, the above criterion becomes,
\begin{equation}
n\lambda_{\rm dB}^3>\zeta \left(\frac{3}{2}\right).
\label{eqn:n_criteria2}
\end{equation}
Upon noting that the average inter-particle distance $d=n^{-\frac{1}{3}}$ and $\zeta(\frac{3}{2})^\frac{1}{3} \sim 1$ we arrive at,
\begin{equation}
\lambda_{\rm dB} \gappeq d.
\end{equation}
Thus, Bose-Einstein condensation coincides with the condition that the particle waves overlap with each other, as depicted in Fig. \ref{fig:bec_schem2}. The individual particles become smeared out into one giant wave of matter, the condensate.

\subsection{Internal energy}

The internal energy of the gas $U$ is determined by the excited states only,
since the ground state possesses zero energy\index{energy!internal}; therefore
we can express $U$ by integrating across the excited state particles as,
\begin{equation}
U=\int \limits_0^\infty E~ N_{\rm ex}(E) ~{\rm d} E.
\label{eqn:U}
\end{equation}
%
%
Upon evaluating this integral below and above $T_{\rm c}$ we find,
\begin{eqnarray}
\displaystyle 
U = 
\begin{cases} \dfrac{3}{2}\dfrac{\zeta(5/2)}{\zeta(3/2)}N k_{\rm B} T \left(\dfrac{T}{T_{\rm c}} \right)^{3/2} & \text{for $T< T_{\rm c}$,}
\\[1em]
 \dfrac{3}{2} N k_{\rm B} T & \text{for $T \gg T_{\rm c}$.}
\end{cases}
\label{eqn:U_cases}
\end{eqnarray}
 The $T \gg T_{\rm c}$ result is consistent with the classical equipartition theorem for an ideal gas, which states that each particle has on average $\frac{1}{2}k_{\rm B} T$ of kinetic energy per direction of motion.  The different behavior for $T < T_{\rm c}$ confirms the presence of a distinct state of matter.
 
 \subsection{Pressure}
 
 The pressure of an ideal gas\index{pressure!ideal gas} is $P=2U/3\mathcal{V}$. From Eq. (\ref{eqn:U_cases}), then for $T\gg T_{\rm c}$ we recover the standard result for a classical ideal gas that $P \propto T / \mathcal{V}$. For $T < T_{\rm c}$, and recalling that $T_{\rm c} \propto 1/\mathcal{V}^{2/3}$, we find that $P \propto T^{5/2}$.  The pressure of the condensate is zero at absolute zero and does not depend on the volume of the box!  A consequence of this is that the condensate has infinite compressibility, as explored in Problem \ref{prob:compressibility}.

\subsection{Heat capacity}
\begin{figure}[t]
\centering
~~(a)~~~~~~~~~~~~~~~~~~~~~~~~~~~~~~~~~~~~~~~~~~~~~~~~~~~~~~~~~~~(b)~~~~~~~~~~~~~~~~~~~~~~~~~~~~~~~~~~~~~~~~~~~~~~~~~~~~~~~~~~
\\
\includegraphics[width=0.9\columnwidth,clip=true]{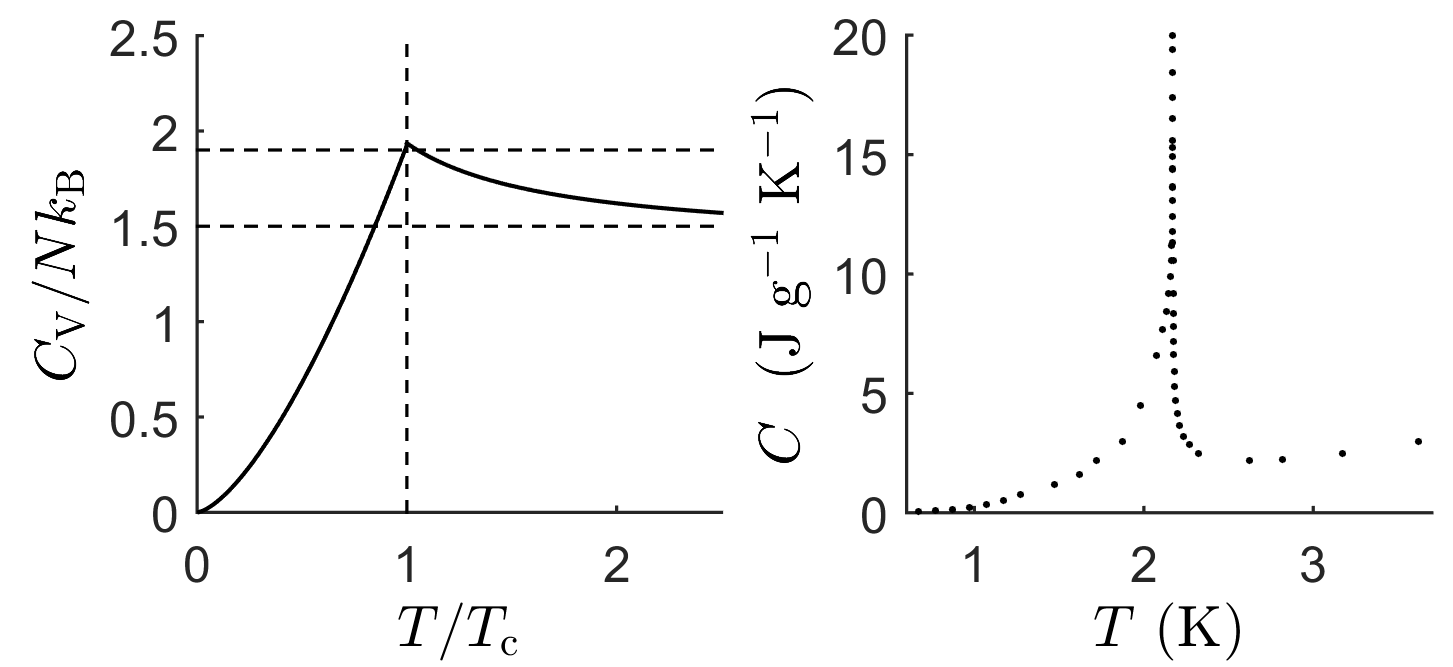}
\caption{(a) Heat capacity $C_{\rm \mathcal{V}}$ of the ideal Bose gas as a function of temperature $T$. (b) Experimental heat capacity data of liquid Helium, taken from \cite{buckingham}, about the $\lambda$-point of 2.2 K.  Both curves show a similar cusped structure.}
\label{fig:specific_heat}
\end{figure}

The heat capacity of a substance is the energy required to raise its temperature by unit amount. At constant volume it is defined as,\index{heat capacity}
\begin{equation}
C_{\rm \mathcal{V}}=\left(\frac{\partial U}{\partial T}\right)_{\rm \mathcal{V}}.
\end{equation}
From Eq. (\ref{eqn:U_cases}) we find,
\begin{eqnarray}
C_{\rm \mathcal{V}} = 
\begin{cases} 1.93 N k_{\rm B} T^{3/2} & \text{for $T< T_{\rm c}$,}
\\[.2em]
\dfrac{3}{2}N k_{\rm B} &\text{for $T\gg T_{\rm c}$.}
\end{cases}
\label{eqn:heat_capacity}
\end{eqnarray}
A more precise treatment, describing the dependence at intermediate temperatures, can be found in Ref. \cite{pitaevskii_2003}.  The form of $C_{\rm \mathcal{V}}(T)$ is depicted in Fig. \ref{fig:specific_heat}, showing a cusp-like dependence around $T_{\rm c}$. In general, discontinuities in the gradient of $C_{\rm \mathcal{V}}(T)$ are signatures of phase transitions between distinct states of matter.   The similarity of this prediction to measured heat capacity curves for Helium about the $\lambda$-point was key evidence in linking helium II to Bose-Einstein condensation. 

\subsection{Ideal Bose gas in a harmonic trap}
\subsubsection{Critical temperature and condensate fraction}
\label{sec:harm_crit}
\begin{figure}[t]
\centering
\includegraphics[width=0.5\columnwidth]{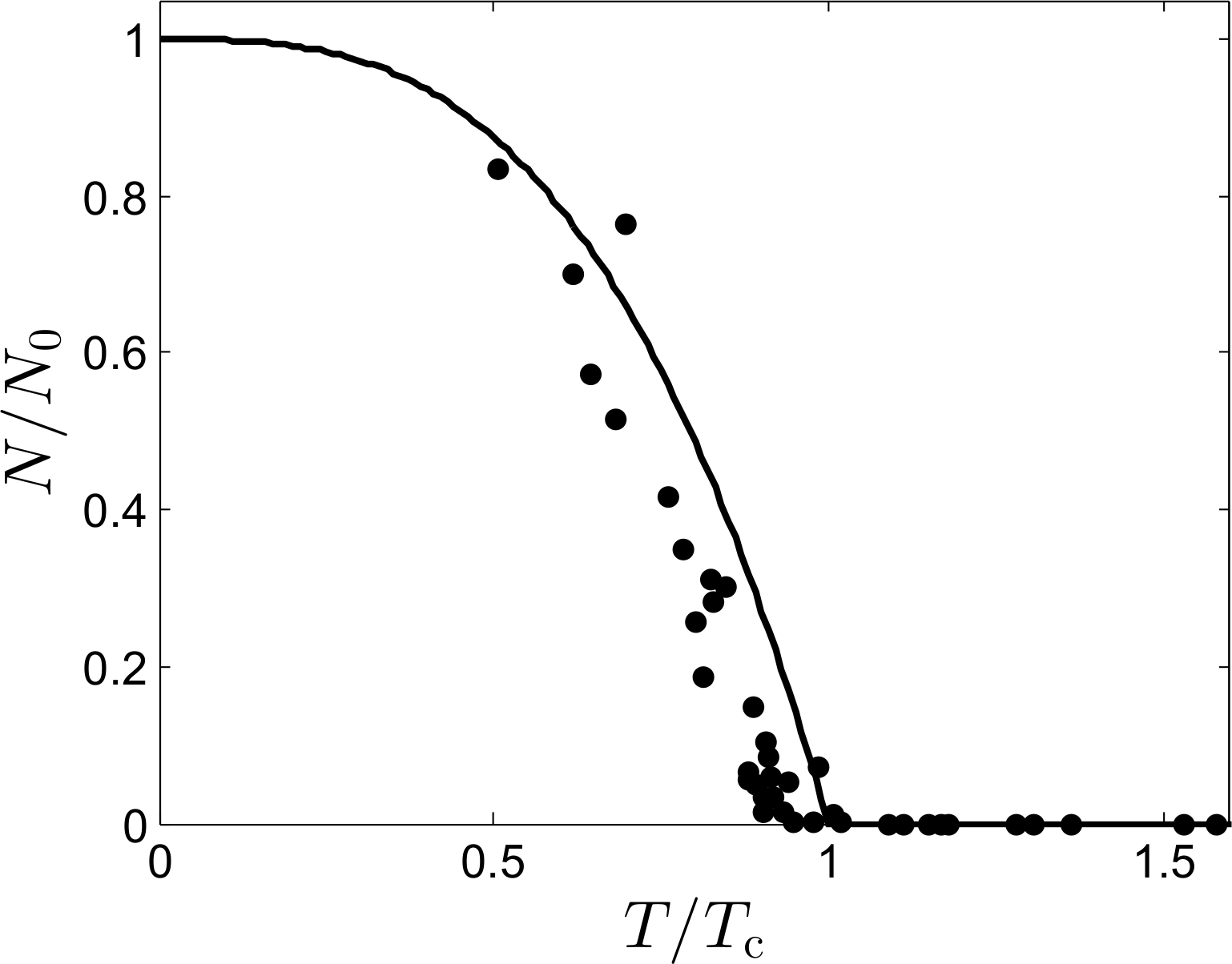}
\label{fig:BEC_in_trap}
\caption{Variation of condensate fraction $N_0/N$ with temperature for a harmonically-trapped BEC, with the ideal-gas predictions (solid line) compared to experimental measurements from Ref. \cite{ensher_1996} (circles), with $T_{\rm c} = 280$ nK.}
\label{fig:BEC_in_trap2}
\end{figure}

In typical experiments, atomic Bose-Einstein condensates 
are confined by harmonic 
(quadratic) potentials, rather than boxes
\footnote{Box-like traps \cite{Gaunt-2013,Chomaz_2015} are also possible,
and allow the condensate to have
uniform density, facilitating comparison with the theory
of homogeneous condensates.}, with the general form, \index{trap!harmonic trap}
\begin{equation}
V(x,y,z)=\frac{1}{2}m\left(\omega_x^2 x^2 + \omega_y^2 y^2 + \omega_z^2 z^2 \right),
\label{eqn:harmonic}
\end{equation}
where $m$ is the atomic mass, and $\omega_x$, $\omega_y$ and $\omega_z$ are trap frequencies \index{trap!trap frequencies} which characterise the strength of the trap in each direction.  Here the density of states is modified, being $g(E)=E^2/(2\hbar^3 \omega_x \omega_y \omega_z)$ in 3D.  This leads, for example, to a critical temperature of the form,
\begin{equation}
T_{\rm c}  = \frac{\hbar}{k_{\rm B}} (\omega_x \omega_y \omega_z)^{1/3} \left[\frac{N}{\zeta(3)} \right]^{1/3},
\label{eqn:harmonic_Tc}
\end{equation}
and for the condensate fraction\index{condensate fraction} to vary with temperature as,
\begin{equation}
\frac{N_0}{N}=1-\left(\frac{T}{T_{\rm c}}\right)^3.
\label{eqn:harmonic_N0}
\end{equation}
These predictions agree well with experimental measurements of harmonically-trapped atomic BECs, as seen in Fig. \ref{fig:BEC_in_trap2}.  This is despite the fact that atomic BECs are not {\em ideal} but feature significant interactions between atoms.

\subsubsection{Density profile}
We can deduce the density profile of the  (non-interacting) condensate in a harmonic trap as follows.  
The ground quantum state in a harmonic trap is the ground harmonic oscillator state\index{harmonic!oscillator state}.  For simplicity, assume a spherically-symmetric trap with  $\omega_x=\omega_y=\omega_z\equiv \omega_r$.  The ground quantum state for a single particle is provided by solving the time-independent Schr\"odinger equation under this harmonic potential, giving the ground harmonic oscillator wavefunction $\psi(r) = \left(\frac{m \omega}{\pi \hbar}\right)^{3/4} e^{-m \omega r^2 / 2 \hbar}$.  The quantity
$|\psi(r)|^2$ represents the probability of finding the particle at position $r$. For a condensate of $N_0$ such particles, with $N_0 \gg 1$, the particle density profile will follow as,
\begin{equation}
n(r) = N_0 |\psi|^2= N_0 (\hbar \ell_r^2)^{-3/2} e^{-r^2 / \ell_r^2},
\end{equation}
where we have introduced the harmonic oscillator length $\ell_r = \hbar/m \omega_r$ \index{harmonic! oscillator length} which characterises the width of the density distribution.  

We can also deduce the density profile of the thermal gas.  Taking the classical limit, the atoms will be distributed over energy according to the Boltzmann distribution $N(E) \propto e^{-E/k_{\rm B} T}$.   The trapping potential $V(r)$ allows us to map energy (potential) to position, leading to a spatial particle distribution, 
\begin{equation}
n(r) = N_{\rm ex} (\hbar \ell_{r,\rm th}^2)^{-3/2} e^{-r^2/\ell_{r,\rm th}^2},
\end{equation}\index{thermal gas}
where $\ell_{r, \rm th}=\sqrt{2k_{\rm B} T/m \omega_r^2}$ characterises the width of the thermal gas and the profile has been normalized to $N_{\rm ex}$ atoms.  For increased temperature, the atoms have higher average energy and climb further up the trap walls, leading to a wider profile.  While the profiles of the ideal condensate and ideal thermal gas are both Gaussian in space, their widths have different functional forms.  In particular, the width of the thermal gas depends on temperature, whereas the condensate width does not.

The typical experimental protocol to form a BEC proceeds by cooling a relatively warm gas towards absolute zero.  Above $T_{\rm c}$ the gas has a broad thermal distribution, which shrinks during cooling.  As $T_{\rm c}$ is under-passed, the condensate distribution forms.  In typical atomic BEC experiments, $\ell_r \ll \ell_{r, \rm th}$, such that this is distinctly narrower than the thermal gas, and the combined density profile is bimodal.  Under further cooling, the condensate profile grows (with fixed width) at the expense of the thermal profile, and for $T \ll T_{\rm c}$ the thermal gas is negligible.  In reality, atomic interactions modify the precise shapes of the density profiles but this picture {\it qualitatively} describes what is observed in experiments (see Figs. \ref{fig:BEC2} and \ref{fig:dfg}).  \index{Bose-Einstein condensate!atomic}

\section{Ideal Fermi gas}
We outline the corresponding behaviour of the ideal Fermi gas\index{Fermi gas!ideal Fermi gas}.  Since (identical) fermions are restricted to up to one per state, Bose-Einstein condensation is prohibited, and the Fermi gas behaves very differently as $T \rightarrow 0$.  At $T=0$ the Fermi-Dirac distribution (\ref{eqn:FD_dist}) reduces to a step function, 
\begin{eqnarray}
f_{\rm FD}(E)=
\begin{cases} 1 & \text{for $E\leq E_{\rm F}$,}
\\
0 &\text{for $E > E_{\rm F}$.}
\end{cases}
\end{eqnarray}
All states are occupied up to an energy threshold $E_{\rm F}$, termed the {\em Fermi energy} (equal to the $T=0$ chemical potential). \index{energy!Fermi}   With this simplified distribution it is straightforward to integrate the number of particles,
\begin{equation}
N=\int N(E) ~{\rm d}E=\int \limits_0^{E_{\rm F}} g(E)~f_{\rm FD}(E)~{\rm d}E = \frac{4 \pi \mathcal{V}}{3}\left(\frac{2 m E_{\rm F}}{h^2} \right)^{3/2},
\end{equation}
where we have used the density of states (\ref{eqn:e_dos}).    Note that the continuum approximation $N=\int g(E)~N(E)~{\rm d}E$ holds for $N \gg 1$ fermions since the unit occupation of the ground state  is always negligible.  Rearranging for the Fermi energy in terms of the particle density $n=N/\mathcal{V}$ gives, 
\begin{equation}
E_{\rm F} = \frac{\hbar^2}{2m} \left(6 \pi^2 n \right)^{2/3}.
\end{equation}
From this we define the Fermi momentum $p_{\rm F}=\hbar k_{\rm F}$ where $k_{\rm F}=(6 \pi^2 n)^{1/3}$ is the Fermi wavenumber.  In momentum space, all states are occupied up to momentum $p_{\rm F}$, termed the {\it Fermi sphere}. 

Similarly, the total energy of the gas at $T=0$ is,\index{energy!internal}
\begin{equation}
U = \int N(E) E ~{\rm d}E = \frac{4 \pi \mathcal{V}}{5}\left(\frac{2 m}{h^2} \right)^{3/2} E_{\rm F}^{5/2} = \frac{3}{5} N E_{\rm F}.
\end{equation}
From the pressure relation for an ideal gas, $P=2U/3\mathcal{V}$, the pressure of the ideal Fermi gas at $T=0$ is,
\begin{equation}
P=\frac{2}{5}n E_{\rm F}.
\end{equation}
This pressure is finite even at $T=0$, unlike the Bose and classical gases, and does not arise from thermal agitation.  Instead it is due to the stacking up of particles in energy levels, as constrained by the quantum rules for fermions.  This {\em degeneracy pressure} prevents very dense stars, such as neutron stars, from collapsing under their own gravitational fields.\index{pressure!degeneracy pressure}

\begin{figure}[t]
\centering
\includegraphics[width=0.75\columnwidth]{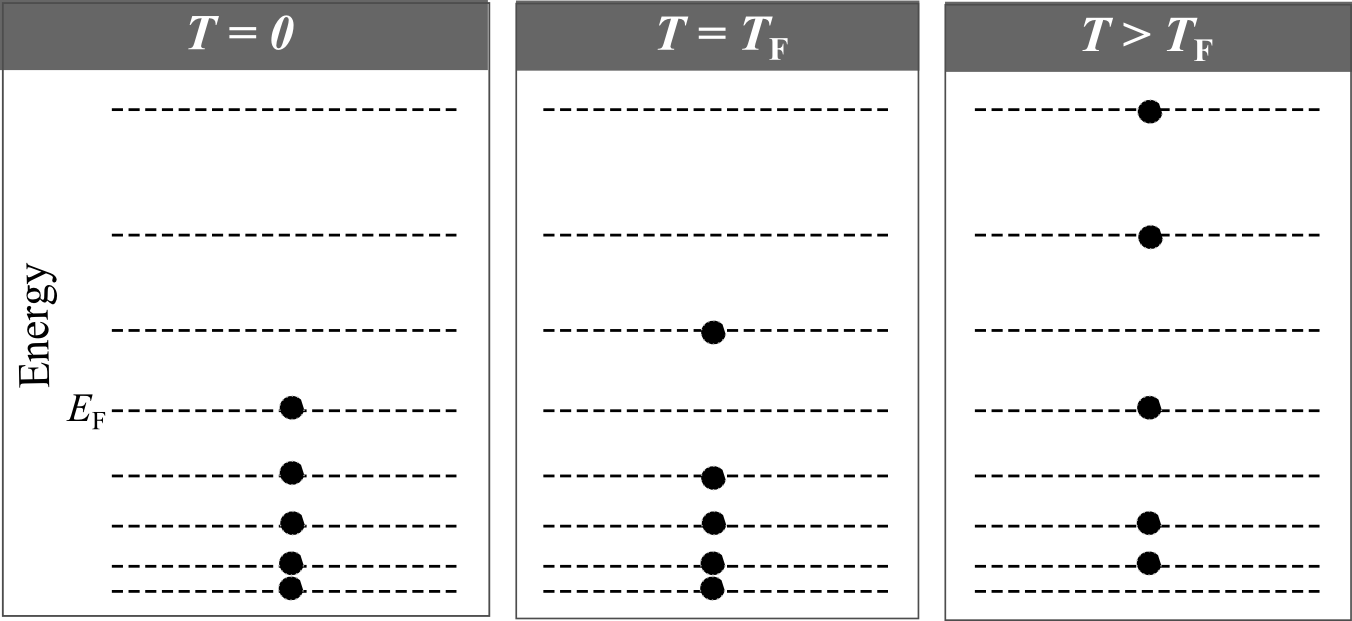}
\caption{Energy level occupations for an ideal Fermi gas.  At $T =0$, there is unit occupation of states up to the Fermi energy.  At $T=T_{\rm F}$, there is some excitation of states around $E=E_{\rm F}$.  For $T \gg T_{\rm F}$, the system approaches the classical limit, with particles occupying many high-energy states. }
\label{fig:fermi_growth}
\end{figure}

As temperature is increased from zero, the step-like Fermi-Dirac distribution becomes broadened about $E=E_{\rm F}$, representing that some high energy particles become excited to energies exceeding $E_{\rm F}$.  It is useful to define the {\em Fermi temperature} $T_{\rm F} = E_{\rm F}/k_{\rm B}$.  \index{temperature!Fermi}At low temperatures $T \sim T_{\rm F}$, only particles in states close to $E_{\rm F}$ can be excited out of the Fermi sphere, and the system is still dominated by the stacking of particles.  For high temperatures $T \gg T_{\rm F}$, there is significant excitation of most particles, thermal effects dominates, and the system approaches the classical Boltzmann result.  The Fermi temperature is associated with the onset of degeneracy, i.e. when quantum effects dominate the system.   These regimes are depicted in Fig. \ref{fig:fermi_growth}.  

Now consider the Fermi gas to be confined in a harmonic trap.  For $T \gg T_{\rm F}$ the gas will have a broad, classical profile.  As $T$ is decreased, the profile will narrow but eventually saturates below $T_{\rm F}$ due to degeneracy pressure.  The width of the Fermi gas at zero temperature is proportional to $N^{1/6} \ell_r$ \cite{butts_1997}, such that, for $N \gg 1$, this cloud is much wider than its classical and Bose counterparts.  This picture is confirmed by the experimental images in Fig. \ref{fig:dfg}.

\section{Summary}
In his 1925 prediction of Bose-Einstein condensation of an ideal gas, Einstein suggested hydrogen, helium and the electron gas were the best candidates for observing Bose-Einstein condensation.  However, the former candidates are no longer gases at the required densities, and the latter (as soon realized) is fermionic.   For over a decade, Bose-Einstein condensation had ``the reputation of having only a purely imaginary character'' \cite{london_1938}, deemed too fragile to occur in real gases with their finite size and particle interactions.  In 1938 Einstein's idea became revived when Fritz London recognized the similarity to the heat capacity curves in Helium  as it entered the superfluid phase.  It took several more decades to cement this link with microscopic theory.  Bose-Einstein condensation is now know to underly superfluid He$^4$ and He$^3$, superconductors and the ultracold atomic Bose gases.  We explore the latter in the next chapter.

\section*{Problems}
\addcontentsline{toc}{section}{Problems}
\begin{prob}
\label{boltzmann1} Consider a system with 6 classical particles, total energy of $6 \epsilon$, and 7 cells with energies $0, \epsilon, 2 \epsilon, 3 \epsilon, 4 \epsilon, 5\epsilon$ and $6 \epsilon$.  Complete the table below by entering the cell populations for each macrostate, the statistical weighting for each macrostate $W$, and the average population per cell $\bar{N}(E)$ (averaged over macrostates).  What is the most probable macrostate?  Plot $\bar{N}(E)$ versus $E$.  It should be evident that the average distribution approximates the Boltzmann distribution, despite the small number of particles.

\small
\setlength\extrarowheight{2pt}
\centering
\begin{tabular}{ x{3.7cm} | x{1cm} | x{1cm} | x{1cm} | x{1.2cm}}
& \multicolumn{3}{ c |}{Macrostates} & \\
\hline
   Cell energy $E$ & 1 & $\hdots$ & 11 & $\bar{N}(E)$ \\  
   \hline   
   6$\epsilon$ & ? & $\hdots$ & ? &  ? \\
   5$\epsilon$ & ? & $\hdots$  & ? & ? \\
   $\vdots$ & $\vdots$ & $\hdots$ & $\vdots$ &  $\vdots$ \\
   $\epsilon$ & ? & $\hdots$ & ? & ?  \\
   0 & ? & $\hdots$ &  ? & ? \\
   \hline  
   Statistical weighting $W$ & ? & $\hdots$ &  ? & \\
\hline
\end{tabular}
\normalsize

\end{prob}

\begin{prob}
\label{boltzmann2}  Consider a system with $N$ classical particles distributed over 3 cells (labelled $1, 2,$ and $3$) of energy $0$, $\epsilon$ and $2\epsilon$.  The total energy is $E=0.5 N \epsilon$.
\begin{itemize}
\item[(a)] ~Obtain an expression for the number of microstates in terms of $N$ and $N_3$, the population of cell 3.  
\item[(b)] ~Plot the number of microstates as a function of $N_2$ (which parameterises the macrostate) for $N=50$. Repeat for $N=100$ and $500$.  Note how the distribution changes with $N$.  What form do you expect the distribution to tend towards as $N$ is increased to much larger values?
\end{itemize}

%
%
%
%
%
%
%
\end{prob}

\begin{prob}
\label{2dbec}
Consider an ideal gas of bosons in two dimensions, confined within a two-dimensional box of volume $\mathcal{V}_{\rm 2D}$.
\begin{itemize}
\item[(a)] ~Derive the density of states $g(E)$ for this two-dimensional system. 
\item[(b)] ~Using this result show that the number of particles can be expressed as,
\begin{equation}
N_{\rm ex}=\frac{2\pi m \mathcal{V}_{2D} k_{\rm B}T}{h^2}\int^\infty_0 \frac{ze^{-x}}{1-ze^{-x}}dx, \nonumber
\end{equation}
where $z=e^{\mu/k_{\rm B}T}$ and $x=E/k_{\rm B}T$.  Solve this integral using the substitution $y=z e^{-x}$.  
\item[(c)] ~Obtain an expression for the chemical potential $\mu$ and thereby show that Bose-Einstein condensation is possible only at $T=0$.
\end{itemize}
\end{prob}

%
%

\begin{prob}
\label{energy}
Equation (\ref{eqn:U_cases}) summarizes how the internal energy of the boxed 3D ideal Bose gas scales with temperature.  Derive the full expressions for the internal energy for the two regimes (a) $T<T_{\rm c}$ (for which $z=1$), and (b) $T\gg T_c$ (for which $z\ll 1$).  Extend your results to derive the expressions for the heat capacity given in Eq. (\ref{eqn:heat_capacity}).
\end{prob}

\begin{prob}
\label{trapped}
Bose-Einstein condensates are typically confined in harmonic trapping potentials, as given by Eq. (\ref{eqn:harmonic}).  Using the corresponding density of states provided in Section \ref{sec:harm_crit}:
\begin{itemize}
\item[(a)] ~Derive the expression for the critical number of particles.
\item[(b)] ~Derive the expression (\ref{eqn:harmonic_Tc}) for the critical temperature.
\item[(c)]  ~Determine the expression (\ref{eqn:harmonic_N0}) for the variation of condensate fraction $N_0/N$ with $T/T_{\rm c}$.
\item[(d)] ~In one of the first BEC experiments, a gas of $40,000$ Rubidium-87 atoms (atomic mass $1.45 \times 10^{-25}$ kg) underwent Bose-Einstein condensation at a temperature of $280$ nK.  The harmonic trap was spherically-symmetric with with $\omega_r=1130$ Hz. Calculate the critical temperature according to the ideal Bose gas prediction.  How does this compare to the result for the boxed gas (you may assume the atomic density as $2.5 \times 10^{18}$ m$^{-3}$).
\end{itemize}
\end{prob}


%

\begin{prob}
\label{prob:compressibility}
 The compressibility $\beta$ of a gas, a measure of how much it shrinks in response to a compressional force, is defined as, 
\begin{equation}
\beta=-\frac{1}{\mathcal{V}}\frac{\partial \mathcal{V}}{\partial P}. \nonumber
\end{equation}
Determine the compressibility of the ideal gas for $T<T_{\rm c}$.  

{\it Hint}: Since $T_{\rm c}$ is a function of $\mathcal{V}$, you should ensure the full $\mathcal{V}$-dependence is present before differentiating.

%
%

\end{prob}

\chapter{Gross-Pitaevskii model of the condensate}
\label{gpe} 


\abstract{The Gross-Pitaevskii equation (GPE)
is a successful and well-established model for describing an atomic 
Bose-Einstein condensate.  Here we introduce 
this model, along with its assumptions.  Throughout the rest of 
this chapter we explore its properties and key time-independent solutions.}

\section{The Gross-Pitaevskii Equation}
\label{sec:gpe}

We  assume that the gas is at zero temperature, such that the thermal gas and thermally-driven excitations of the condensate are non-existent.  This is valid for $T \ll T_{\rm c}$, which is often satisfied in BEC experiments.  In any real gas, the particles also interact with each other, deviating from the ideal gas predictions of Chapter 2. Particle interactions amplify the fluctuations in any quantum field (so-called ``quantum fluctuations''); these excite particles out of the ground state and deplete the condensate.  An exact description of $N$ interacting quantum particles would proceed by parameterising the system by an $N$-body wavefunction, ${\bf \Psi}({\bf r}_1, {\bf r}_2,...{\bf r}_N, t)$, which obeys the many-body Schr\"odinger equation.  However, the complexity of this approach makes it intractable for modelling more than a few particles, let alone the thousands or millions 
typical of an atomic BEC.  

Fortunately, the interactions in atomic BECs are weak; this is due 
to their extreme diluteness and the weak forces between 
neutral atoms.  As such, quantum fluctuations have a weak effect 
on the condensate, and will be ignored.  Then, and assuming a large number 
of particles ($N \gg 1$), the many-body wavefunction
can be approximated
by an effective single-particle wavefunction, $\Psi({\bf r},t)$.
Given the physical picture of the condensate as a giant matter wave 
(see Section \ref{sec:overlap}), it is natural to describe it 
via a single wavefunction.  This {\em macroscopic wavefunction} 
is a complex field that can be written as,
\begin{equation}
\Psi({\bf r},t)=\sqrt{n({\bf r},t)}\exp\left[i S({\bf r},t) \right],
\end{equation}
where $n$ and $S$ are the density and phase distributions of the condensate, and is normalized to $N$ atoms, i.e.,\index{normalization}
\begin{equation}
\int |\Psi|^2~{\rm d}^3{\bf r}=N.
\label{eqn:norm}
\end{equation}

In the absence of interactions, this wavefunction would be governed by the single-particle Schr\"odinger equation, $i \hbar \partial \Psi/\partial t =[-(\hbar^2/2m)\nabla^2+V({\bf r},t)]\Psi$, where $\nabla^2$ is the Laplacian operator and $V({\bf r},t)$ is the potential acting on the wavefunction (which, in general, may depend on position and time).  However, the governing equation must be modified to account for the interactions between atoms.  The gas is sufficiently dilute that three-body (and higher) interactions are typically negligible.  The dominant interactions are elastic two-body interactions arising from van der Waals forces between the neutral atoms.  For two atoms at positions ${\bf r_1}$ and ${\bf r_2}$ this interaction is well-described by the contact (hard-sphere) interaction,\index{interactions!contact interaction}
\begin{equation}
\mathcal{U}({\bf r_1}-{\bf r_2})=g \delta ({\bf r_1}-{\bf r_2}),
\end{equation}
where $\delta$ is Dirac's delta function, and the coefficient $g$ is given by,
\begin{equation}
g=\frac{4\pi \hbar^2 a_{\rm s}}{m}.
\label{eqn:ints}
\end{equation}
Here $a_{\rm s}$ is the {\em s-wave scattering length}, a quantity used in atomic physics for characterising the interactions of atoms in the low energy limit (for a detailed description see, e.g. Ref. \cite{pethick_2008}).\index{scattering length}  For the two most common BEC atomic species, $^{87}$Rb and $^{23}$Na, $a_{\rm s}=5.8$ and $2.8$nm, respectively.  While the true interaction potential between two atoms is more complicated, 
its detailed shape is unimportant provided that
$a_{\rm s} \ll d$, where $d$ is
the average interparticle distance (or, equivalently, $n a_{\rm s}^3 \ll 1$).  Furthermore, within this picture, the condition for weak interactions is $a_{\rm s} \ll \lambda_{\rm dB}$.

Taking into account these interactions,
 the mean-field wavefunction $\psi({\bf r},t)$ can be shown to satisfy a modified Schr\"odinger equation called the {\em Gross-Pitaevskii equation},\index{Gross-Pitaevskii equation} \index{Gross-Pitaevskii equation!time-dependent} 
\begin{equation}
i \hbar \frac{\partial \Psi}{\partial t}=
-\frac{\hbar^2}{2m}\nabla^2 \Psi
+ V({\bf r},t) \Psi  + g  |\Psi |^2 \Psi.
\label{eqn:gp1}
\end{equation}
\noindent
The formal derivation of the GPE is beyond our scope but can be found in, e.g. \cite{annett_2004,pethick_2008,pitaevskii_2003}. 
The first two terms on the right-hand side are familiar from the Schr\"odinger equation, accounting for kinetic and potential energy.  The cubic term $g \vert \Psi \vert^2 \Psi$ arises from the atomic interactions and makes  the equation {\em nonlinear}. Similar {\em Nonlinear Schr\"odinger Equations (NLSEs)} arise in optics, plasma physics and water waves. In one spatial dimension, the NLSE has special mathematical properties, such as soliton solutions and infinite conservation laws (see Chapter 4).  The physical interpretation of the nonlinear term is that, at a given point in space, there is an energy contribution arising from the 
mean-field interactions of all the atoms in the immediate vicinity.  
The quantity $g$ depends on the given atomic species and can
be positive or negative.  Experimentalists can also control sign and magnitude of $g$  using Feshbach resonances.\index{Feshbach resonance}  Here magnetic fields are used to couple the two-body scattering to a bound state; when this coupling is
close to some resonant magnetic field, huge changes in the two-body scattering properties are possible.  For $g>0$ the interactions are {\em repulsive}, for $g<0$ the interactions are attractive, and for $g=0$ there are no interactions (and the equation reduces to the Schr\"odinger equation).   The case of repulsive interaction is the most
studied, so, unless we explicitly specify the sign of $g$, we take $g>0$ hereafter. 

The GPE can also be extended to take thermal and quantum effects into account, 
and further information can be found in Refs. 
\cite{Proukakis_2008,Blakie_2008,finite_temp_book}.

\subsection{Mass, Energy and Momentum}
The total mass of the condensate is $M=m N$, where $N$ is provided by the normalization condition on $\Psi$, Eq. (\ref{eqn:norm}).  \index{mass}

The energy is,\index{energy!condensate}
\begin{equation}
E=\int \left[ \frac{\hbar^2}{2m}| \nabla \Psi |^2+V |\Psi|^2
+\frac{g}{2}| \Psi |^4 \right] {\rm d}^3 {\bf r} = E_{\rm kin}+E_{\rm pot}+E_{\rm int}.
\label{eqn:energy_int}
\end{equation}
The terms represent (from left to right) 
{\em kinetic energy} $E_{\rm kin}$, {\em potential energy} $E_{\rm pot}$ and {\em interaction energy} $E_{\rm int}$. Providing that the potential $V$ is independent of time, then the energy $E=E_{\rm kin}+E_{\rm pot}+E_{\rm int}$ is conserved
during the time evolution of the condensate.

It can be useful, particularly when determining the energy numerically, to define $\Psi=\Psi_{\rm r}+i\Psi_{\rm i}$, where $\Psi_{\rm r}$ and $\Psi_{\rm i}$ are the real and imaginary parts of the wavefunction.  Then, the $|\nabla \Psi|^2$ term in the energy can be expressed in a more convenient form,  $|\nabla \Psi|^2=(\nabla \Psi_{\rm r})^2+(\nabla \Psi_{\rm i})^2$.

Meanwhile the momentum of the condensate is,\index{momentum!condensate}
\begin{equation}
\mathcal{P}=\frac{i \hbar}{2} \int \left(\Psi \nabla \Psi^* - \Psi^* \nabla \Psi  \right){\rm d}^3{\bf r}.
\end{equation}

\section{Time-independent GPE}

Time-independent solutions of the GPE satisfy,
\begin{equation}
\Psi({\bf r},t)=\psi({\bf r})e^{-i\mu t/\hbar},
\label{eqn:time-indep}
\end{equation}
where $\mu$ is a constant called the {\em chemical potential}.\index{chemical potential}  The exponential term represents the freedom for the phase to freely
 evolve with time, uniformly across the system, while the density $n({\bf r},t)=|\psi({\bf r})|^2$  is unaffected.   Inserting Eq.~(\ref{eqn:time-indep})
into Eq.~(\ref{eqn:gp1}), we obtain the {\em time-independent GPE} for the time-independent wavefunction $\psi({\bf r})$,\index{Gross-Pitaevskii equation!time-independent}
\begin{equation}
\mu \psi=-\frac{\hbar^2}{2m}\nabla^2 \psi + V({\bf r}) \psi + g |\psi|^2 \psi.
\label{eqn:tigp1}
\end{equation}
Note that the potential $V$ must be independent of time here. Solutions of the time-independent GPE are stationary solutions of the system, and the lowest energy solution is the ground state of the BEC.  
$\psi({\bf r})$ is  real for the simple solutions that we discuss in this Chapter.

The chemical potential is the eigenvalue of time-independent GPE, and direct integration leads to the expression,\index{chemical potential}
\begin{equation}
\mu=\frac{1}{N} (E_{\rm kin}+E_{\rm pot}+2 E_{\rm int}).
\label{eqn:mu}
\end{equation}
In the absence of interactions, this reduces to the energy per particle, consistent  with the eigenvalue of the time-independent Schr\"odinger equation.  More generally, the chemical potential is defined as $\mu=\partial E/\partial N$.

\section{Fluid dynamics interpretation}
\label{sec:fluid}

There is a deep link between the GPE and fluid dynamics. 
Indeed, we can picture the condensate as a fluid,
characterised by its density and velocity distributions.
From the earlier relation, $\Psi({\bf r},t)=\sqrt{n({\bf r},t)} e^{i S({\bf r},t)}$ (known in this context as the {\em Madelung transform}) \index{Madelung transform}the number density follows as $n({\bf r},t)=|\Psi({\bf r},t)|^2$. From this relation
 we have also the {\em mass  density}\index{mass}
$\rho({\bf r},t)$ conventionally used in fluid dynamics, $\rho({\bf r},t)=m~n({\bf r},t)$.

The fluid velocity field ${\bf v}({\bf r},t)$ is defined from the phase via,\index{fluid velocity}
\begin{equation}
{\bf v}({\bf r},t)=\frac{\hbar}{m}\nabla S({\bf r},t).
\label{eqn:vel}
\end{equation}

Using the Madelung transform $\Psi=\sqrt{n}e^{iS}$ and the above velocity relation, we find that the energy integral of Eq. (\ref{eqn:energy_int}) can be written as, \index{energy!condensate}
\begin{equation}
E=\int \left[ \frac{\hbar^2}{2m}\left(\nabla \sqrt{n}\right)^2+\frac{mnv^2}{2}+Vn+\frac{gn^2}{2}\right]{\rm d}^3 {\bf r}.
\end{equation}
The first two terms comprise the kinetic energy.  The first of these is the quantum kinetic energy.  It arises due to the zero-point motion of confined particles, and vanishes for a uniform system.  The second term is the conventional kinetic energy associated with the flow of the fluid.

Inserting the Madelung transform into the GPE, and separating real and 
imaginary terms, we obtain two equations. 
The first is the classical 
{\em continuity equation},\index{fluid equations!continuity equation}
\begin{equation}
\dfrac{\partial n}{\partial t} + \nabla \cdot(n {\bf v})=0.
\label{eqn:cont1}
\end{equation}
The continuity equation expresses conservation of the number of atoms (or, when written
in terms of $\rho({\bf r},t)$, conservation of mass).
By integrating the equation over a given volume, we see that,
if the number of atoms changes in that volume, it is because fluid has moved
in or out of it.

The second equation is,
\begin{equation}
\displaystyle m \frac{\partial {\bf v}}{\partial t} =- \nabla \left(\frac{1}{2}mv^2+ V+ gn - \frac{\hbar^2}{2m} \frac{\nabla^2 \sqrt{n}}{\sqrt{n}}\right). \label{eqn:quasiEuler2}
\end{equation}
The $\nabla^2 \sqrt{n} / \sqrt{n}$ term is termed the {\it quantum pressure} term (see below). \index{pressure!quantum pressure}  \index{fluid equations!Euler equation}
With some manipulation, we can write this in the equivalent form,
\begin{equation}
mn \left(
\frac{\partial {\bf v}}{\partial t}+ ({\bf v} \cdot \nabla) {\bf v} 
\right)
=-\nabla (P+P') -n \nabla V,
\label{eqn:quasiEuler1}
\end{equation}
where $P$ and $P'$ are respectively 
the {\em pressure} and the {\em quantum pressure}, \index{pressure} \index{pressure!quantum pressure}
\begin{equation}
P=\frac{g n^2}{2},
\qquad
P'=-\frac{\hbar^2}{4 m} n
\nabla^2 (\ln n).
\label{eqn:pressure}
\end{equation}
Equations (\ref{eqn:cont1}) and (\ref{eqn:quasiEuler2})  (or, equivalently, 
Eqs. (\ref{eqn:cont1}) and (\ref{eqn:quasiEuler1})) are known as the 
{\it superfluid hydrodynamic equations}.  They can also be written in 
index notation\footnote{In index notation, Eqs. (\ref{eqn:cont1}) and 
(\ref{eqn:quasiEuler1}) are 
$\dfrac{\partial n}{\partial t} + \dfrac{\partial (n v_j)}{\partial x_j}=0$ 
and $mn \left(
\dfrac{\partial v_k}{\partial t}+ v_j \dfrac{\partial v_k}{\partial x_j} \right)
=-\dfrac{\partial P}{\partial x_k} - \dfrac{\partial P'_{jk}}{\partial x_j}
-n \dfrac{\partial V}{\partial x_k}$,
where $v_j$ is the $j$th Cartesian component ($j=1,2,3$)
of the velocity $\bf v$, we have
assumed summation over repeated indices, and where the components 
$P'_{jk}$ of the quantum stress tensor $P'$ are 
$P'_{jk}=-\dfrac{\hbar^2}{4 m} n \dfrac{\partial^2 (\ln{n})}{\partial x_j 
\partial x_k}.$}.

Notice that the pressure depends only on the density. This property
makes the condensate a {\em barotropic fluid}; as a consequence,
surfaces of constant
pressure are also surfaces of constant density.  The quantum pressure is a pure quantum effect, and
vanishes if we set Planck's constant equal to zero.
It has the same origin as the quantum kinetic energy, i.e. zero point motion, which creates a pressure that opposes any `squashing' or
`bending' of the condensate.  In a uniform condensate the quantum pressure 
is zero because $n$ is constant.

Equation~(\ref{eqn:quasiEuler1}) is very
similar to the classical Euler equation for an inviscid fluid. 
To understand the relation between
condensates and classical fluids,
we compare the relative importance of pressure and quantum pressure.
Using Eqs.~(\ref{eqn:pressure}), we estimate that the order 
of magnitude of $P$ and $P'$ are respectively 
$P \sim gn^2$ and 
$P' \sim \hbar^2 n/m \xi^2$, where
$\xi$ is the length scale of the variations of $n$.  Then $P'/P \sim \hbar^2/(m n g \xi^2)$, and hence
the quantum pressure becomes negligible ($P' \ll P$) in the limit of length 
scales larger than $\xi$.  If in addition, the trapping potential is 
absent ($V=0$) then Eq.~(\ref{eqn:quasiEuler1}) become negligible, and the 
equation reduces to the classical Euler equation, which describes the motion 
of a classical fluid without viscosity.

The lengthscale in question is provided by the {\em healing length}, defined as,\index{healing length}
\begin{equation}
\xi = \frac{\hbar}{\sqrt{g m n}}.
\label{eqn:healing_eq}
\end{equation}
The typical value of the healing length in atomic BECs is 
$\xi \sim 10^{-6}~\rm m$; 
for superfluid helium ($^4$He) the healing length is
much smaller, $\xi \sim 10^{-10}~\rm m$.


%
%

\section{Stationary solutions in infinite or semi--infinite homogeneous systems}
\label{sec:infinite}

In experiments, atomic condensates are confined by bowl-like trapping 
potentials $V({\bf r})$.  Condensates are therefore small 
(typically of the order of $10^{-5}$ or $10^{-4}~{\rm m}$)
and inhomogeneous (the density depends on the position). 
However, many general properties of atomic condensates can be understood
from the simpler scenario of a homogeneous condensate in an 
infinitely-sized or semi-infinitely-sized system. The homogeneous
condensate is also a useful model of superfluid helium, 
as the sizes of the samples of
$^4$He typically used in experiments range from $10^{-2}$ 
to $10^{-1}~{\rm m}$, many
orders of magnitude larger than the healing length.
A homogeneous condensate would not be stable 
for $g<0$ (as we see later) and so we consider $g>0$ 
for now.

\subsection{Uniform condensate}
For $V=0$ (uniform condensate of infinite extent), the stationary solution 
is uniform, and the time-independent GPE becomes, 
\begin{equation}
\mu \psi= g |\psi|^2 \psi.
\end{equation}

\noindent
The solution is then, 
\begin{equation}
\psi=\psi_0=\sqrt{\mu/g}.
\end{equation}

\noindent
The corresponding number and mass densities are, respectively,
\begin{equation}
n=n_0=| \psi_0|^2=\mu/g,
\qquad
\rho=\rho_0=m \mu/g.
\end{equation}

\subsection{Condensate near a wall}
\label{sec:healing}

Consider a {\em one-dimensional hard wall} defined by,\index{healing profile}
\begin{eqnarray}
V(x) = 
\begin{cases} \infty &{\rm for} ~~x < 0, \nonumber
\\
 0 &{\rm for}~~ x \geq 0. \nonumber
\end{cases}
\end{eqnarray}

\noindent
No atoms exist in the region $x<0$ (since this would require infinite energy), and so 
the boundary condition at $x=0$ is $\psi(0)=0$.
Away from the wall (in the positive $x$ direction) the condensate
must recover
its bulk form, giving the second boundary condition that
$\psi(x) \to \psi_0 =\sqrt{\mu/g}$ for $x \to \infty$.
In the semi-infinite region $x \geq 0$ the one-dimensional (1D)
time-independent GPE is,
\begin{equation}
\mu \psi=-\frac{\hbar^2}{2m}\frac{\partial^2 \psi}{\partial x^2} + g | \psi
|^2 \psi.
\label{eqn:healing_gpe}
\end{equation}
The solution of this equation which satisfies the boundary conditions is,
\begin{equation}
\psi(x)=\psi_0~{\rm tanh} \left(\frac{x}{\xi}\right).
\label{eqn:healing_profile}
\end{equation}
The meaning of the healing length $\xi$ is now
apparent: it is the characteristic minimal distance over which $\psi$ 
changes spatially. \index{healing length} The `healing' profile is supported at a wall 
by the balance between the kinetic energy term in the GPE and 
the interaction term.  Denoting the spatial scale of the variation in 
the wavefunction as $\xi$,  these terms are of the order of 
$\hbar^2/m\xi^2$ and $g n_0$, respectively.  Equating these terms and 
rearranging leads to $\xi=\hbar/\sqrt{m n_0 g}$, the healing length as 
defined in Eq. (\ref{eqn:healing_eq}).  Note that the healing length 
is sometimes defined with a $\sqrt{2}$ in the denominator.

\begin{figure}[t]
\centering
\includegraphics[width=0.65\textwidth]{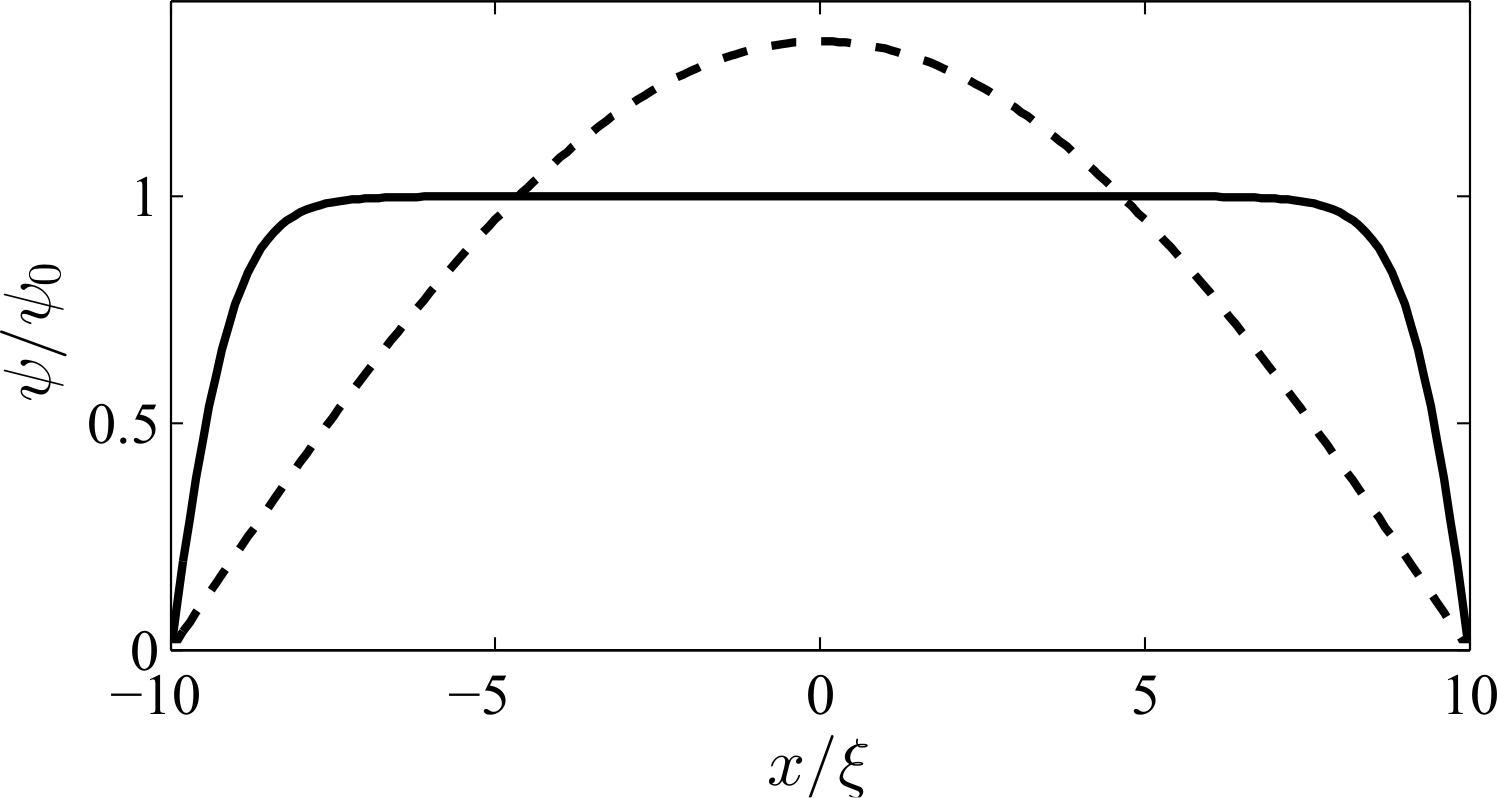}
\caption{Condensate wavefunction $\psi$ (in units of $\psi_0$) as a function of position $x$ (in units of $\xi$) within a 1D infinite square well of width (here with width $20\xi$).  Shown are the profiles for  a non-interacting 
condensate ($g=0$) and a repulsively-interacting ($g>0$) condensate. Note how the wavefunction ``heals' at each boundary according to Eq. (\ref{eqn:healing_profile}), recovering its bulk density at
a distance from the wall of the order of few times the healing length $\xi$. }
\label{fig:healing}
\end{figure}

In an infinite square well of width $L_0$, which is much wider than the healing length ($L_0 \gg \xi$), we then expect the wavefunction to `heal' at each boundary, according to Eq. (\ref{eqn:healing_profile}), and reach the bulk value in the centre of the well.   This is shown in Fig.~\ref{fig:healing}. 

It is interesting to compare this to the case of $g=0$, for which the ground state is given by the well-known solution of the Schr\"odinger equation for a particle in an infinite well,  $\psi \sim \sin(\pi x/L_0)$.   Clearly the interactions between the atoms broaden and flatten the density 
profile by increasing the energetic cost of concentrating
atoms in one place.


\section{Stationary solutions in harmonic potentials}
\label{sec:harmonic}
Atomic condensates are typically confined by harmonic potentials which may, in general, be anisotropic in space.  For simplicity here we start by considering 
a spherically-symmetric harmonic trap, \index{trap!harmonic trap}
\begin{equation}
V({\bf r})=\frac{m}{2}\omega_r^2 r^2,
\end{equation}
\noindent
where $r^2=x^2+y^2+z^2$.
The characteristic length scale of this potential is the {\em harmonic oscillator length}, \index{harmonic! oscillator length}
\begin{equation}
\ell_r=\sqrt{\hbar/m \omega_r}.
\end{equation}

There is no general analytic solution for the ground state (lowest energy)
solution of the BEC in a harmonic trap; usually the ground state
is found by numerically solving
Eq.~(\ref{eqn:tigp1}).
However, there
exist useful analytic results for certain regimes which we describe below.
It is useful to work in terms of the {\em interaction parameter}\footnote{More generally, for an anisotropic harmonic trap, the corresponding interaction parameter is $N a_{\rm s}/\bar{\ell}$, where $\bar{\ell}=\sqrt{\hbar/m \bar{\omega}}$ and $\bar{\omega}=(\omega_x \omega_y \omega_z)^{1/3}$ is the geometric mean of the trap frequencies.}, $N a_{\rm s}/\ell_r$.  Below we distinguish the following cases:
no interactions, strong repulsive interactions
($Na_{\rm s}/\ell_r \gg 1$) and weak interactions ($|N a_{\rm s}/\ell_r | \ll 1$).\index{interactions!interaction parameter}


\subsection{No interactions}

In the absence of atomic interactions ($g=0$) 
the time-independent GPE reduces to the Schr\"odinger equation,

\begin{equation}
\mu \psi=-\frac{\hbar^2}{2m}\nabla^2 \psi + \frac{m \omega_r^2 r^2}{2} \psi.
\label{eqn:gpe-non-interacting}
\end{equation}
The ground state harmonic oscillator solution is well-known to be a 
three-dimensional Gaussian wave function,\index{harmonic!oscillator state}
\begin{equation}
\psi({\bf r})=\frac{N^{1/2}}{\pi^{3/4} \ell_r^{3/2}}
\exp \left(-\frac{r^2}{2 \ell^2}\right).
\label{eqn:psi-non-interacting}
\end{equation}

\noindent
Using Eq. (\ref{eqn:energy_int}), one can show that this  
has the expected 3D harmonic oscillator energy $E=\dfrac{3}{2}N \hbar \omega_r$.

\subsection{Strong repulsive interactions}
\label{sec:strong_ints}

Let the interactions be strongly repulsive, satisfying $Na_{\rm s}/\ell_r \gg 1$.  We expect 
a condensate profile which is significantly broadened and flattened due to the repulsive interactions.  
An analytic solution is found if we 
neglect the $\nabla^2 \psi$-term in the GPE;
this is known as the {\em Thomas-Fermi approximation}. \index{Thomas-Fermi!approximation}
The time-independent GPE simplifies to,

\begin{equation}
\mu \psi=g \vert \psi \vert^2 \psi+V\psi.
\end{equation}

\begin{figure}[t]
\centering
(a)~~~~~~~~~~~~~~~~~~~~~~~~~~~~~~~~~~~~~~~~~~~~~~~~~~~~~~~(b)~~~~~~~~~~~~~~~~~~~~~~~~~~~~~~~~~~~~~~~~
\\
\includegraphics[width=0.5\columnwidth]{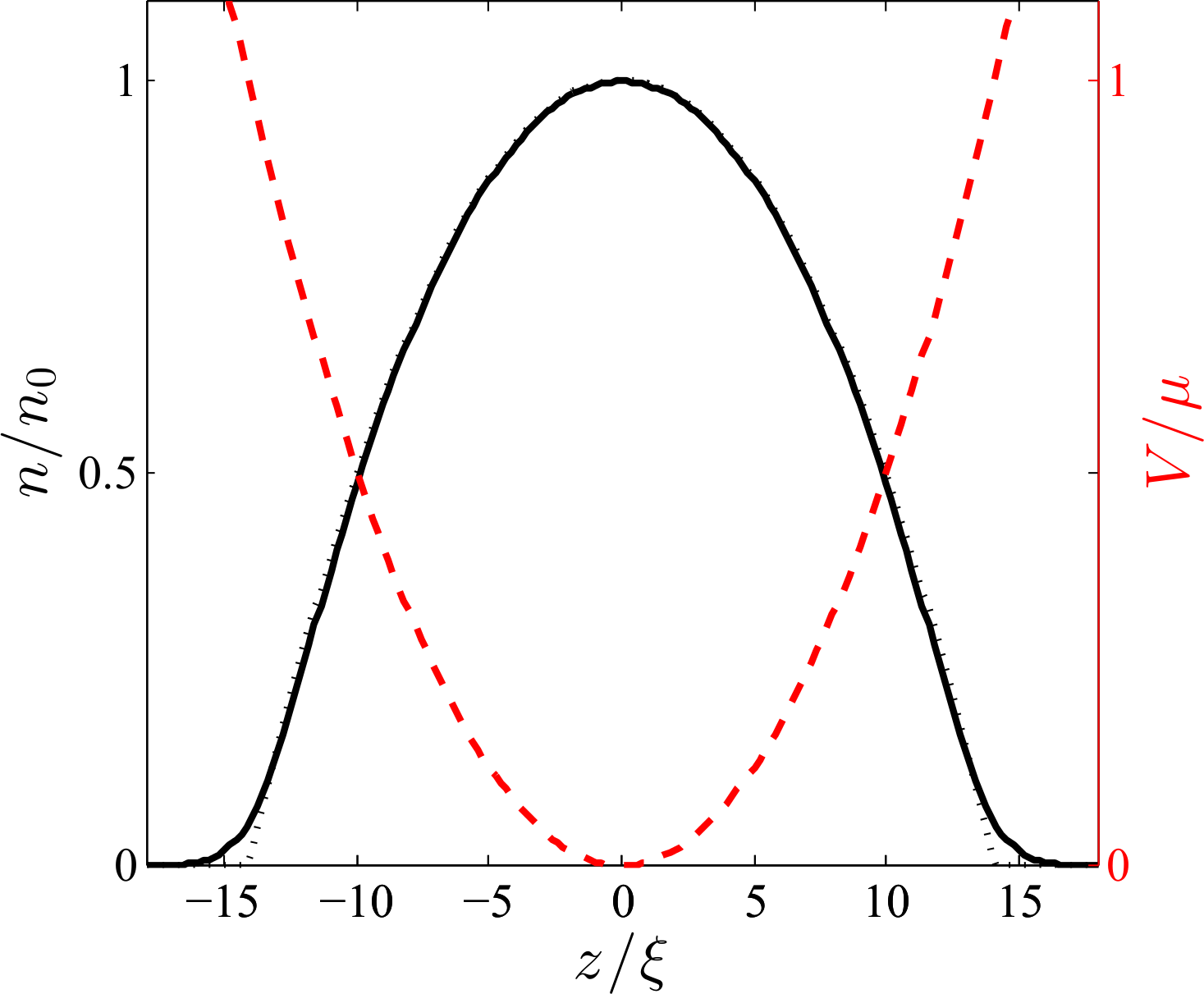}~~~~
\includegraphics[width=0.36\columnwidth]{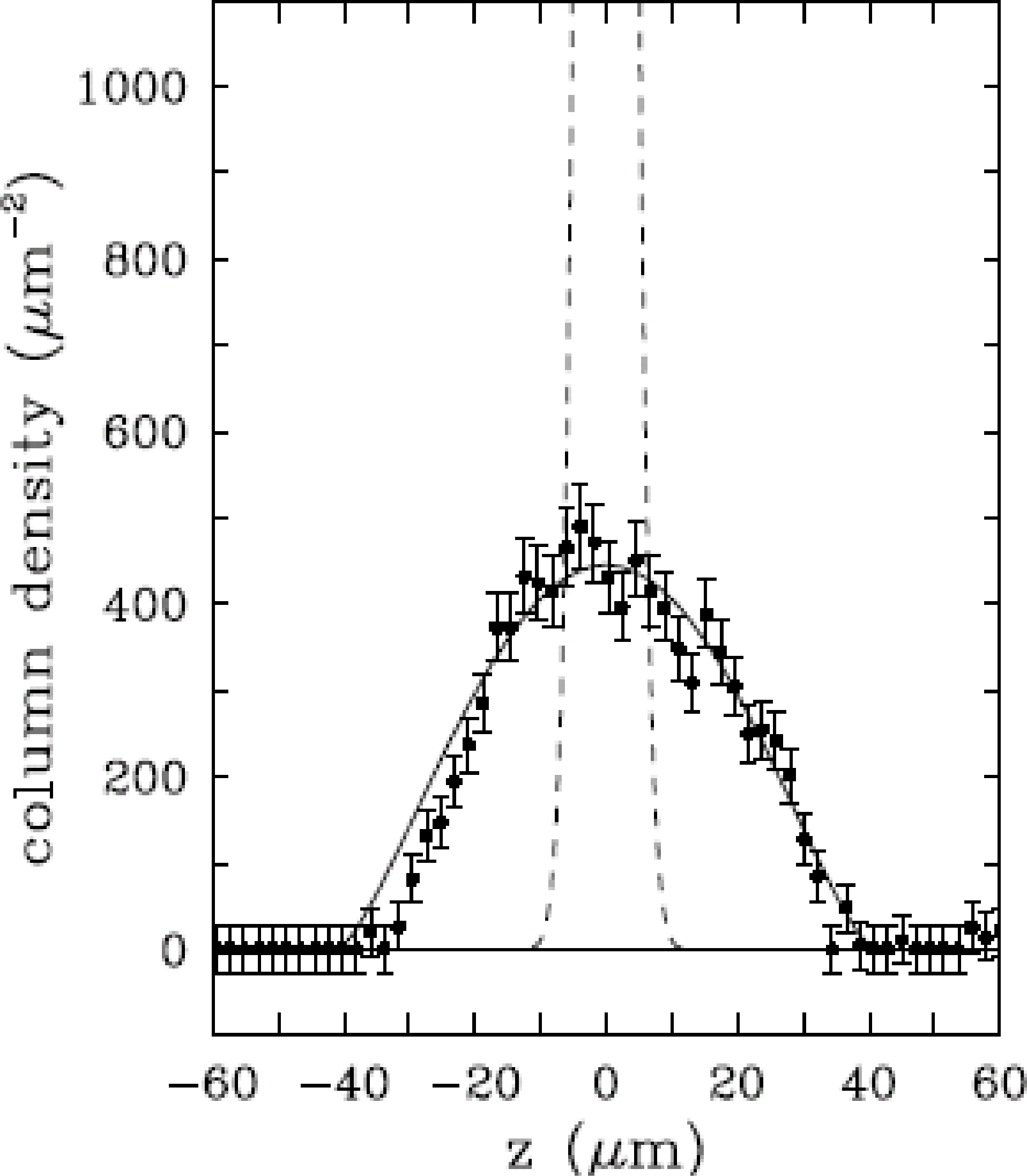}
\caption{(a) Density profile $n(z)$ plotted versus position $z$
(in units of the healing length $\xi$).
The agreement between the
analytic Thomas-Fermi density profile (dotted black line) and the
numerically-determined solution of the GPE (solid black line) 
is so good that the lines 
overlap everywhere but in the tails near $z \approx \pm 15 \xi$.
The harmonic trapping potential $V(z)$ is indicated by the dashed red line. 
(b) An experimental density profile, compared to the Thomas-Fermi
prediction (solid line) and the non-interacting prediction (dashed line).  Reprinted figure with permission from \cite{dalfovo_1999}. Copyright 1999 by the American Physical Society.
}
\label{fig:fermithomas}
\end{figure}

\noindent
Substituting $n=\vert \psi \vert^2$ and $V(r)=\frac{1}{2}m\omega^2 r^2$,
we obtain $n(r)=(2\mu-m \omega_r^2 r^2)/2g$.
Density cannot be negative, so we assume that $n(r)=0$ if
$2 \mu \leq m \omega_r^2 r^2$.
The last equality defines the {\em Thomas-Fermi radius} $R_r$, which satisfies, \index{Thomas-Fermi!radius}

\begin{equation}
\mu=\frac{1}{2}m\omega_r^2 R_r^2.
\end{equation}
We conclude that the Thomas-Fermi density profile is,
\begin{equation}
n(r)=
\left\{
\begin{array}{lr}
\dfrac{\mu}{g}\left(1-\dfrac{r^2}{R_r^2} \right) = \dfrac{m\omega_r^2 (R_r^2 - r^2)}{2g}  &  {\rm ~if~} r \le R_r,\\[.5em]
0                         &  {\rm ~if~} r > R_r,
\end{array}
\right.
\end{equation}

\noindent
and has the shape of an inverted parabola.
Provided that
$Na_{\rm s}/\ell_r \gg 1$, the Thomas-Fermi solution is an excellent
approximation of the solution of the GPE determined numerically, and
compares well with experimental data,
as shown in Fig.~\ref{fig:fermithomas}.  Note, however, the slight deviation from the true numerical solution close to the condensate's edge; 
here the gradient terms, neglected within the Thomas-Fermi model, become significant.

The application of the normalization condition, Eq. (\ref{eqn:norm}), 
to the above solution and manipulation of the resulting expression
leads to useful relations for the chemical potential and the
energy of the condensate in terms of the number of atoms $N$,\index{chemical potential} \index{energy!condensate}
\begin{equation}
\mu = \dfrac{\hbar \omega_r}{2}\left(\dfrac{15 N a_{\rm s}}{\ell_r} \right)^{2/5}, \quad E=\dfrac{5}{7}\mu N.
\label{eqn:tf_quantities}
\end{equation}
The latter is obtained from the relation $\mu=\partial E/\partial N$.  Since $Na_{\rm s}/\ell_r\gg 1$, it is evident that in the Thomas-Fermi regime the chemical potential and energy per particle are considerably greater than the typical trap energy $\hbar \omega_r$.

In the more general case where the harmonic potential is anisotropic in space, $V(x,y,z)=m(\omega_x^2 x^2+\omega_y^2 y^2 + \omega_z^2 z^2)/2$,
the Thomas-Fermi boundary is an ellipsoidal surface satisfying the equation,

\begin{equation}
\frac{x^2}{R_x^2}+\frac{y^2}{R_y^2}+\frac{z^2}{R_z^2}=1,
\end{equation}

\noindent
where the three Thomas-Fermi radii $R_x$, $R_y$ and $R_z$ satisfy,
\begin{equation}
\mu=\frac{1}{2}m\omega_x^2 R_x^2=\frac{1}{2}m\omega_y^2 R_y^2=
\frac{1}{2}m\omega_z^2 R_z^2.
\end{equation}  
In this anisotropic case, it is most convenient to write the density profile as,
\begin{equation}
n(x,y,z)=
\left\{
\begin{array}{lr}
\dfrac{\mu}{g} \left(1-\dfrac{x^2}{R_x^2}-\dfrac{y^2}{R_y^2}-\dfrac{z^2}{R_z^2} \right)  &  {\rm, within~the~ellipsoid},\\
0                         &  {\rm  elsewhere}.
\end{array}
\right.
\label{eqn:3dtf}
\end{equation}
From this the anisotropic versions of the chemical potential and energy, Eqs. (\ref{eqn:tf_quantities}), can be determined.

\subsection{Weak interactions}
\label{sec:weak_ints}

The following variational approach determines an approximate solution of
the time-independent GPE in a harmonic potential when the interactions
(either positive or negative) are weak, that is $\vert Na_{\rm s}/\ell_r \vert  < 1$. \index{variational method}

In the limiting case $g=0$ we know that the exact wavefunction is the Gaussian harmonic oscillator ground state, Eq. (\ref{eqn:psi-non-interacting}). \index{harmonic!oscillator state}
For weak interactions we assume the following trial wavefunction, 
or {\em ansatz},
which is  Gaussian in
shape but has variable width $\sigma \ell_r$, 

\begin{equation}
\psi({\bf r})=\left(\frac{N}{\pi^{3/2} \sigma^{3} \ell_r^{3}}\right) ^{1/2}
\exp\left(-\frac{r^2}{2\sigma^2 \ell_r^2}\right).
\label{eqn:variat}
\end{equation}

\noindent
where $\sigma$ is our {\em variational parameter}.
If $g=0$ then $\sigma=1$, i.e. we recover the exact non-interacting result.

\begin{figure}[b]
\centering
\includegraphics[width=0.55\columnwidth]{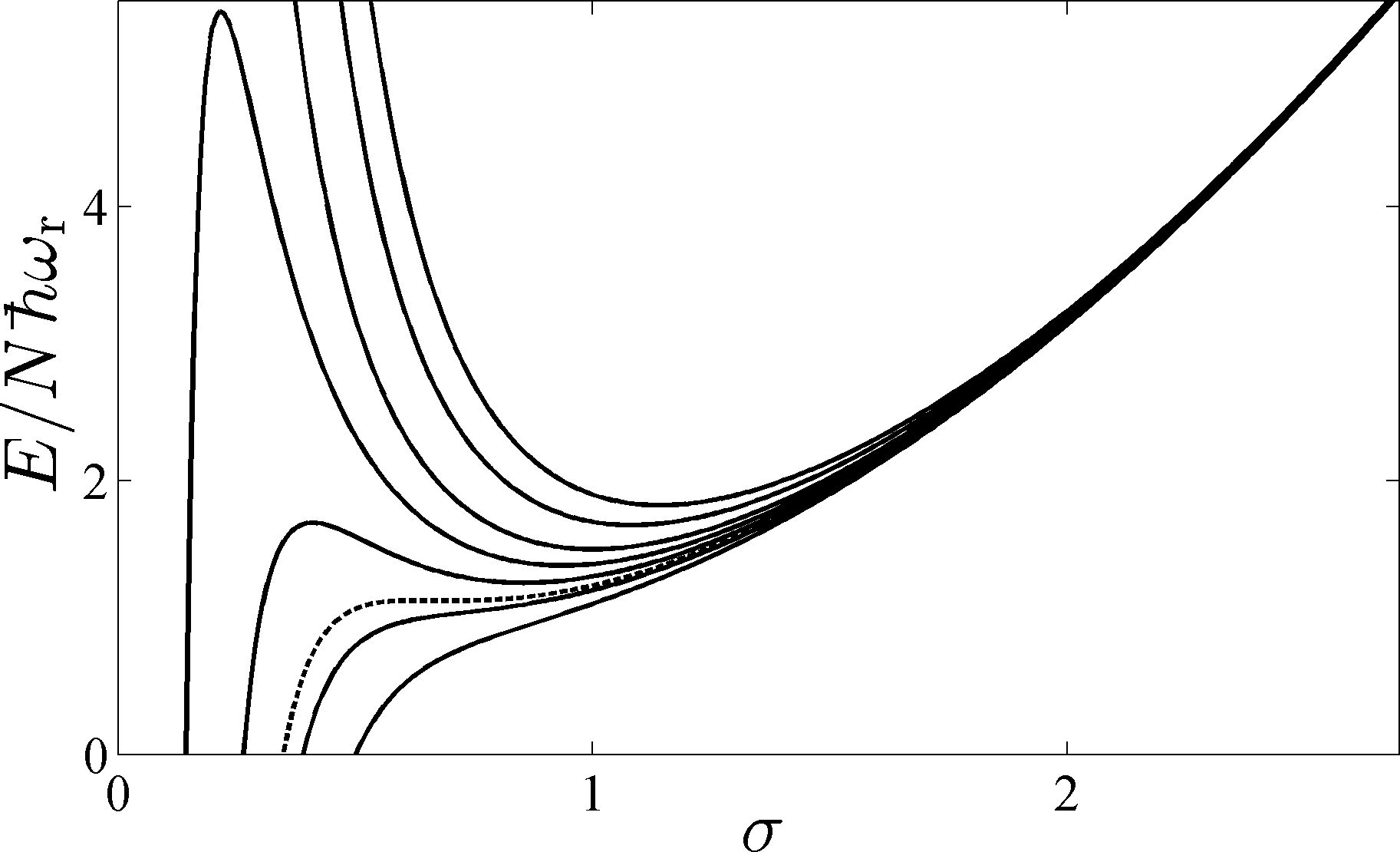}
\caption{Energy $E$ (in units of $N \hbar \omega_r$)
versus $\sigma$ according to
Eq.~(\ref{eqn:variational_energy}) for various values of the interaction parameter 
$N a_{\rm s}/\ell_r$ corresponding, from top to bottom, to $[-1, -0.75, -0.67, -0.5,-0.25, 0, 0.5, 1]$.  $N a_{\rm s}/\ell_r=-0.67$ (dashed line) marks the critical point for the onset of collapse.
} 
\label{fig:variational_3D}
\end{figure}

Using the energy integral (\ref{eqn:energy_int}), the energy
of the ansatz is, 

\begin{equation}
E(\sigma)=\hbar \omega_r N \left[\frac{3}{4\sigma^2}+
\frac{3 \sigma^2}{4}+\frac{1}{\sqrt{2\pi}}\left(\frac{N a_s}{\ell_r}\right)
\frac{1}{\sigma ^3}\right].
\label{eqn:variational_energy}
\end{equation}

From left to right, the terms in the bracket represent kinetic energy,
potential energy and interaction energy.  For a given system
(i.e. for specific values of $N$, $\omega$, $a_{\rm s}$ and $\ell_r$),
Eq.~(\ref{eqn:variational_energy}) tells us how the energy varies
with $\sigma$.  The variational solution is defined as the variational state with the lowest energy, i.e. the minimum of $E(\sigma)$; the corresponding width is denoted
$\sigma_{\rm min}$.  Figure \ref{fig:variational_3D} plots $E(\sigma)$ for various values of the interaction parameter $N a_{\rm s}/\ell$.  The behaviour is different depending on whether the interactions are repulsive or attractive:

\begin{itemize}

\item{} For {\bf repulsive interactions} ($g>0$), $E(\sigma)$ diverges to infinity for both
$\sigma \rightarrow 0$ (due to the positive kinetic and interaction energies) and  $\sigma \rightarrow \infty$ (due to the potential energy), with
a global minimum in-between, corresponding to the variational ground state.  If $g=0$, $\sigma_{\rm min}=1$, corresponding to
the non-interacting Gaussian solution. For increasing $g$,
$\sigma_{\rm min}$ increases, i.e., the condensate becomes wider.

\item{} For {\bf attractive interactions} ($g<0$), 
$E(\sigma)$ now diverges to 
minus infinity as $\sigma \rightarrow 0$. This is due to the dominance of the negative interaction energy in this limit.    The lowest energy solution is thus a wavepacket of zero width, 
i.e. an unstable collapsed state!\footnote{In reality, the BEC does 
not quite collapse to zero width; at high densities, repulsive inter-atomic 
forces kick-in which cause the condensate to then explode outwards, 
an effect
termed the {\em bosenova}.}  
However, for small $|N a_{\rm s}/\ell_r|$, a {\em local minimum} 
exists in $E(\sigma)$ at non-zero width, representing a stable condensate 
of finite size.  For larger $|N a_{\rm s}/\ell_r|$, the local minimum shifts to smaller widths; the attractive interactions cause the condensate to become narrower and more peaked.  However, beyond some critical attractive interactions, the local minimum disappears and no stable 
solutions exist.  In other words, all states collapse to zero width.
The variational method predicts collapse to occur for $N a_{\rm s}/\ell_r\leq -0.67$; this is close to the 
experimentally measured value of $N a_{\rm s}/\ell_r \le -0.64$.  \index{collapse}
This tendency to collapse is the reason why repulsive condensates are more common and why we have
avoided discussing condensates with attractive interactions so far.

\end{itemize}

Note that the above-assumed Gaussian profile is just an approximation.  In the presence of repulsive interactions, the true condensate profile (e.g. as obtained by numerical solution of the GPE) is broader than a Gaussian (becoming more Thomas-Fermi like for increasing repulsive interactions), while for attractive interactions the shape is narrower and more peaked.

\subsection{Anisotropic harmonic potentials and condensates of reduced dimensionality}
\label{sec:reduced_dims}
\begin{figure}[t]
\centering
\includegraphics[width=0.9\columnwidth]{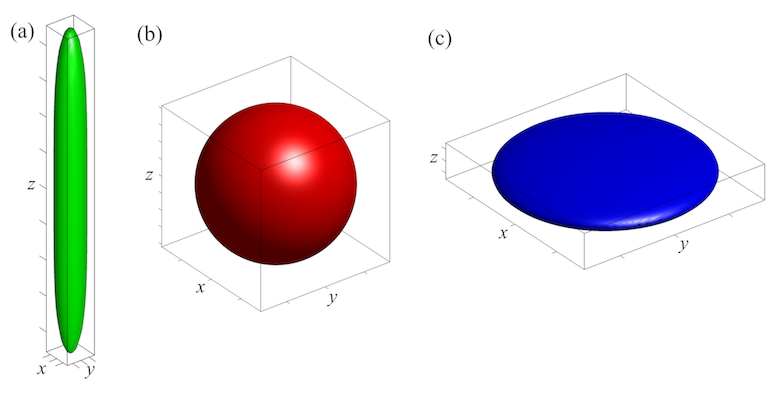}
\caption{The three most common condensate shapes that can be formed 
in an axisymmetric harmonic potential: 
(a) a cigar condensate ($\omega_x, \omega_y > \omega_z$), (b) a spherical condensate ($\omega_x=\omega_y=\omega_z$), and (c) a pancake condensate ($\omega_x, \omega_y < \omega_z$).}
\label{fig:BEC_shape}
\end{figure}

The shape of the condensate is determined by the shape of the trapping
potential.  A spherical harmonic potential induces a
spherical condensate.  It is also common to encounter elongated, or {\em cigar-shaped}, condensates and flattened, or {\em pancake-shaped}, condensates.  The former case is achieved if the condensate is more tightly trapped in two directions, e.g. $\omega_x, \omega_y > \omega_z$, and the latter case, if it is more tightly trapped in one direction, e.g. $\omega_z > \omega_x, \omega_y$.  These shapes are illustrated in Fig. \ref{fig:BEC_shape}. 

By making these trap anisotropies more extreme, it is possible to engineer condensates of reduced dimensionality.  \index{dimensional reduction}Consider first a highly elongated trap ($\omega_x, \omega_y \gg \omega_z$).  If the transverse trapping potential (which is of energy $\hbar (\omega_x \omega_y)^{1/2}$) is much larger than the condensate energy scale (the chemical potential, $\mu$), then excitations of the condensate in the $x$ and $y$ directions are highly suppressed, and the only significant dynamics occur in the $z$ direction.  The system has become effectively one-dimensional.  An effectively two-dimensional condensate can be realized for $\omega_x,\omega_y \ll \omega_z$ and $\hbar \omega_z \gg \mu$.  

In these limits the condensate can be described by suitable one-dimensional and two-dimensional GPEs.  The reduction of the full three-dimensional GPE to these forms is straightforward, as we now outline for a one-dimensional system.  Assuming the above criteria for an effectively one-dimensional condensate, we take the following ansatz for the condensate wavefunction, \index{Bose-Einstein condensate!one-dimensional}
\begin{equation}
\psi(x,y,z,t)=\psi_z(z,t) G_x(x)G_y(y).
\label{eqn:phi}
\end{equation}
In other words, we have decomposed $\psi$ into independent components along $x$, $y$ and $z$.  Under the criterion $\hbar (\omega_x \omega_y)^{1/2} \gg \mu$ then the $x$ and $y$ component will be ``locked'' into the respective ground harmonic oscillator states, which are represented by the Gaussian functions,
\begin{equation}
G_x(x)=\frac{1}{(\pi \ell_x^2)^{1/4}} e^{-x^2/2 \ell_x^2},
\qquad
G_y(y)=\frac{1}{(\pi \ell_y^2)^{1/4}} e^{-y^2/2 \ell_y^2},
\end{equation}
where $\ell_x=\sqrt{\hbar/m \omega_x}$ and $\ell_y=\sqrt{\hbar/m \omega_y}$ denote the harmonic oscillator lengths along $x$ and $y$.  The time-dependence now only appears in the axial wavefunction, $\psi_z$.  Note that $\psi_z$ is normalized to the number of atoms, i.e. $\int |\psi_z|^2~{\rm d}z=N$; as a result the transverse wavefunctions are both normalized to unity (leading to their pre-factors).

To obtain a 1D GPE, one proceeds by inserting the wavefunction ansatz
 (\ref{eqn:phi}) into the 3D GPE and manipulating.  Since,
\begin{equation}
\frac{{\rm d}^2 G_x(x)}{{\rm d}x^2}=
\left(
\frac{x^2}{\ell_x^4}- \frac{1}{\ell_x^2}
\right) G_x(x),
\end{equation}
\noindent
and similarly for $G_y(y)$, each term in the GPE acquires a $G_x(x)G_y(y)$ factor.  To eliminate these factors, one multiplies the equation through by $G_x^* G_y^*$  (where $^*$ denotes complex conjugate) and integrates over all $x$ and $y$.  It is helpful to note that $\int_{-\infty}^{\infty} e^{-x^2}{\rm d}x=\sqrt{\pi}$.  This leads to the following one-dimensional GPE for $\psi_z(z)$,
\begin{equation}
\mu_{\rm 1D} \psi_z=
-\frac{\hbar^2}{2m} \frac{{\rm d}^2 \psi_z}{{\rm d}z^2} 
+g_{\rm 1D} \vert \psi_z \vert^2 \psi_z + \frac{1}{2}m \omega_z^2 z^2 \psi_z.
\label{eqn:reduced-gp}
\end{equation}
\noindent
Here $g_{\rm 1D}$ and $\mu_{\rm 1D}$ are the effective one-dimensional interaction strength and chemical potential, defined as,
\begin{equation}
g_{{\rm 1D}}=\frac{g}{2 \pi \ell_x \ell_y},
\qquad
\mu_{{\rm 1D}}= \mu -\frac{\hbar \omega_x}{2} -\frac{\hbar \omega_y}{2}.
\end{equation}
Note that the trap geometries are often cylindrically symmetric, with $\omega_x=\omega_y$; this symmetry can simplify the integration steps.

Following similar arguments for an effectively two-dimensional condensate, one obtains the effective two-dimensional GPE for the two-dimensional wavefunction $\psi_{xy}(x,y,t)$,
\begin{eqnarray}
\mu_{\rm 2D} \psi_{\perp}&=&
-\frac{\hbar^2}{2m} \left(\frac{{\rm d}^2 \psi_{\perp}}{{\rm d}x^2}+\frac{{\rm d}^2 \psi_{\perp}}{{\rm d}y^2}\right) 
+g_{\rm 2D}  \vert \psi_{\perp} \vert^2  \psi_{\perp}+ \frac{1}{2}m(\omega_x^2 x^2+\omega_y^2 y^2) \psi_{\perp},\nonumber \\
g_{{\rm 2D}}&=&\frac{g}{\sqrt{2 \pi} \ell_z},
\qquad
\mu_{{\rm 2D}}= \mu -\frac{\hbar \omega_z}{2}. \nonumber
\end{eqnarray}
In this case the two-dimensional wavefunction is normalized according 
to $\int |\psi_{\perp}|^2~{\rm d}x{\rm d}y=N$. \index{Bose-Einstein condensate!two-dimensional}

In these one- and two-dimensional cases, the system energy is still described 
according to Eq. (\ref{eqn:energy_int}), with the gradient operator replaced 
by its one- and two-dimensional equivalents, and the integration taken over 
one and two dimensions, respectively.  Moreover, the same analysis techniques 
presented for three-dimensional stationary solutions, e.g. the Thomas-Fermi 
approximation and the Gaussian variational  approach, can be employed.  
In particular, the 1D GPE provides a simplified platform to study many 
generic properties of condensates, and, for example, its stationary solutions 
under hard-wall and periodic boundaries are 
well-established \cite{carr_2000a,carr_2000b}. 
Note, however, that the system stability can be significantly affected 
by the dimensionality of the system, for example, collapse under attractive 
interactions does not occur within the 1D GPE, as will be discussed 
further in Chapter 4.

\section{Imaging and column-integrated density}
\label{sec:image}

\begin{figure}[t]
\centering
\includegraphics[width=0.5\columnwidth]{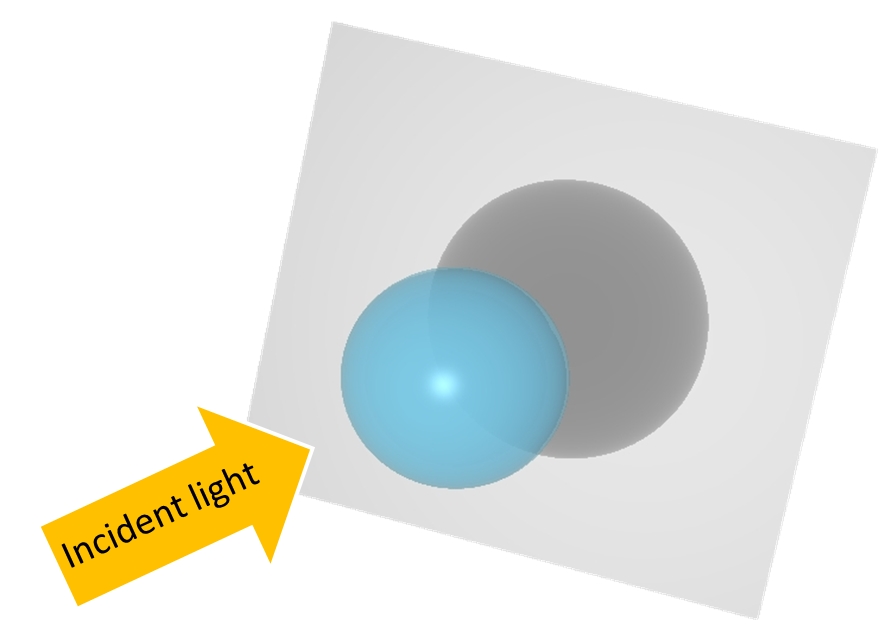}
\caption{Absorption imaging of a BEC.  Laser light incident upon the BEC 
creates a shadow behind it, whose darkness is proportional to the 
column-integrated density of the BEC.}
\label{fig:abs_imaging}
\end{figure}

The most common approach to image a condensate is via optical absorption imaging.  
The condensate is illuminated by an uniform light beam from one side.  The atoms absorb a proportion of the light such that a two-dimensional shadow is cast behind the condensate; this is recorded by  camera as shown in Figure \ref{fig:abs_imaging}, forming an {\em absorption image} of the condensate. \index{imaging!absorption imaging} Examples are the images
in Fig.~\ref{fig:dfg}.  Importantly, the darkness of the shadow is proportional 
to the atomic density, integrated along the direction light is 
travelling in\footnote{Fortunately, the atomic density is so low that 
scattering of the light beam is negligible and so the light effectively 
takes a direct path through the condensate.};
we call this the {\em column-integrated density}.      \index{imaging!column-integrated density}

To enable comparison between experimental absorption images and
theoretical models, one must relate three-dimensional
wavefunctions to the corresponding
two-dimensional column-integrated density profiles.  
Assuming imaging in the $z$-direction, the column-integrated density $n_{\rm CI}$ is,

\begin{equation}
n_{\rm CI}(x,y)=\int_{-\infty}^{\infty} n(x,y,z)~{\rm d}z.
\end{equation}

\section{Galilean invariance and moving frames}
A condensate in a homogeneous ($V=0$) system satisfies the GPE, 
\begin{equation}
i \hbar \frac{\partial \Psi}{\partial t}=
\left(-\frac{\hbar^2}{2m}\nabla^2+ g  |\Psi|^2 \right) \Psi.
\label{eqn:1dgpe}
\end{equation}
The stationary solution is $\Psi_0=\sqrt{n_0}\exp\left[-i\mu t/\hbar\right]$, corresponding to a static (${\bf v}=0$) condensate.  Now let us imagine, instead, that this condensate is moving with uniform velocity $v_0$ in the positive $x$ direction, say.  We can construct this moving solution as, \index{Gross-Pitaevskii equation!moving frame} 
\begin{equation}
\Psi=\Psi_0(x-v_0 t,y,z) \exp\left[i\frac{mv_0 x}{\hbar}-\frac{mv_0^2 t}{2 \hbar} \right].
\label{eqn:moving_solution}
\end{equation}
Note that the density remains $n_0$ throughout.  This is a demonstration of {\em Galilean invariance}, i.e. that the laws of physics are the same in all {\em inertial frames} (frames moving at fixed relative speed to each other). This is true only if the system is translationally invariant, i.e. the potential is the same everywhere.  

Above, we imagine the condensate flowing at speed $v_0$ relative to the static observer (the {\em lab frame}).  Instead, we can take the observer to be moving with the condensate.  We can then write the {\em moving frame GPE},
\begin{equation}
i \hbar \frac{\partial \Psi}{\partial t}=
\left(-\frac{\hbar^2}{2m}\nabla^2+ g  |\Psi|^2 + i \hbar v_0 \frac{\partial}{\partial \tilde{x}}\right) \Psi,
\label{eqn:moving_gpe}
\end{equation}
where $\tilde{x}$ is the $x$-coordinate in the moving frame and the Laplacian is evaluated in terms of the moving frame coordinates.  In this moving frame, the flowing condensate solution of Eq. (\ref{eqn:moving_solution}) is actually a stationary solution.  It can be useful to work in the moving frame when modelling flows of condensates.

\section{Dimensionless variables}
\label{sec:dimensionless}

The typical numbers which appear in the GPE equation
are very small and cumbersome, for example the reduced Planck's constant is
$\hbar =1.055 \times 10^{-34} ~{\rm J~s}$.  When numerically solving the GPE to model a condensate, it would be better if the numbers which we compute were of order unity; this minimises the role of floating point errors which are inherent to modern digital computation.  Another problem is that not all the parameters which appear in the
GPE are independent: identifying the truly independent parameters reduces
the number of numerical simulations which are needed to understand the nature
of the solution.
It is therefore useful to 
introduce {\em dimensionless variables} and write the GPE in simpler
{\em dimensionless form}. \index{dimensionless variables} \index{Gross-Pitaevskii equation! dimensionless}
To illustrate the
procedure, we consider two examples: homogeneous and harmonically-trapped
condensates. 

Before we start, we notice
for the sake of generality that we are free to introduce the chemical
potential $\mu$ in the time-dependent GPE by letting, in analogy with
Eq.~(\ref{eqn:time-indep}), 

\begin{equation}
\Psi({\bf r},t)=\psi({\bf r},t)e^{-i\mu t/\hbar},
\label{eqn:time-dep}
\end{equation}

\noindent
where now $\psi({\bf r},t)$ depends also on $t$; in other words,
the exponential term takes
care of part of (but not all of) the time dependence of the wavefunction.
The resulting time-dependent GPE is,

\begin{equation}
i \hbar \frac{\partial \psi}{\partial t}=
-\frac{\hbar^2}{2m}\nabla^2 \psi
+ g \vert \psi \vert^2 \psi + V \psi - \mu \psi.
\label{eqn:gp2}
\end{equation}

\subsection{Homogeneous condensate}

In the absence of trapping ($V=0$), the governing equation is,

\begin{equation}
i \hbar \frac{\partial \psi}{\partial t}=
-\frac{\hbar^2}{2m}\nabla^2 \psi
+ g \vert \psi \vert^2 \psi - \mu \psi.
\label{eqn:gpe3}
\end{equation}
We have seen that
the wavefunction of a uniform condensate at rest is
$\psi_0=\sqrt{\mu/g}$, corresponding to the number density
$n_0=\mu/g$. We have also seen that the characteristic minimum distance 
over which the wavefunction varies is the healing length 
$\xi=\hbar/\sqrt{m \mu}$. Therefore
the quantities $n_0$ and $\xi$ are
convenient units of density and length. Similarly, it is apparent from 
Eq.~(\ref{eqn:time-indep}) or Eq.~(\ref{eqn:time-dep}) 
that $\tau=\hbar/\mu$ is the
natural unit of time. These remarks suggest the introduction of the
following dimensionless variables (hereafter denoted by primes),

\begin{equation}
x'=\frac{x}{\xi}, \qquad y'=\frac{y}{\xi}, \qquad z'=\frac{z}{\xi}, \\
\end{equation}

\noindent
(in other words, ${\bf r'}={\bf r}/\xi$), and,

\begin{equation}
t'=\frac{t}{\tau}, \qquad \psi'=\frac{\psi}{\psi_0}.
\end{equation}
To begin substituting these new variables into the GPE, we need to develop relations for their derivatives.  Using the chain rule,
\begin{equation}
\frac{\rm d}{{\rm d} x}=\frac{1}{\xi}\frac{\rm d}{{\rm d}x'},
\quad 
\frac{\rm d}{{\rm d}y}=\frac{1}{\xi}\frac{\rm d}{{\rm d}y'},
\quad 
\frac{\rm d}{{\rm d}z}=\frac{1}{\xi}\frac{\rm d}{{\rm d}z'},
\quad 
\frac{\rm d}{{\rm d}t}=\frac{1}{\tau}\frac{\rm d}{{\rm d}t'}.
\end{equation}

\noindent
Hence the gradient and Laplacian operators acting on the
primed variables are defined as,
\begin{equation}
\nabla=\frac{1}{\xi}\nabla',
\quad 
\nabla^2=\frac{1}{\xi^2}\nabla'^2.
\end{equation}
Introducing these relations, Eq.~(\ref{eqn:gpe3}) becomes the following {\em dimensionless GPE}, 
\begin{equation}
i \frac{\partial \psi'}{\partial t'}=
-\frac{1}{2}\nabla'^2 \psi' 
+ \vert\psi' \vert^2 \psi'  -\psi'.
\label{eqn:gpe4}
\end{equation}

\noindent
This equation contains no parameters - it has
been simplified to its mathematical 
essence\footnote{
In the literature, after transforming the GPE into dimensionless
form, it is common to drop the primes.
}.  
These units are often 
termed {\em natural} or {\em healing length} units.  \index{units!healing length}

\subsection{Harmonically-trapped condensate}
Here we assume that the condensate is confined by a spherical harmonic trap $V=m \omega_r^2 r^2/2$.  Then the governing equation is,
\begin{equation}
i \hbar \frac{\partial \psi}{\partial t}=
-\frac{\hbar^2}{2m}\nabla^2 \psi
+ g \vert \psi \vert^2 \psi +V \psi - \mu \psi.
\label{eqn:gpe5}
\end{equation}
In this case the natural
units of length and time are based on the harmonic oscillator length $\ell_r=\sqrt{\hbar/m \omega_r}$ and the inverse of the trap frequency, $\omega_r^{-1}$. We set ${\bf r'}={\bf r}/\ell$ (that is to say
$x'=x/\ell$, $y'=y/\ell$ and $z'=/\ell$)
and $t'=t/\tau$, where $\tau=1/\omega_r$. 

It is conventional with these units to define the dimensionless wavefunction $\psi'$ as being normalized to unity, i.e.,
\begin{equation}
\int \vert \psi' \vert^2 {\rm }{\rm d}^3{\bf r'}=1.
\end{equation}
Comparing to Eq. (\ref{eqn:norm}) and noting that ${\rm d}^3{\bf r}=\ell_r^3 {\rm d}^3{\bf r'}$, it follows that $\psi=(N/\ell_r^3)^{1/2} \psi'$.

Introducing these relations into Eq.~(\ref{eqn:gpe5}) we arrive at the dimensionless form,

\begin{equation}
i \frac{\partial \psi'}{\partial t'}=
-\frac{1}{2}\nabla'^2 \psi'
+ C \vert \psi' \vert^2 \psi' +\frac{r'^2}{2}\psi' - \mu' \psi'.
\label{eqn:gpe6}
\end{equation}

\noindent
where $\mu'=\mu/\hbar \omega_r$ and,
\begin{equation}
C=\frac{4 \pi a_s N}{\ell}.
\end{equation}

\noindent
is a dimensionless interaction parameter.  These units are often termed {\em harmonic oscillator units}.  For anisotropic harmonic traps, the harmonic units can be defined instead in terms of one of the trap frequencies or their geometric mean, $\bar{\omega}=(\omega_x \omega_y \omega_z)^{1/3}$. \index{units!harmonic oscillator}

\section*{Problems}
\addcontentsline{toc}{section}{Problems}

\begin{prob}
\label{dimensions}

(a) Using the normalization condition, determine the dimensions of the
wavefunction $\Psi$ in
S.I. units (metres, kilograms, seconds).\\
(b) Verify that all terms of the GPE have the same dimension.\\
(c) Show that $g \vert \Psi\vert^2$ has dimension of energy.
\end{prob}


\begin{prob}
\label{spherical_bec}
Consider a BEC in the Thomas-Fermi limit confined within a three-dimensional spherical
harmonic trap. \\
(a) Normalize the wavefunction, and hence determine an expression for the
Thomas-Fermi radius $R_r$ in terms of $N$, $a_{\rm s}$ and $\ell_r$.  \\
(b) Determine an expression for the peak density in terms of $N$ and $R_r$.\\
(c) Find an expression for the ratio $R_r/\ell_r$, and comment on its behaviour for large $N$.\\
(d) What is the energy of the condensate?
\end{prob}

\begin{prob}
\label{variational}
Derive the expression for the variational energy of a three-dimensional trapped condensate, Eq. (\ref{eqn:variational_energy}).  Repeat in two dimensions (for a potential $V(x,y)=m\omega_r^2 (x^2+y^2)/2$) and in one dimension (for a potential $V(x)=m\omega_r^2 x^2/2$).  For each case plot $E/N\hbar \omega_r$ versus the variational width $\sigma$, for some different values of the interaction parameter $N a_{\rm s}/\ell_r$.  What effect does dimensionality have on the shape of the curves?  How do this change the qualitative behaviour described in Section \ref{sec:weak_ints}?
\end{prob}

\begin{prob}
\label{absorption}
Consider a BEC in the non-interacting limit with wavefunction

\begin{equation}
\psi(x,y,z)=
\sqrt{n_0} ~e^{-x^2/2 \ell_x^2}e^{-y^2/2 \ell_y^2}e^{-z^2/2 \ell_z^2},
\end{equation}

\noindent
where $n_0$ is the peak density and $\ell_x$, $\ell_y$ and $\ell_z$ are 
the harmonic oscillator lengths in three Cartesian directions.  
The BEC is imaged along the $z$-direction.  Determine the form of the 
column-integrated density $n_{\rm CI}(x,y)$.
Hint: $\int^{\infty}_0 e^{ax^2}=\frac{1}{2}\sqrt{\pi/a}$.
\end{prob}

\begin{prob}
\label{moving}
Consider a 1D uniform static condensate with $V(x)=0$.  Obtain an expression for the energy $E$ in a length $L$ of the condensate, in terms of $n_0$, $g$ and $L$.

Now consider the condensate to be flowing with uniform speed $v_0$, by constructing a solution according to Eq. (\ref{eqn:moving_solution}).  Show that the solution satisfies the 1D GPE, and confirm that the velocity field of this solution is indeed $v(x)=v_0$.  What is the corresponding energy for the flowing condensate, and how does it differ from the static result?  Finally, what is its momentum?
\end{prob}

\begin{prob}
\label{dimensionless}
Consider a homogeneous condensate. Identify dimensionless variables so
that the dimensionless GPE is,

\begin{equation}
i \frac{\partial \psi'}{\partial t'}=
-\nabla'^2 \psi' 
+ \vert\psi' \vert^2 \psi'  -\psi',
\label{eqn:gpe5}
\end{equation}

\noindent
i. e., without the $1/2$ factor as in Eq.~(\ref{eqn:gpe4}).

\end{prob}

%
%
%
%
%
%
%
%
%
%
%
%
%

\chapter{Waves and Solitons}
\label{intro} 

\abstract{In the previous chapter we considered the shape of steady
state condensates, either homogeneous or confined by trapping
potentials. We have seen that the condensate described
by the GPE is a special kind
 of fluid, similar to the idealized Euler fluid without viscosity
that appears in classical fluid dynamics textbooks. Not surprisingly for a fluid, the
dynamics of the condensate exhibit a variety of interesting time-dependent phenomena,
from sound waves and shape oscillations, to solitons and vortices.}

\section{Dispersion relation and sound waves}
\label{sec:sound-waves}

\subsection{Dispersion relation}
\label{sec:dispersion}

Of particular importance are the behaviour of perturbations to the ground state (either homogeneous or in a trap).  This includes {\em sound waves}, i.e., small-lengthscale density perturbations of the ground state which
oscillate periodically, as illustrated in Fig.~(\ref{fig:waves}).   \index{sound} We now derive the behaviour of these perturbations for a homogeneous condensate.  
The governing equation of motion is the GPE as it appears in
either Eq.~(\ref{eqn:gp1}) or Eq.~({\ref{eqn:gp2}) with $V=0$.  
We consider the latter.  Assuming one-dimensional motion along the $x$ direction, the GPE is,

\begin{equation}
i \hbar \frac{\partial \psi}{\partial t}=
-\frac{\hbar^2}{2m} \frac{\partial^2 \psi}{\partial x^2}
+ g  |\psi |^2 \psi - \mu \psi.
\label{eqn:gp3}
\end{equation}

\noindent
We know that, for a homogeneous condensate, the steady solution
of this equation is the uniform state $\psi_0=\sqrt{\mu/g}$, with number density
$n_0=\vert \psi_0 \vert^2=\mu/g$.  
We perturb this uniform state by assuming a wavefunction with the form,
\begin{equation}
\psi(x,t)=\psi_0 + \epsilon \psi_1(x,t) + \epsilon^2 \psi_2(x,t) 
+ \cdots,
\label{eqn:perturb}
\end{equation}

\noindent
where $\epsilon \ll 1$ is a small parameter and the functions $\psi_1$,
$\psi_2$, etc, must be determined. 
Substituting Eq.~(\ref{eqn:perturb}) into 
Eq.~({\ref{eqn:gp3}), noting that temporal and spatial derivatives of the steady uniform background $\psi_0$ are zero, and neglecting terms which are quadratic or of higher
order in $\epsilon$, we obtain,

\begin{equation}
i \hbar \frac{\partial \psi_1}{\partial t}=
-\frac{\hbar^2}{2m}\frac{\partial^2 \psi_1}{\partial x^2}
+ \mu \psi_1^* +\mu \psi_1.
\label{eqn:gp6}
\end{equation}

\noindent
This is the {\em linearized equation of motion for the perturbations}.
We look for travelling wave solutions of the general form,

\begin{equation}
\psi_1(x,t)=A e^{i(kx-\omega t)}+ B^* e^{-i(kx-\omega t)},
\label{eqn:wave_solution}
\end{equation}

\noindent
where $A$ and $B$ are complex amplitudes which depend on the
initial condition\footnote{We write the amplitude of the second term
as $B^*$ rather than $B$ for mathematical convenience.}, 
$k$ is the {\em wavenumber}
and $\omega$ the {\em angular frequency} of the wave. 
Substituting into Eq.~(\ref{eqn:gp6}),
we find that non-trivial (non-zero) solutions for $A$ and $B$ exist only if,
\begin{equation}
\omega= \sqrt{\left(\frac{\hbar k^2}{2m}\right)^2 +\frac{n_0 g}{m} k^2}.
\label{eqn:dispersion}
\end{equation}
This is called the {\em dispersion relation}, or sometimes the {\em Bogoliubov dispersion relation} after Nikolay Bogoliubov who first derived it\index{dispersion relation}.  It relates the wave's angular frequency $\omega$ to its wavenumber $k$, or equivalently, its period $2\pi /\omega$ to its wavelength $2 \pi/k$.  

\begin{figure}[b]
\centering
\includegraphics[width=0.99\columnwidth,angle=0]{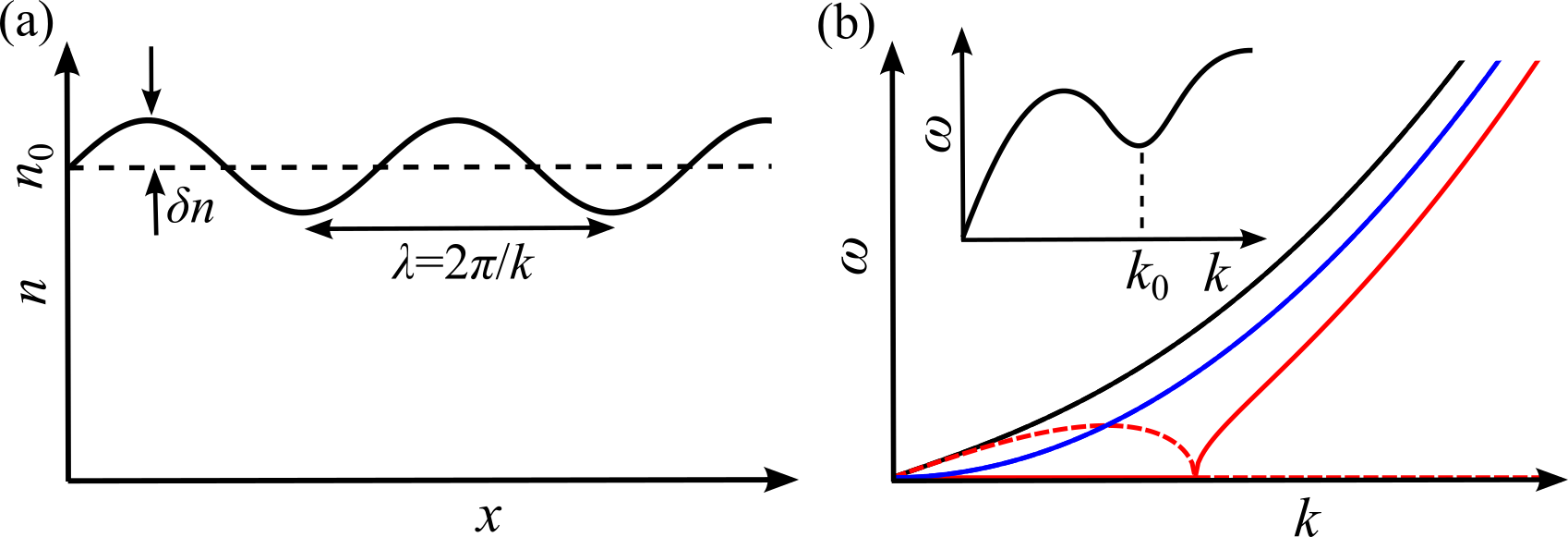}
\caption{(a) One dimensional sound waves, that is, sinusoidal perturbations of the background density $n_0$, of wavelength $\lambda$ and amplitude
$\delta n_0(x,t) \ll n_0$. (b): The dispersion relation $\omega(k)$ of the homogeneous (weakly-interacting) condensate, according to Eq. (\ref{eqn:dispersion}), for $g>0$ (black line), $g=0$ (blue line) and $g<0$ (red line).  Solid lines plot the real part of $\omega$ and dashed lines plot the imaginary part.  For $g<0$, $\omega$ becomes imaginary for small $k$; everywhere else $\omega$ is real.  Inset: In helium II, the dispersion relation has a different and distinct shape, featuring a maxon (local maximum) and roton (local minimum).  The roton wavenumber is indicated as $k_0$.}
\label{fig:waves}
\end{figure}

Notice that the
phase velocity of the wave\footnote{The phase velocity of a wave is the rate at which its phase propagates in space.} $v_{\rm ph}=\omega/k$ depends, in general, on $k$. Suppose that 
the initial condition at $t=0$ is a generic
wave packet, i.e., the superposition of 
different plane waves with different amplitudes and phases; 
since these waves move
at different phase velocities, the wave-packet spreads
out as it propagates, or {\em disperses}.

Consider the behaviour of the dispersion relation $\omega(k)$ for different regimes of interactions, as plotted in Fig. \ref{fig:waves}(b).  
\begin{itemize}
\item  In the {\bf absence of interactions} ($g=0$) the dispersion relation reduces to
$\hbar \omega = \hbar^2 k^2/2 m$, in other words the wave behaves
like a free particle of momentum $p=\hbar k$ and energy $\hbar \omega$.  Note that $\omega$ is real; then the exponential terms in the solution of Eq. (\ref{eqn:wave_solution}) have imaginary exponents and so describe a temporally-oscillating solution.  
\item For {\bf repulsive interactions} ($g>0$), this free-particle behaviour ($\omega \sim k^2$) is recovered  in the limit of large $k$/short waves.  However, for low $k$/long waves, the dispersion relation is linear in $k$.  This linear behaviour is characteristic of sound waves - see below.  As for $g=0$, the angular frequencies are real. \index{sound}
\item For {\bf attractive interactions}, the situation is fundamentally different. For $g<0$ and in the regime of sufficiently small $k$, $\omega^2$ is negative and, correspondingly, $\omega$ becomes complex.  Then these exponential terms develop {\em real} and {\em positive} exponents, such that they exponentially increase in amplitude over time.  This signifies the dynamical instability of the homogeneous attractively-interacting condensate - small perturbations are not stable and grow out of control.  In fact, this instability is due to the collapse instability we've already described for an attractive condensate; here the condensate prefers to collapse rather than stay as a uniform density profile.\index{collapse}
\end{itemize}

In helium II the shape of the dispersion curve is somewhat different (see Fig. \ref{fig:waves}(inset)), due to the strong inter-atomic interactions in the system.  The dispersion relation is linear for small $k$, but then features a local maximum, termed the {\em maxon}, and a local minimum, termed the {\em roton}.   At even higher $k$ the dispersion relation flattens off.  

\subsection{Sound waves}
For repulsive interactions ($g>0$) and in the limit of small $k$/long waves, the above dispersion relation predicts waves whose angular frequency increases linearly with wavenumber.  This is characteristic of sound waves.  The phase velocity of these waves is $v_{\rm ph}=\omega/k \approx \sqrt{n_0 g /m}$, which is approximately constant for all wavelengths.  This defines the speed of sound, \index{sound!speed of sound}
\begin{equation}
c=\sqrt{\frac{n_0 g}{m}}.
\label{eqn:c}
\end{equation}
The physical interpretation of these waves is readily obtained using
the Madelung transform. By perturbing the 
state of uniform density $n=n_0=\mu/g$, we obtain
the one-dimensional wave equation,
\begin{equation}
\frac{\partial^2}{\partial t^2} \delta n=c^2 \frac{\partial^2}{\partial x^2}
\delta n,
\end{equation}

\noindent
where $\delta n(x,t) \ll n_0$ are density perturbations about the background condensate (here taken to be the homogeneous condensate), as
shown in Fig.~\ref{fig:waves}. The wave solution which
we have found is one-dimensional - the wave propagates along $x$ - but
can be
easily generalized to two and three dimensions. 

In a trapped condensate ($V \neq 0$) the speed of sound will vary
with the position due to the spatial dependence of the density.  
The speed of sound is less near the edge of the condensate
where the density tends to zero.

The prediction of sound waves was tested experimentally in Ref. \cite{andrews_1997} by using a laser beam 
to initially and suddenly ``punch'' a hole in the density at the centre of a condensate, much like a stone being thrown into a pond.  This generated low amplitude ripples, i.e.,
sound waves, which travelled outwards along the condensate, as shown in Fig. \ref{fig:sound}(a).  The speed of the waves was found to follow the square-root of the density, as seen in Fig. \ref{fig:sound}(b) and in agreement with the prediction of 
Eq.~(\ref{eqn:c}).  

\begin{figure}[h]
\centering
\includegraphics[width=0.99\columnwidth,angle=0]{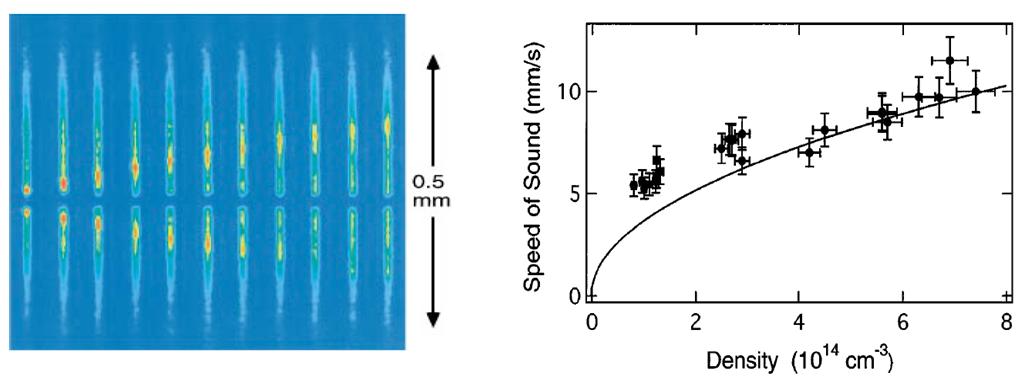}
\caption{Left: Experimental creation of sound waves after suddenly 
generating a `hole' in the centre of the condensate using a laser beam.  
Shown is the column-integrated density profile of the condensate taken at regular intervals of time.  
Yellow, red and blue are respectively large, medium and small
values of the density. 
Right: The measured speed of sound $c$ (points) as a function of the
background number density, in agreement
with Eq.~(\ref{eqn:c}) (solid line).  Reprinted figures with permission from \cite{andrews_1997}. Copyright 1997 by the American Physical Society.}
\label{fig:sound}
\end{figure}

\section{Landau's criterion and the breakdown of superfluidity}
\label{sec:landau}

Under some perturbation the condensate can become excited.  Here we develop a simple yet powerful criterion for excitations to develop, as developed by Landau \cite{Nozieres_1999}.  Consider a homogeneous ground state condensate, into which an impurity (e.g. another atom) of mass $M$ enters with initial velocity ${\bf v_i}$.  Let us imagine that the impurity imparts an excitation of the condensate with energy $\hbar \omega$ and momentum $\hbar {\bf k}$; this subsequent velocity of the impurity is ${\bf v_f}$.  The initial energy (relative to the static background condensate) is just the initial kinetic energy of the impurity, $Mv_{\rm i}^2/2$, while the final energy after generating the excitation is $Mv_{\rm f}^2/2 + \hbar \omega$, where $v_{\rm i}=|{\bf v_i}|$ and $v_{\rm f}=|{\bf v_f}|$.  Applying conservation of energy gives,
\begin{equation}
\frac{1}{2}Mv_{\rm i}^2 = \frac{1}{2}M v_{\rm f}^2 +\hbar \omega.
\label{eqn:coe}
\end{equation}
Similarly, applying conservation of momentum before and after the event gives,
\begin{equation}
M{\bf v_i} = M {\bf v_f}+\hbar {\bf k}.
\label{eqn:mom}
\end{equation}
Inserting Eq. (\ref{eqn:mom}) into Eq. (\ref{eqn:coe}) and simplifying gives,
\begin{equation}
\hbar \omega = \frac{1}{2}M v_{\rm i}^2 -\frac{1}{2M}\left(M {\bf v_i} - \hbar {\bf k} \right)^2 = \hbar {\bf k} \cdot {\bf v_i}-\frac{\hbar^2 k^2}{2M}.
\end{equation}

If $M$ is sufficiently large, the second term at the right 
hand side can be neglected; for excitations to be energetically favoured, 
the initial velocity $v_{\rm i}$ then has to satisfy,
 \index{Landau criterion} \index{critical velocity}
 \begin{equation}
 v_{\rm i} \geq \left| \frac{{\bf k} \cdot {\bf v_i}}{k} \right| \geq 
\frac{\omega}{k}.
 \end{equation}
One can instead write this as,
\begin{equation}
v_i \geq v_{\rm c},
\end{equation}
where $v_{\rm c}$ is termed the critical superfluid velocity,
\begin{equation}
v_{\rm c} = {\rm min}\left( \frac{\omega}{k}\right).
\end{equation}
This is a defining property of superfluidity.  For $v_{\rm i}<v_{\rm c}$ the impurity propagates with no damping, i.e. as a superfluid, while for $v_{\rm i} \geq v_{\rm c}$ excitations the motion becomes dissipated by the transfer of energy and momentum to the fluid.  This marks the breakdown of superfluidity.  

The function $\omega=\omega(k)$ (the dispersion relation)
is typically a non-trivial function of $k$.
For a weakly-interacting and homogeneous condensate, with dispersion relation given by Eq. (\ref{eqn:dispersion}), this gives,
\begin{equation}
v_{\rm c}=c.
\end{equation}
For $v<c$ the atom moves with no damping or hindrance, a defining characteristic of superfluidity. \index{superfluidity} For $v>c$ damping can occur through the creation of condensate excitations.

\section{Collective modes}
\label{sec:shape_oscillations}

In a trapped condensate of finite size, sound waves should have a wavelength considerably smaller than the condensate size (or, equivalently, the angular frequency of the wave should be considerably larger than the trap frequency); this, for example, is clearly satisfied in the experimental images in Fig. \ref{fig:sound}(left).  However, if the wavelength of the density perturbations becomes of the order of the condensate size, then these excitations involve a motion of the whole system.  These are the {\em collective modes}.  \index{collective modes}

There is a wide family of collective modes which are supported under harmonic trapping.  Here we consider the simplest and most common types, illustrated in Fig. \ref{fig:collective_modes}: 
\begin{itemize}
\item The {\bf dipole mode} corresponds to an oscillation of the condensate's centre-of-mass about the trap centre.  In a harmonic trap, this oscillation occurs at the trap frequency in the respective direction.  This mode is not affected by $g$ since, in a harmonic trap, the centre-of-mass motion is decoupled from the internal dynamics.  For this reason, this mode is often excited experimentally to measure the trap frequency.  
\item The {\bf monopole mode} involves contraction-expansion oscillations of the condensate, which are in-phase across the directions.
\item The {\bf quadrupole mode} also involves contraction-expansion oscillations, but where the oscillation in one direction is in anti-phase to that in the other directions.  Both the quadrupole and monopole modes are sensitive to interactions.
\end{itemize}

\begin{figure}[b]
\centering
\includegraphics[width=0.8\columnwidth,angle=0]{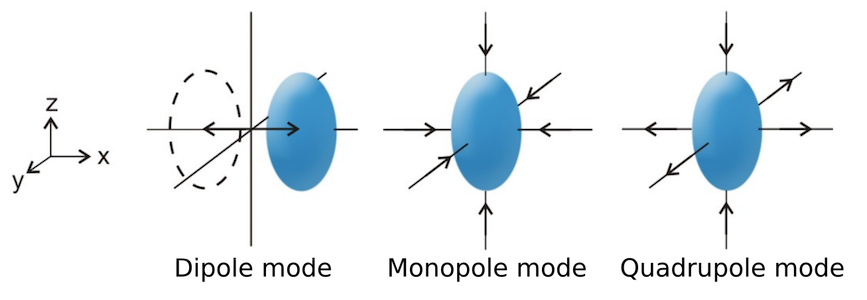}
\caption{The three common collective modes of a harmonically-trapped condensate.}
\label{fig:collective_modes}
\end{figure}

\subsection{Scaling solutions}
\index{scaling solutions}
Experimentally, the collective modes are typically induced by forming a stationary condensate, and then suddenly changing the harmonic trap.  To induce the dipole mode, the trap can be suddenly translated in space; the condensate finds itself up the trap wall and begins to undergo centre-of-mass oscillations about the trap centre.  To induce the monopole, quadrupole or similar modes, the trap frequencies can be suddenly changed in time.  This scenario, in the absence of centre-of-mass motion, is the one we consider here.  A similar methodology can be used to account for the centre-of-mass dynamics.

We consider a condensate which is at equilibrium at $t=0$, with the trap frequencies suddenly changed for $t >0$. We follow the approach introduced in Ref. \cite{dum_1996}.
We can model the ensuing oscillations of the condensate through the 
hydrodynamical description (Section \ref{sec:fluid}), along with the 
Thomas-Fermi approximation (Section \ref{sec:strong_ints}).  \index{Thomas-Fermi!solutions}
Recalling the hydrodynamic equations, Eqs. (\ref{eqn:cont1},\ref{eqn:quasiEuler1}), dropping terms which depend on the gradients of density, and introducing a general harmonic potential, leads to,
\begin{eqnarray}
\dfrac{\partial n}{\partial t} + \nabla \cdot(n {\bf v})&=&0,  \label{eqn:hydro1}
\\
\displaystyle m \frac{\partial {\bf v}}{\partial t} + \nabla \left(\frac{1}{2}mv^2+ \frac{1}{2}m(\omega_x^2 x^2+ \omega_y^2 y^2 + \omega_z^2 z^2) + gn\right)&=&0. \label{eqn:hydro2}
\end{eqnarray}
The equilibrium solution of the condensate at $t=0$ is found by setting ${\bf v}$ and the time-derivatives to zero; then the second equation reduces to,
\begin{equation}
\nabla \left(\frac{1}{2}m(\omega_x^2 x^2+ \omega_y^2 y^2 + \omega_z^2 z^2)+gn \right)=0.
\end{equation}
Integrating over space and rearranging gives,
\begin{equation}
n=\frac{2\mu-m(\omega_x^2 x^2+ \omega_y^2 y^2 + \omega_z^2 z^2)}{2g},  \textrm{for} ~n\geq 0,
\end{equation}
where $\mu$ arises as the integration constant.  This is the 
equilibrium density profile for the condensate in the Thomas-Fermi limit, as obtained 
in Section \ref{sec:strong_ints}.
  We may also write this in the form,
\begin{equation}
n=n_0 \left( 1-\frac{x^2}{R_{x,0}^2}- \frac{y^2}{R_{y,0}^2}-\frac{z^2}{R_{z,0}^2}\right),  {\rm for}~n(x,y,z) \geq 0,
\label{eqn:tf2}
\end{equation}
where $R_{j,0}=\sqrt{2 \mu/m \omega_j}$, with $j=x,y,z$, are the Thomas-Fermi radii \index{Thomas-Fermi!radius} and $n_0$ is the central density of the condensate at $t=0$.  Applying the usual normalization condition $\int n(x,y,z)~{\rm d}^3 {\bf r}=N$ \index{normalization} gives the expression for the central density,
\begin{equation}
n_0 = \frac{15 N}{8 \pi R_{x,0} R_{y,0} R_{z,0}}.
\end{equation}

Following a sudden change in the trap frequencies, $\omega_j \rightarrow \omega_j(t)$, the condensate profile becomes time-dependent. We consider the time-dependent density to maintain the same general shape throughout but where its dimensions become scaled over time.  This is accounted for by making the radii time-dependent, $R_j \rightarrow R_j(t)$.  If we further introduce {\em scaling parameters} $b_j(t)=R_j(t)/R_{j,0}$ then we can write the time-dependent profile as,
\begin{eqnarray}
n(x,y,z,t)=\frac{n_0}{b_x b_y b_z}\left(1-\frac{x^2}{b_x^2 R_{x,0}^2}-\frac{y^2}{b_y^2 R_{y,0}^2}-\frac{z^2}{b_z^2 R_{z,0}^2}\right),
\label{eqn:tf3}
\end{eqnarray}
This is known as the {\em scaling solution}.   The modified pre-factor accounts for the time-dependence of the central density.  The initial conditions of the dynamics are,
\begin{equation}
b_j(t=0)=1, \quad \dot{b}_j(t=0)=0,
\label{eqn:ics}
\end{equation}
where the dot represents the time derivative.

To satisfy the continuity equation, the velocity field which matches this density must be of the form,
\begin{equation}
{\bf v}({\bf r},t)=\frac{1}{2} \nabla \left[\alpha_x(t)x^2+\alpha_y(t)y^2+\alpha_z(t) z^2 \right], \quad \alpha_j=\frac{\dot{b}_j}{b_j}.
\end{equation}

One proceeds (although the derivation is beyond our scope) to introduce the time-dependent density and velocity distributions into the Thomas-Fermi hydrodynamic equations (\ref{eqn:hydro1},\ref{eqn:hydro2}).  This leads to three coupled equations of motion for the scaling variables $b_j(t)$,
\begin{equation}
\ddot{b}_j+\omega_j(t)^2 b_j -\frac{\omega_{j,0}^2}{b_j b_x b_y b_z}=0,
\label{eqn:scaling}
\end{equation}
where $\omega_{j,0}$ is the initial trap frequency in the $j$th direction.  Remarkably, these equations involve the scaling variables $b_i$ and the trap frequencies, only.  What is also remarkable is that the same scaling equations of motion arise for a Gaussian ansatz, a justifiable approximation for weak interactions.     As such these scaling equations have a much wider coverage than the strongly-interacting Thomas-Fermi limit.

For a cylindrically symmetric trap $V(r,z)=m(\omega_r^2 r^2 + \omega_z^2 z^2)/2$, where $r^2=x^2+y^2$, this description reduces to two equations of motion,
\begin{equation}
\ddot{b}_r+\omega_r(t)^2 b_r -\frac{\omega_{r,0}^2}{b_r^3 b_z}=0, \quad \ddot{b}_z+\omega_z(t)^2 b_z -\frac{\omega_{z,0}^2}{b_r^2 b_z^2}=0.
\label{eqn:scaling_rz}
\end{equation}
To demonstrate the collective mode dynamics, we solve these two ordinary differential equations numerically for Thomas-Fermi condensate initially confined to a spherically-symmetric trap with $\omega_r=\omega_z=2 \pi \times 50$Hz.  To induce a monopole mode, we reduce the trap frequencies by $10\%$ for $t>0$.  As seen in Fig. \ref{fig:modes2}(a), the widths increase initially, and continue to oscillate around a new, larger equilibrium width.  Characteristic of a monopole mode, the oscillations are in phase along $r$ and $z$.  Meanwhile, a quadrupole mode is generated by simultaneously increasing $\omega_z$ and decreasing $\omega_r$ (both by $10\%$).  As seen in Fig. \ref{fig:modes2}(b), the condensate initially expands radially and shrinks axially, and continues to oscillate in anti-phase.  

\begin{figure}[t]
\centering
\includegraphics[width=0.98\columnwidth,angle=0]{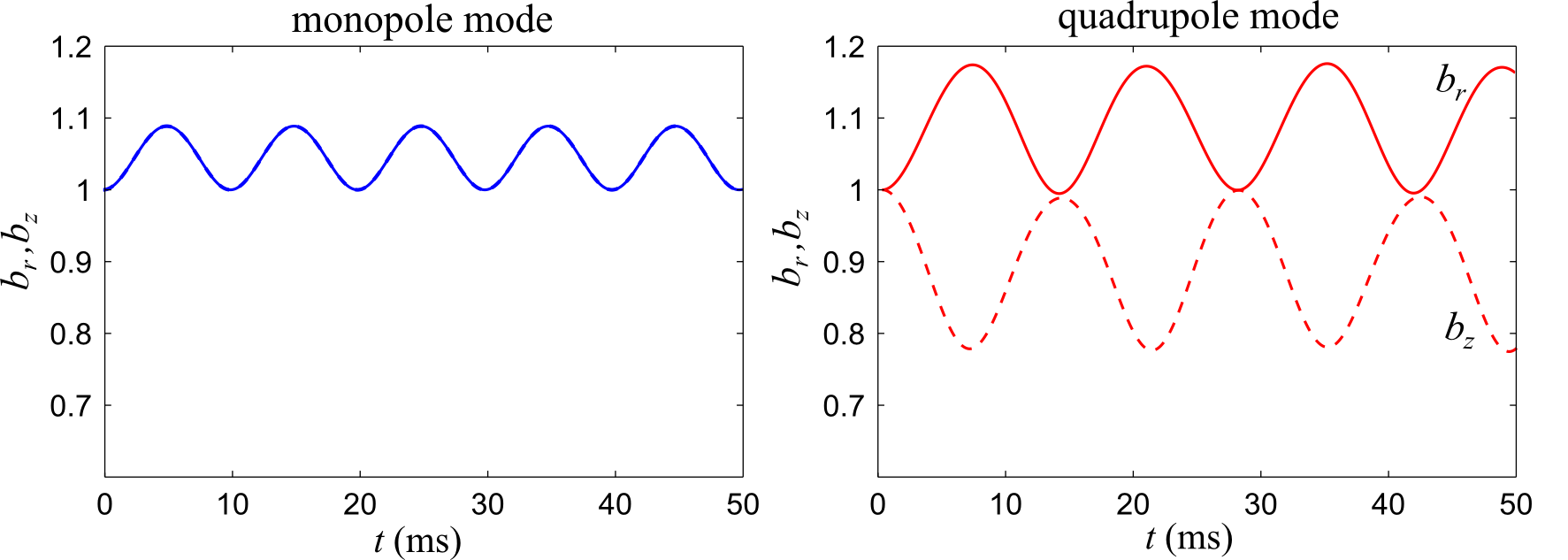}
\caption{A monopole mode and a quadrupole mode of a condensate in a cylindrically-symmetric harmonic trap.  Shown are the radial and axial scaling parameters as a function of time, i.e. $b_r(t)$ and $b_z(t)$, as solved according to the Thomas-Fermi scaling equations of motion, Eqs. (\ref{eqn:scaling_rz}).  Note that for the monopole mode, the two curves lie on top of each other.}
\label{fig:modes2}
\end{figure}

The above scaling equations of motion are valid for arbitrarily large mode amplitudes (providing the Thomas-Fermi approximation is maintained).  In the limit of perturbatively small-amplitude modes (e.g. by linearizing about the equilibrium condensate, similar to Section \ref{sec:sound-waves} for a homogeneous system), one can determine the frequency of the collective modes analytically \cite{pitaevskii_2003}.  Under cylindrical symmetry,  the mode frequencies obey,
\begin{equation}
\omega_{\rm M,Q}^2= \omega_r^2 \left(2+\frac{3}{2}\lambda^2 \pm \frac{1}{2}\sqrt{16-16\lambda^2+9 \lambda^4} \right),
\end{equation}
where the ``+'' refers to the monopole mode frequency, $\omega_{\rm M}$, and the ``-'' to the quadrupole mode frequency, $\omega_{\rm Q}$, and $\lambda=\omega_z/\omega_r$ is the {\em trap ratio}.  For an approximately spherical trap ($\lambda \approx 1$) this gives,
\begin{equation}
\omega_M \approx \sqrt{5} \omega_r, \quad \omega_Q \approx \sqrt{2}\omega_r.
\end{equation}
These are in close agreement with the frequencies of the oscillations in Fig. \ref{fig:modes2}.

These scaling predictions give excellent agreement with the mode dynamics observed in experiments. These modes play an important role in this field.  They are straightforward to generate experimentally and can be measured to high accuracy., and provide a versatile means to test theoretical models and assumptions.  According to these predictions, the modes persist forever since the condensate has no viscosity.  In reality, thermal dissipation causes the modes to decay over time, although usually on a much longer timescale than the oscillations themselves.    

\subsection{Expansion of the condensate}
\index{expansion}
A particular case of these scaling dynamics is when the trap is suddenly switched off and the condensate is allowed to expand freely.  This is routinely performed in BEC experiments since some expansion of the gas is often necessary to enable imaging of small features such as dark solitons and vortices. 

For $\omega_r(t)=\omega_z(t)=0$ the cylindrically-symmetric scaling equations (\ref{eqn:scaling_rz}) reduce to,
\begin{equation}
\ddot{b}_r=\frac{\omega_{r,0}^2}{b_r^3 b_z}, 
\quad \ddot{b}_z=\frac{\omega_{z,0}^2}{b_r^2 b_z^2}.
\label{eqn:scaling_rz2}
\end{equation}
Replacing the time variable with  $\tau = \omega_z t$ and introducing the initial trap ratio, $\lambda=\omega_{z,0}/\omega_{r,0}$, gives,
\begin{equation}
\frac{{\rm d}^2 b_r}{{\rm d} \tau^2}=\frac{1}{b_r^3 b_z }, \quad \frac{{\rm d}^2 b_z}{{\rm d} \tau^2}=\frac{\lambda^2}{b_r^2 b_z^2 }.
\end{equation}

It is possible to obtain analytic expressions for $b_r(t)$ and $b_z(t)$ for the case of a cigar-shaped condensate, $\lambda \ll 1$ \cite{dum_1996}.  We proceed by expanding the solutions in powers of $\lambda^2$, i.e.,
\begin{align}
b_r(\tau)=1+\alpha_r(\tau) \lambda^2 + \beta_r(t) \lambda^4+..., \quad b_z(\tau)=1+\alpha_z(\tau) \lambda^2+ \beta_z(\tau) \lambda^4+...  \nonumber
\end{align}
To lowest order in $\lambda$, the axial dynamics satisfy,
\begin{equation}
\dfrac{{\rm d}^2 b_z}{{\rm d} \tau^2}=0, \quad b_z(\tau)=1.
\end{equation}
In obtaining this solution for $b_z$ we have applied the initial conditions in Eq. (\ref{eqn:ics}).  Employing this result, we find the radial dynamics satisfy,
\begin{equation}
\dfrac{{\rm d}^2 b_r}{{\rm d} \tau^2} = \dfrac{1}{b_r^3}, \quad  b_r(t)=\sqrt{1+\tau^2}.
\end{equation}  
Continuing to second order, we find that the axial expansion satisfies,
\begin{equation}
b_z(\tau)=1+\lambda^2 [\tau \textrm{arctan} \tau - \ln \sqrt{1+\tau^2}]+\mathcal{O}(\lambda^4).
\label{eqn:exp2B}
\end{equation}

The expansion develops very differently in the two directions.  The radial size increases rapidly at first, whereas the axial spreading is weak, suppressed by the $\lambda^2$ factor.  We define the aspect ratio of the condensate as $R_r/R_z=R_{r,0}b_r/R_{z,0}b_z$.   Initially, $R_r/R_z=\omega_z/\omega_r=\lambda$.  Over time the aspect ratio is modified by the scaling dynamics as $\lambda b_r/b_z$.  Using the above analytic expressions, it can be shown that in the limit of large $\tau$, the aspect ratio approaches the value,
\begin{equation}
\left(\frac{R_r}{R_z}\right)_{\tau \rightarrow \infty}=\frac{2}{\pi \lambda}.
\end{equation}
In other words the condensate reverses its aspect ratio.  These predictions agree accurately with experimental observations of condensate expansion, as seen in Fig. \ref{fig:exp}.

\begin{figure}[t]
\centering
\includegraphics[width=0.5\columnwidth,angle=0]{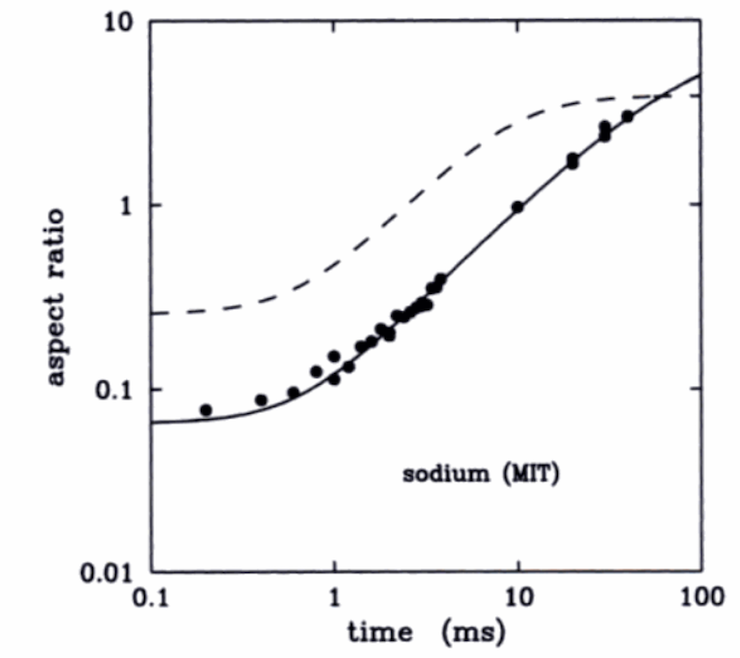}
\caption{The evolution of the aspect ratio $R_r/R_z$ during expansion of a cylindrically-symmetric condensate, from the experiment of Ref. \cite{mewes_1996}.   The circles are the experimental data points. The condensate evolves from its initial cigar shape ($R_r/R_z < 1$) to a pancake shape ( $R_r/R_z>1$). The numerical solution of the scaling equations (\ref{eqn:scaling_rz}) and the analytic predictions for $\lambda \ll 1$ are indistinguishable (appearing as the solid black line).  The dot-dashed line is the corresponding prediction for non-interacting atoms.  The figure is reproduced from Ref. \cite{dalfovo_1999b} with permission of the Societa Italiana di Fisica.}
\label{fig:exp}
\end{figure}

\section{Solitons}
\label{sec:solitons}
\index{solitons}
In one-dimension and in the absence of an external potential, the GPE is,
\begin{equation}
i\hbar \frac{\partial \psi}{\partial t}=
-\frac{\hbar^2}{2m} \frac{\partial^2 \psi}{\partial x^2} 
+g \vert \psi \vert^2 \psi,
\label{eqn:1dgpe}
\end{equation}
where the variables and parameters take their 1D definitions.  This is a form of the 1D {\em nonlinear Schr\"odinger equation}. This equation is well-studied in the context of nonlinear optics. It has the special property of being {\em integrable} such that its solutions possess an infinite set of conserved quantities (integrals of motion).  The simplest of these quantities (and those with a clear physical interpretation for our system) are the norm $N$, \index{solitons!norm} the momentum $\mathcal{P}$ and the energy $E$, \index{solitons!energy} \index{solitons!momentum}
\begin{eqnarray}
N&=&\int \limits_{-\infty}^{+\infty}|\psi|^2 {\rm d}x, \label{eqn:sol_norm}
\\
P&=&\frac{i\hbar}{2} \int \limits_{-\infty}^{+\infty}\left(\psi \frac{\partial \psi^*}{\partial x} -\psi^* \frac{\partial \psi}{\partial x}\right) {\rm d}x, \label{eqn:sol_mom}
\\
E&=&\int \limits_{-\infty}^{+\infty}\left(\frac{\hbar^2}{2m}\left|\frac{\partial \psi}{\partial x} \right|^2 + \frac{g}{2}|\psi|^4 \right) {\rm d}x. \label{eqn:sol_en}
\end{eqnarray}\index{solitons!integrals of motion} 

It is due to these special properties that Eq. (\ref{eqn:1dgpe}) supports solutions known as {\em solitons}.  
Solitons are nonlinear waves which
arise in many areas of physics, from fluids to optics to plasmas \cite{dauxois_2006}.  Solitons have three characteristic properties \cite{drazin_1989}: \index{solitons}
\begin{itemize}
\item They have a permanent, unchanging form.
\item They are localized in space.
\item They emerge unscathed from collisions with other solitons.   
\end{itemize}
Their permanent form is due to dispersion being perfectly balanced by their
nonlinearity; as a consequence, solitons propagate without 
spreading out in space.  This makes them analogous to particles, and motivated their particle-like name `'solitons''.  

The soliton solutions of the nonlinear Schr\"odinger equation were obtained in the pioneering works of Zakharov and Shabat, using a technique called the {\em inverse scattering transform} (see Ref. \cite{ablowitz_1981} for more information).  Depending on the sign of the interaction parameter $g$, 
{\em dark solitons} and {\em bright solitons} are supported, as we see next. Dark and bright solitons were first studied in the context of nonlinear optics; there they correspond to a dip and a peak in an optical intensity field, respectively, giving rise to their names.  

Condensates are, in reality, three-dimensional and feature trapping potentials, and so ``solitons'' therein are not strictly solitons.  However, they show the key solitonic properties, and so we continue to use the term ``soliton''.  In other literature, they are often referred to as ``solitary waves''.  
 
\section{Dark solitons}
\label{subsec:dark-solitons}

\subsection{Dark soliton solutions}
{\em Dark solitons} are supported for repulsive interactions ($g>0$).  
These waves consist of a localized density dip with a phase jump across it, and propagate along at speed $u$.  The speed can exist in the range $0<u\leq c$, where $c$ is the speed of sound.  A broad review of dark solitons in condensates is given in Ref. \cite{frantzeskakis_2010}.
\index{solitons!dark solitons}

The general dark soliton solution to Eq. (\ref{eqn:1dgpe}) with $g>0$ is,
\begin{equation}
\psi(x,t)=\sqrt{n_0} \left\{B~{\rm tanh}\left[\frac{B(x-ut)}{\xi}\right] + i\frac{u}{c}\right\}\exp\left(-\frac{i \mu t}{\hbar}\right),
\label{eqn:grey-soliton}
\end{equation}
where $B=\sqrt{1-u^2/c^2}$.  The density and phase profiles of some dark solitons are shown in Fig. \ref{fig:grey-dark-soliton}.   The density depression and the phase profile vary with speed.  Note that the soliton width is always of the order of the healing length $\xi$.\index{solitons!solutions}

A dark soliton state is an excited state; the ground state is the soliton-free homogeneous density.  In the lab frame (the frame at rest with the background condensate), the soliton is a moving solution.  However, dark solitons become stationary solutions in the moving frame, e.g. the $u/c=0.5$ soliton is a stationary (excited) state in the frame moving at the same speed.

\begin{figure}[t]
\centering
\includegraphics[width=0.9\columnwidth,angle=0]{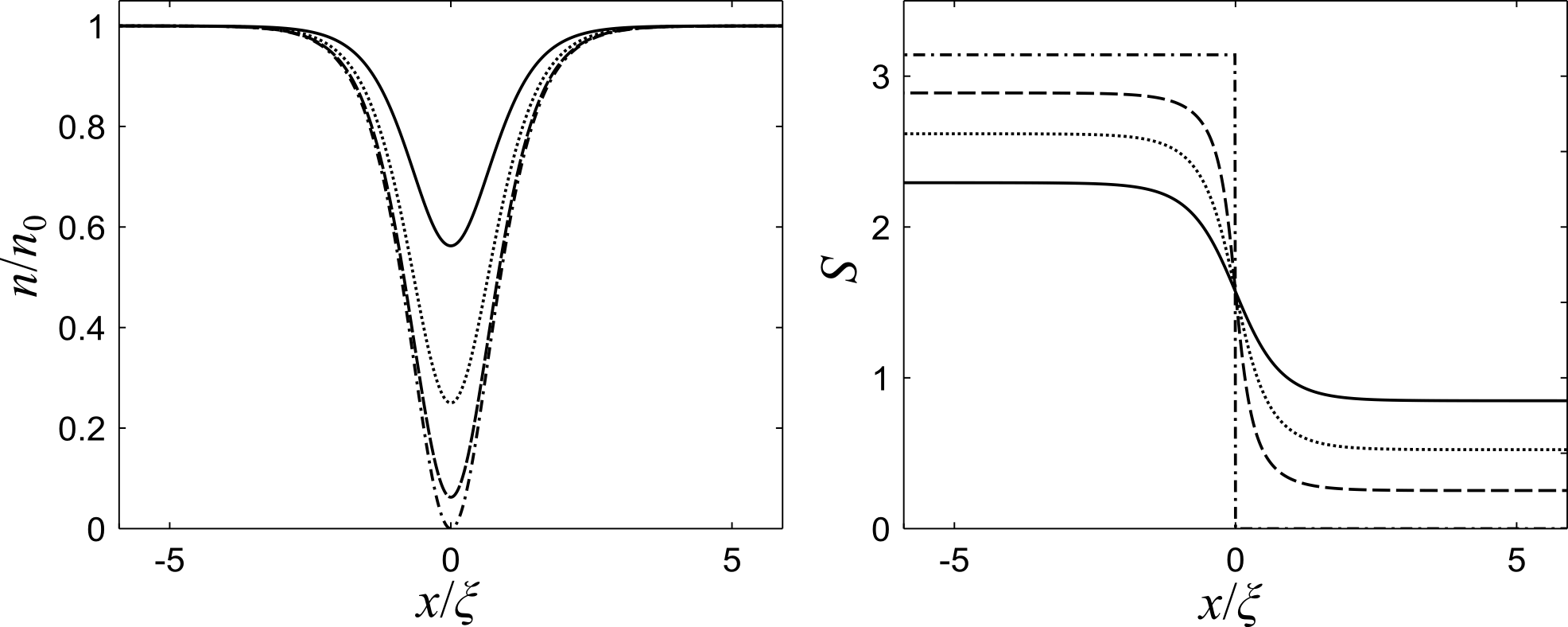}
\caption{Density $n(x)$ and phase $S(x)$ profiles of dark solitons with various speeds: $u/c=0$ (dot-dashed line), $u/c=0.25$ (dashed line), $u/c=0.5$ (dotted line) and $u/c=0.75$ (solid line).  Density is scaled in terms of the background density of the homogeneous system, $n_0$, and position in terms of the healing length, $\xi$. 
}
\label{fig:grey-dark-soliton}
\end{figure}

If $u=0$ then one obtains the stationary {\em black soliton}, whose density profile is,
\begin{equation}
n(x)=n_0{\rm tanh}^2(x/\xi).
\end{equation}
The density of the black soliton goes to zero at its centre, and
the phase jump is a sharp step of $\pi$.  At the opposite speed extreme, the $u=c$ dark soliton has zero density depth and no phase slip, i.e. it is indistinguishable from the background.  For more general speeds, the {\em soliton depth}, that is, the maximum depth of the soliton density depression, follows from Eq. (\ref{eqn:grey-soliton}) as,
\begin{equation}
n_{\rm d}=n_0(1-u^2/c^2).
\label{eqn:nd}
\end{equation}

The phase slip profile across the soliton also varies with speed.  We define the total phase slip $\Delta S$ as the difference between the phases at $\pm \infty$, i.e. $\Delta S=S(x=-\infty)-S(x=\infty)$.  The dark soliton solution becomes $\psi \to \sqrt{n_0} (1+iu/c)$ for $x \to \infty$, and $\psi \to \sqrt{n_0} (-1+iu/c)$ for $x \to - \infty$. Therefore, as we move from $x=\infty$, through the origin, to $x=-\infty$, the phase of 
$\psi$ (that is, the angle $S$
between ${\rm Re}(\psi)$ and ${\rm Im}(\psi)$) 
changes from $S=S_1$ at the point $(1,u/c)$ on the 
complex plane $\psi$,
to $S=\pi/2$ at
$(0,u/c)$, to $S=\pi-S_1$ at $(-1,u/c)$.  Hence the change of the phase
from $x=\infty$ to $x=-\infty$ is $\Delta S=2{\rm arccos}(u/c)$.
Taking the limit $u/c \to 0$, we conclude that the phase jump is $\Delta S=\pi$, as we have said. 
A dark soliton is therefore a 1D {\em phase defect}:
a discontinuity of the quantum mechanical phase (Fig. 4.7).


\begin{figure}[t]
\centering
\includegraphics[width=0.6\columnwidth,angle=0]{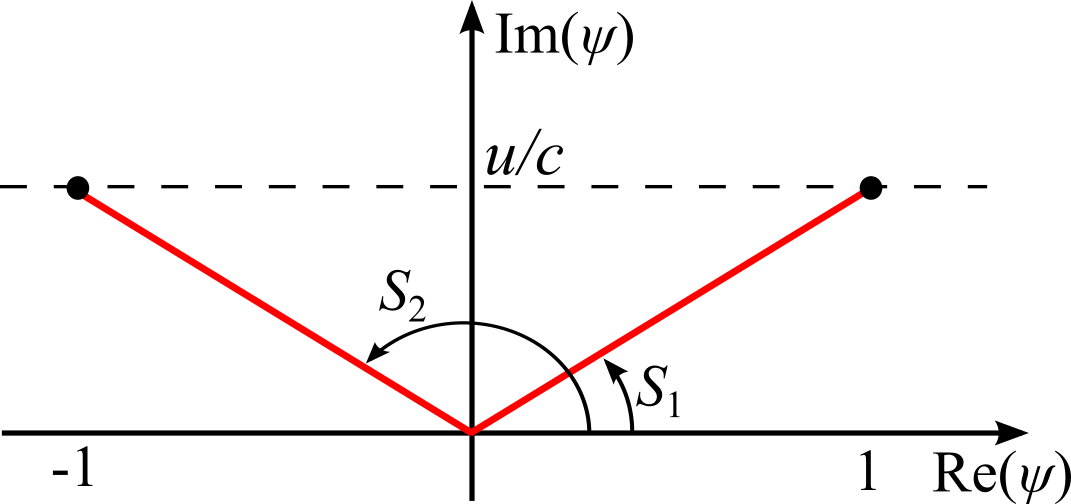}
\caption{Phase jump from $x \to \infty$ (at $(1,u/c)$) to $x \to -\infty$
(at $(-1,u/c)$) corresponding to a dark soliton. 
}
\label{fig:dark-soliton-phase}
\end{figure}

\subsection{Particle-like behaviour}
\label{sec:ds_particle}

The energy and momentum of a dark soliton are $E_{\rm s}=\frac{4}{3}n_0\hbar c B^3$ and $\mathcal{P}_{\rm s}=-2 \hbar n_0 u B/c+2\hbar n_0 \arctan(B c / u)$, as obtained in Problem \ref{dark_soliton_integrals}.  \index{solitons!integrals of motion}Here we show that the dark soliton behaves like a classical particle.  Differentiating  its energy $E_{\rm s}$ and momentum $\mathcal{P}_{\rm s}$ with respect to speed $u$ gives,
\begin{eqnarray}
\frac{{\rm d}E_{\rm s}}{{\rm d}u}=-\frac{4n_0 \hbar u B}{c}, \quad \frac{{\rm d}P_{\rm s}}{{\rm d}u}=-\frac{4n_0 \hbar B}{c}. \label{eqn:ds_ps}
\end{eqnarray}
Note that the energy and momentum both {\it decrease} as the soliton gets faster!
Using these results and the chain rule we can then form,
\begin{equation}
\frac{{\rm d}E_{\rm s}}{{\rm d}\mathcal{P}_{\rm s}}=\frac{{\rm d}E_{\rm s}}{{\rm d}u} \frac{{\rm d}u}{{\rm d}\mathcal{P}_{\rm s}}=u.
\end{equation}
This result informs us that the dark soliton behaves like a classical particle. Its effective mass is defined as $m_{\rm s}=\dfrac{{\rm d}\mathcal{P}_{\rm s}}{{\rm d}u}$, which we know from Eq. (\ref{eqn:ds_ps}) is,
\begin{equation}
m_{\rm s}=-\frac{4n_0 \hbar B}{c}.
\end{equation}

The dark soliton behaves as a classical particle with {\em negative mass}.  This is not surprising given that the dark soliton is an {\em absence of atoms}.   We can also estimate the ratio of the soliton mass to the atomic mass,
\begin{equation}
\frac{|m_{\rm s}|}{m}\sim \frac{4 \hbar n_0}{mc} \sim 4\xi n_0.
\end{equation}
$\xi n_0$ is the number of atoms within a $\xi$-sized length of the system and typically $\xi n_0 \gg 1$ such that the soliton is considerably more massive than an atom.

In the limit of slow solitons ($B \approx 1$) the energetics of the soliton reduces to a particularly simple form.  Taking the Taylor expansion of the soliton energy, Eq. (\ref{eqn:sol_en}, about $v=0$ and up to terms in $v^2$ gives,
\begin{eqnarray}
E(v)&=&E(0)+\frac{{\rm d}E}{{\rm d}u}\Big|_0 u+\frac{1}{2}\frac{{\rm d}^2E}{{\rm d}u^2}\Big|_0 u^2+\mathcal{O}(v^3)
\\
&=&\frac{4}{3}n_0 \hbar c - \frac{2n_0 \hbar u^2}{c}.
\label{eqn:ds_en1}
\end{eqnarray}
Introducing the soliton mass in this limit, $m_{\rm s}=-4n_0 \hbar/c$, we obtain,
\begin{equation}
E(v)=E_0+\frac{1}{2}m_{\rm s}u^2,
\end{equation}
which is the form for a classical particle moving in free space with {\em rest mass} $E_0$ and kinetic energy $m_{\rm s}u^2/2$.  This relation shows that, due to the negative effective mass, slower solitons have greater energy.  Conversely, if the soliton loses energy (due to some dissipative processes) it will speed up!

\subsection{Collisions}
\label{sec:ds_collisions}
\index{solitons!collisions}
Solutions of Eq. (\ref{eqn:1dgpe}) for arbitrary numbers of solitons can be obtained analytically using the inverse scattering transform \cite{ablowitz_1981}, and the two soliton solution of the GPE can be found in Ref. \cite{frantzeskakis_2010}.  The collisions of two dark solitons, with equal speeds, are shown in Figure \ref{fig:dark-soliton-collisions} over different incident speeds.    Notice how the solitons emerge from the collision with unchanged form and speed, one of the fundamental properties of solitons.  For low incoming speed ($u/c<0.5$), the solitons appear to bounce, while for higher speeds ($u/c\geq 0.5$) they appear to pass through each other.  The only overall effect of the collision on the outgoing solitons is a shift in their position, relative to how they would have moved in the absence of another soliton; this is known as the {\em phase shift} during the collision.  


\begin{figure}[t]
\centering
\includegraphics[width=0.9\columnwidth,angle=0]{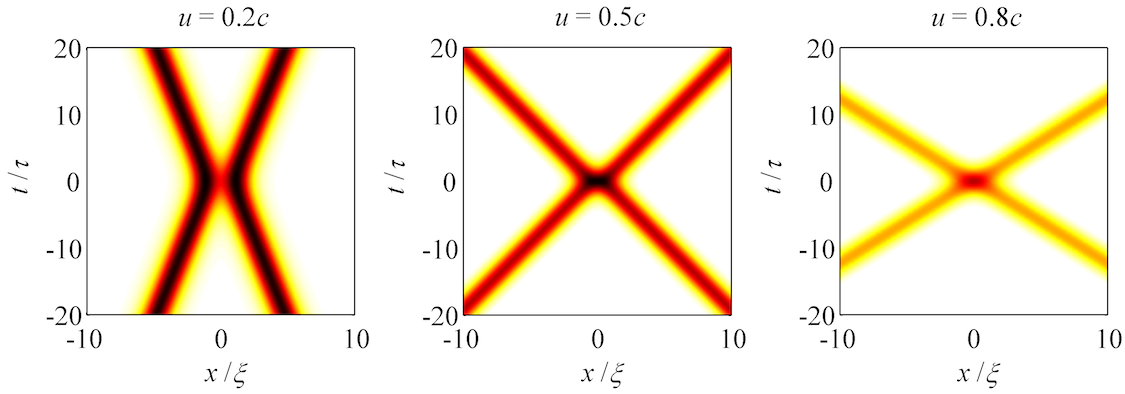}
\caption{Collisions of two dark solitons at different incoming speeds.  White represents the background density $n_0$ and darker shades represents lower densities.  Note that the time axis is centred on the soliton collision.
}
\label{fig:dark-soliton-collisions}
\end{figure}

\subsection{Motion in a harmonic trap}
The special integrable properties that give rise to the soliton solution of Eq. (\ref{eqn:grey-soliton}) hold only if the trapping potential $V(x)$ is zero or uniform in space.  So what happens if a non-uniform potential, typical of real condensates, is applied to the system?  The dark soliton then moves through a continuously changing background density; the soliton, in turn, must adjust to its new surroundings, and in doing so it emits energy in the form of sound waves.  Remarkably, harmonic traps are special in that this decay is prohibited (in fact, the harmonic trap focuses the emitted sound energy back into the soliton).  This stabilizes the soliton, and we find that the trapped soliton retains much of the key soliton properties, albeit with modified dynamics due to the trapping potential. \index{solitons!oscillations}

Figure \ref{fig:dark-soliton-oscillations} shows a simulation of the 1D GPE for a dark soliton in a condensate under a harmonic trap $V(x)=m \omega_x^2 x^2/2$.  The dark soliton is started off away from the trap origin as a black ($v=0$) soliton.  The soliton accelerates towards the trap centre and over-shoots, climbs up the far trap wall and decelerates until it becomes stationary.  This motion repeats, such that the soliton oscillates sinusoidally in the trap.   The motion is akin to a classic harmonic oscillator.  The oscillations continue with uniform amplitude due to the absence of any dissipation in the system.  The soliton oscillation induces a weak ``wobbling'' of the condensate. Note that if the soliton is black and at the trap centre, then it is stationary.  Indeed, this state is the first excited state of the trapped condensate. If there were some dissipation acting on the soliton to reduce its energy, its oscillation amplitude will {\em grow} due to its negative effective mass (in contrast to a conventional damped oscillator).  This effect is termed {\em anti-damping}.   

\begin{figure}[t]
\centering
\includegraphics[width=0.9\columnwidth,angle=0]{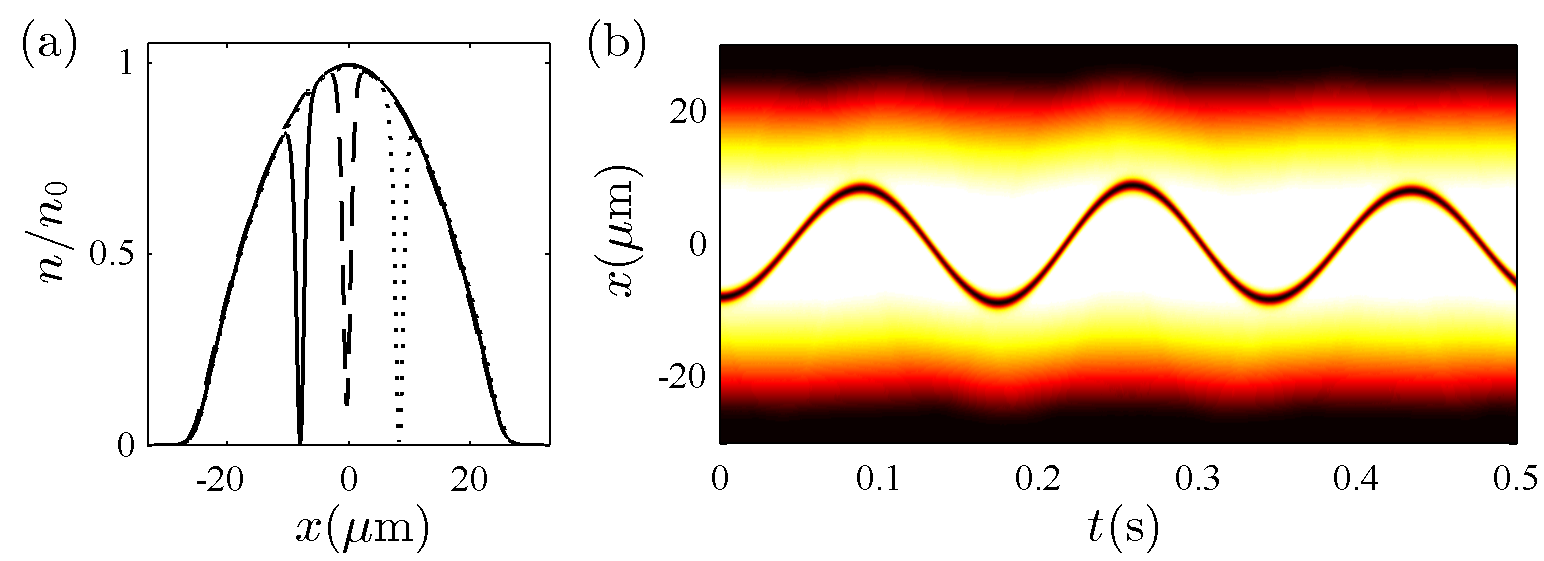}
\caption{A dark soliton oscillating in a harmonically-trapped condensate, simulated by the 1D GPE.  The trap is $V(x)=m \omega_x^2 x^2/2$, with trap frequency $\omega_x=2\pi \times 8$Hz.  (a) shows the density profile at three times, while (b) shows a space-time plot of the condensate density (the dark soliton appearing as the oscillating dark line).}
\label{fig:dark-soliton-oscillations}
\end{figure}

One might expect the soliton to oscillate at the frequency of the trap.  For the simulation in Figure \ref{fig:dark-soliton-oscillations} the trap frequency is $\omega_x=2 \pi \times 8$Hz, whereas the soliton has a frequency $\omega_{\rm s} = 2 \pi / T_{\rm s} \approx 2 \pi \times 5.5$Hz, where $T_{\rm s}\approx 0.18$s is the observed soliton period.  We can interpret this difference as follows.   Assuming that the background density is slowly-varying in space, we can define the soliton energy as per the homogeneus system (see top of Section \ref{sec:ds_particle}) but where the uniform density is replaced by its local value, $n(x)$.  We then obtain,
\begin{equation}
E_{\rm s}(u,x)=\frac{4}{3}\frac{\hbar}{\sqrt{m}g}\left(n(x)g - m u^2 \right)^3.
\end{equation}  
Again, we take the condensate profile to follow the Thomas-Fermi form, $n(x)g=\mu-V(x)=\mu-m\omega^2 x^2/2$.  Inserting into the above equation gives,
\begin{equation}
E_{\rm s}(u,x)=\frac{4}{3}\frac{\hbar}{\sqrt{m}g}\left(\mu - \frac{1}{2}m\omega_x^2 x^2 - m u^2 \right)^3.
\end{equation}  
We proceed to expand this expression for slow solitons ($u/c\ll1$) and close to the origin, via a two-dimensional Taylor series,
\begin{eqnarray}
E_{\rm s}(u,x)&=&E(0,0)+x \frac{\partial E}{\partial x}\Big|_{(0,0)} + u\frac{\partial E}{\partial u}\Big|_{(0,0)} \nonumber \\
&+&\frac{1}{2} \left[u^2 \frac{\partial^2 E}{\partial u^2}\Big|_{(0,0)} +2ux \frac{\partial^2 E}{\partial x \partial z}\Big|_{(0,0)} +x^2 \frac{\partial^2 E}{\partial x^2}\Big|_{(0,0)}  \right]+\mathcal{O}(x^3,u^3). \nonumber
\end{eqnarray} 
This leads to the result,
\begin{equation}
E_{\rm s}(u,x)=E_0 + \frac{1}{2}m_{\rm s}u^2+ \frac{1}{4}m_{\rm s}\omega_x^2 x^2,
\end{equation}
where $m_{\rm s}=-4\hbar n_0c$ is the low-speed soliton effective mass and $E_0=4\hbar \mu^{3/2}/4\sqrt{m}g$ is the rest mass (as used for Eq. (\ref{eqn:ds_en1})).  For a mass $m$ obeying the classic harmonic oscillator $\ddot{x}+\omega_x^2 x=0$,  the corresponding expression is $E=E_0+m u^2/2+m\omega_x^2 x^2/2$.  By comparison we see that the soliton behaves like an oscillator with effective frequency,
\begin{equation}
\omega_{\rm s}=\frac{\omega_x}{\sqrt{2}}.
\end{equation}
This result was first predicted in Ref. \cite{busch_2000}.  For the example in Fig. \ref{fig:dark-soliton-oscillations}, this predicts $\omega_{\rm s}=2 \pi \times 5.66$Hz and $T_{\rm s}=0.177$s, which is in excellent agreement with the simulations.   It has also been found to agree well with experimental observations.\index{solitons!oscillations}

\subsection{Experiments and 3D effects}

Dark solitons were first created in condensates in experiments \cite{burger_1999,denschlag_2000}, although they were short-lived.  A more recent experiment, working in a highly-elongated, effectively one-dimensional geometry and at very cold temperature, generated dark solitons which persisted for several seconds, equivalent to tens of oscillations in the trap \cite{becker_2008}.  The dynamics were in good agreement with the predictions of the 1D GPE.  The eventual disappearance of the soliton was attributed to thermal dissipation acting on the soliton; this causes the soliton to lose energy and anti-damp, eventually becoming indistinguishable from the rest of the condensate. 

The typical approach to generate a dark soliton, as used in the above experiments, is to first form a condensate in the trap, and then briefly illuminated a portion using masked laser light.  Due to the atom-light interaction, the illuminated part of the condensate develops a different phase to the   un-illuminated part, such that an effective 1D step in the phase is created.  This then evolves into one or more dark solitons.  In 3D, these solitons appears as stripes of low density, aligned perpendicular to their axis of propagation, and a phase step along this axis, as illustrated in Fig. \ref{fig:snake} (left).

If the condensate is too wide, the 3D dark soliton is not dimensionally stable.  The soliton stripe becomes unstable to transverse perturbations, causing a bending of the soliton stripe, known as the {\em snake instability}.  \index{solitons!snake instability} \index{solitons!in 3D}The soliton stripe gets torn apart into vortex rings, which are stable excitations in 3D condensates.  This decay is illustrated in Fig. \ref{fig:snake}(right).  To prevent the snake instability, the condensate should be quasi-one-dimensional (with a transverse size of the order or less than the healing length $\xi$ - see Section \ref{sec:reduced_dims}). At the crossover between 1D and 3D, it is possible to form 
{\em solitonic vortices}, which have combined properties of dark solitons 
and vortices \cite{Brand_2002,Serafini-2015}. \index{vortex!solitonic vortex}

\begin{figure}[t]
\centering
\includegraphics[width=0.7\columnwidth,angle=0]{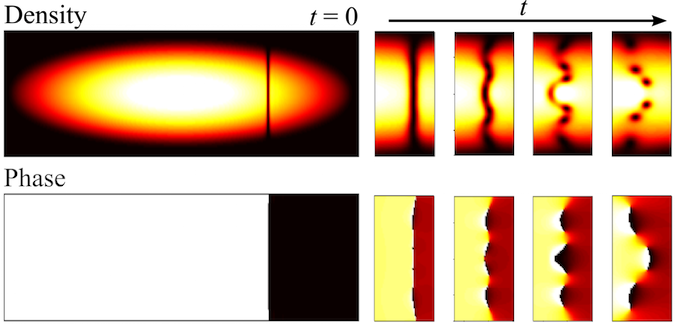}
\caption{Two-dimensional density and phase images through a 3D condensate, which has a 3D dark soliton positioned initially to one side (left).  The initial phase appears as a step profile.  If the condensate is too wide, the soliton undergoes the snake instability (right), leading to its decay into vortex rings (which appear as vortex-antivortex pairs in this 2D image).  }
\label{fig:snake}
\end{figure}

\section{Bright solitons}
\label{sec:bright-solitons}

For attractive interactions ($g<0$) the 1D GPE (\ref{eqn:1dgpe}) supports {\em bright solitons}.  In contrast to dark solitons, these are self-trapped condensates in which the attractive interactions overcome wavepacket dispersion. We saw in Section \ref{sec:dispersion} that the homogeneous condensate in 1D with attractive interactions is unstable.  Actually the stable ground state is a bright soliton.  A detailed review of bright solitons in condensates can be found in Ref. \cite{billam_2012}. \index{solitons!bright solitons}

The general solution for a single bright soliton, containing $N$ atoms and moving at speed $u$, is,
\begin{equation}
\psi(x,t)=\sqrt{\frac{N}{2\xi_{\rm s}}} \textrm{sech}\left(\frac{x-ut}{\xi_{\rm s}} \right)\exp\left[if(x,t) + i\phi \right],
\label{eqn:bs}
\end{equation} 
where $\xi_{\rm s}=2\hbar^2/m|g|N$ characterises the soliton width, $\phi$ is a global phase offset, and
\begin{equation}
f(x,t)=\frac{mux}{\hbar}-\frac{t}{\hbar}\left(\frac{mu^2}{2}-\frac{\hbar^2 }{2m\xi_{\rm s}^2} \right),
\end{equation}\index{solitons!solutions}
is a time- and space-dependent phase factor.  The soliton maintains a sech-squared density profile, shown in Fig. \ref{fig:1D_energy}(a), as it propagates.  For stronger attractive interactions and/or more atoms, the soliton is narrower, indicative of a stronger binding effect.    For a dark soliton, the density profile of the soliton is related to its speed; for bright solitons, the density profile and speed are decoupled, and a bright soliton can take on any speed $u$.

To understand the manner is which the soliton is supported, we take a variational approach.  As an ansatz for the soliton solution, we adopt a
Gaussian wavepacket of width $\ell$, normalised to $N$ atoms, i.e., \index{variational method}
\begin{equation}
\psi(x)=\frac{N^{1/2}}{\pi^{1/4}\ell^{1/2}}
\exp\left(-\frac{x^2}{2 \ell^2}\right).
\label{eqn:ansatz2}
\end{equation}
Using Eq.~(\ref{eqn:energy_int}), the energy-per-particle $E/N$ of this wavepacket is,
\begin{equation}
\frac{E(\ell)}{N}=\frac{\hbar^2}{4 m \ell^2}+\frac{gN}{2\sqrt{2\pi} \ell}.
\label{eqn:en3}
\end{equation}
Consider the form of $E/N$ for two regimes of interactions (illustrated in Fig. \ref{fig:1D_energy}(a)):
\begin{itemize}
\item For $g \geq 0$, $E/N$ decreases monotonically with $\ell$.  All states are prone to expanding and there is no stationary 
state.  For $g=0$ the expansion is driven by dispersion, while for $g>0$ the repulsive interactions also contribute to the expansion\footnote{If a harmonic potential is included, then a positive $x^2$ term is added to $E(\ell)/N$ which then does support an energy minimum, representing the ground trapped condensate.}.
\item For $g<0$ there exists 
a local minimum in $E(\ell)/N$,
implying that a stable stationary 
state exists.  This is due to a delicate balance between the dispersive term, which scales like $1/\ell^2$ and dominates for small $\ell$, and the attractive nonlinear term, which scales like $-1/\ell$ and dominates elsewhere.  
\end{itemize}

For $g<0$ the variational width is found by locating the position of the energy minimum.  Differentiating $E/N$ with respect to $\ell$, setting to zero and 
rearranging gives the width of the variational solution, $\ell_{\rm v}=\frac{\sqrt{2\pi}\hbar^2}{m|g|N}$, which agrees well with the true solution - see Figure \ref{fig:1D_energy}(a). 

\begin{figure}[t]
\centering
\includegraphics[width=0.8\columnwidth]{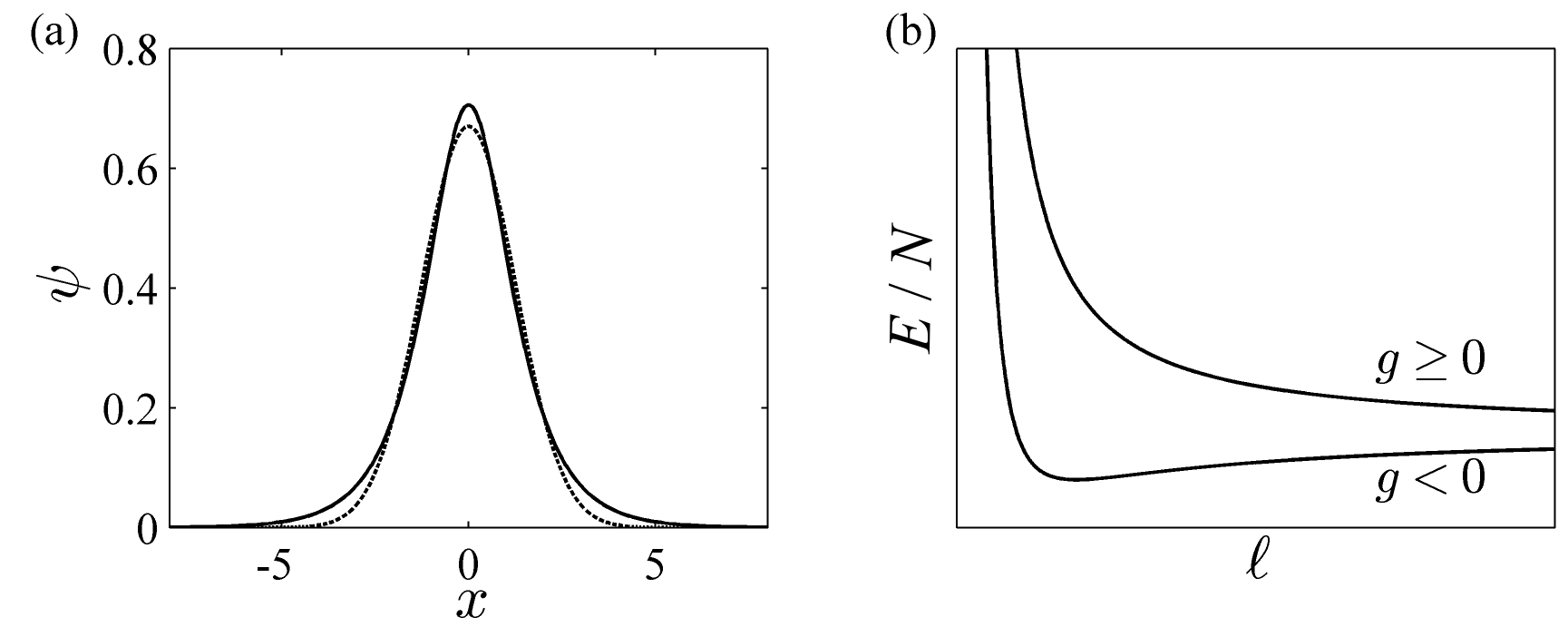}
\caption{(a) Density profile $n(x)=|\psi|^2$ of the bright soliton solution of Eq. (\ref{eqn:bs}), taking $u=0$ (solid line).   The variational solution of the Gaussian ansatz (\ref{eqn:ansatz2}) is in good agreement (dashed line).  (b)  The energy-per-particle $E/N$ of the Gaussian ansatz, given by Eq. (\ref{eqn:en3}), versus the width of the ansatz, $\ell$, for two regimes of interactions.  
} 
\label{fig:1D_energy}
\end{figure}

\subsection{Collisions}

Being a self-contained condensate, a bright soliton has a global phase, $\phi$, and this significantly affects the manner in which bright solitons interact.   Dark solitons, in contrast, have no such phase freedom.  

To get some insight, consider two bright solitons, each with $N$ atoms.  Soliton 1 begins at position $-x_0$ (with $x_0>0$) and propagates to the right with speed $u$, while soliton 2 begins at position $x_0$ and propagates to the left with the same speed. Their  individual solutions are,
\begin{eqnarray}
\psi_1(x,t)&=&\sqrt{\frac{N}{2\xi_{\rm s}}} {\rm sech}\left(\frac{x+x_0-ut}{\xi_{\rm s}}\right) e^{i f(x,t)} e^{i\phi_1}, 
\\
\psi_2(x,t)&=&\sqrt{\frac{N}{2\xi_{\rm s}}} {\rm sech}\left(\frac{x-x_0+ut}{\xi_{\rm s}}\right) e^{i f(x,t)}e^{i\phi_2}.
\label{eqn:phis}
\end{eqnarray}
Note that due to the symmetric configuration, both solitons have the same time- and space-dependent phase factor $e^{i f(x,t)}$, but we allow for different global phase offsets, $\phi_1$ and $\phi_2$.  

Assuming that the solitons are well-separated we can construct their superposition as $\psi'=\psi_1+\psi_2$\footnote{The superposition theorem does not apply
to the GPE since it is a nonlinear equation;
constructing a superposition is only a valid approximation if the density is 
low.  This condition
is satisfied here since we are concerned with the weak overlap between well-separated solitons.}.  We proceed to calculate the density profile of this superposed state, $|\psi'|=|\psi_1+\psi_2|^2$,
\begin{eqnarray}
|\psi'(x,t)|^2&=&\left||\psi_1| e^{if(x,t)} e^{i\phi_1} +|\psi_2| e^{if(x,t)} e^{i\phi_2} \right|^2
\\
&=& |\psi_1|^2+|\psi_2|+|\psi_1||\psi_2|\left(e^{i \Delta \phi}+e^{-i\Delta \phi} \right),
\end{eqnarray}
where we have introduced the {\em relative phase} $\Delta \phi = \phi_2 - \phi_1$.  Using the identity $\cos \theta = (e^{i\theta}+e^{-i\theta})/2$ we obtain,
\begin{eqnarray}
|\psi'(x,t)|^2= |\psi_1|^2+|\psi_2|+2 |\psi_1||\psi_2| \cos \Delta\phi.
\end{eqnarray}
Let us see how this affects the overlap of the two solitons
 by calculating the density at their midpoint (the origin). Introducing the form of $\psi_1$ and $\psi_2$ from Eq. (\ref{eqn:phis}) and setting $x=0$ gives,
\begin{equation}
|\psi'(0,t)|^2=\frac{N}{2 \xi_{\rm s}} \left[2{\rm sech}^2\left(\frac{x_0-ut}{\xi_{\rm s}}\right) +2 \cos \Delta \phi ~{\rm sech}^2\left(\frac{x_0-ut}{\xi_{\rm s}}\right) \right].
\end{equation}
If $\Delta \phi=0$ then the density at the midpoint reinforces (constructive interference), in other words, the solitons overlap with each other.  However, for $\Delta \phi =\pi$ the midpoint density is forced to zero (destructive interference), and the overlap of the solitons is prohibited.  For intermediate values of the relative phase, the overlap varies smoothly between these extremes. \index{interference}

\begin{figure}[b]
\centering
\includegraphics[width=0.85\columnwidth]{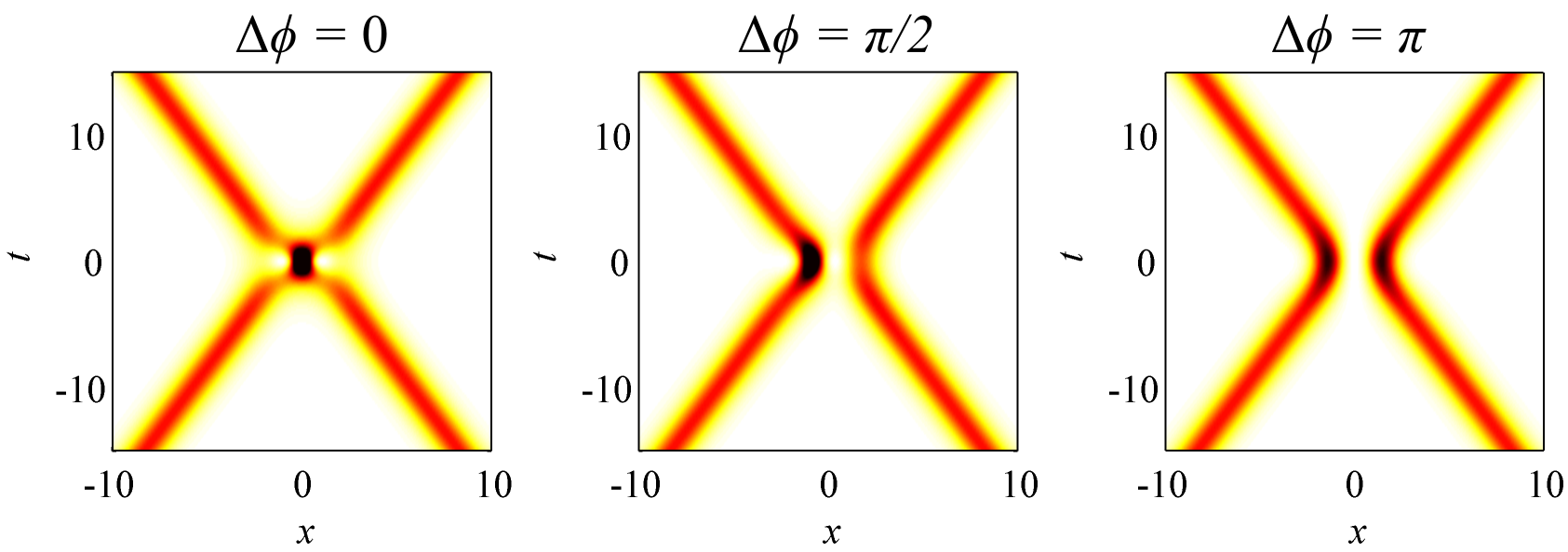}
\caption{Density profile $n(x,t)=|\psi(x,t)|^2$ during the collision of two bright solitons, with speed $u=0.2$ and for different relative phases. 
} 
\label{fig:bright_soliton_collisions}
\end{figure}

Figure \ref{fig:bright_soliton_collisions} shows the collisions for different relative phases.  True to their solitonic character, the solitons emerge unscathed from the collision, barring a shift.  The role of relative phase becomes clear: for $\Delta \phi=0$ the solitons merge at the point of collision, while for $\Delta \phi=\pi$ overlap is prohibited and they appear to bounce.  In between the collision becomes asymmetric.  Despite these different behaviours during the collision, it is remarkable that the outgoing solitons are independent of $\Delta \phi$.  \index{solitons!collisions}

\subsection{Experiments and 3D effects}
The addition of a harmonic potential to the 1D GPE (\ref{eqn:1dgpe}) breaks the integrability of the system, and true soliton solutions no longer exist.  The state adopts behaviours of a trapped condensate, such as collective modes.  The ground state becomes narrower and more peaked than the soliton solution, and in the limit of a very strong trap, the ground state tends towards the Gaussian ground harmonic oscillator state. As such the width of the wavepacket varies between the two limiting cases, $\xi_{\rm s}$ and $l_x$, respectively. However, the solutions continue to show soliton-like behaviour: an initially off-centre soliton will oscillate in the trap (the dipole mode) with unchanged form, and two solitons will collide repeatedly and emerge unscathed.  

In reality, bright solitons are 3D objects, and this introduces the collapse instability discussed in Section \ref{sec:weak_ints}.  
This physical effect is not modelled within the 1D GPE.  This difference can be understood as follows.  For a generalized static attractively-interacting condensate of characteristic size $\ell$ in $\mathcal{D}$ dimensions, its kinetic energy (due to zero-point motion) scales as $\mathcal{D}/\ell^2$ and its interaction energy scales as $-1/\ell^\mathcal{D}$.  For $\mathcal{D}=1$ we recover the behaviour discussed above - the kinetic energy term always wins in the $\ell \rightarrow 0$ limit, negating a collapse instability.  For $\mathcal{D}=3$, however, the negative interaction term always dominates in this limit, such that the packet can lower its energy by shrinking, i.e. a collapse instability. $\mathcal{D}=2$ is a borderline case where stability to collapse depends on the system parameters.  \index{collapse}

A 3D bright soliton can be formed within a {\em waveguide potential}, $V(r,z)=m \omega_r^2 r^2/2$, which has tight harmonic confinement in the transverse directions but is untrapped along $z$. Then a 3D bright soliton can form, which is self-trapped along $z$.  This state is stable to collapse up to a critical interaction strength $N |a_{\rm s}|/l_r\approx 0.7$. \index{solitons!in 3D}

Bright solitons were first formed with condensates in 2002 \cite{strecker_2002, khaykovich_2002}.  They are typically generated as follows.  A stable repulsive condensate is first formed in a highly-elongated harmonic trap.  The interaction strength is then tuned to being attractive by means of a magnetic Feshbach resonance.  In these early experiments, the critical number of atoms was exceeded, driving a collapse.  Out of the collapse one or more bright solitons formed.  More recent experiments form bright solitons by choosing parameters which avoid the collapse instability.  The weak axial trap is either kept on, in which case the soliton/s oscillate axially, or is switched off, such that the 
solitons propagate freely.  An experimental proof of bright solitons is shown in Fig. \ref{fig:bs_exp}.  For repulsive interactions it was seen that the condensate expanded over time, while for attractive interactions it was seen to maintain its shape, characteristic of a soliton.   More recent experiments have studied the collisions of bright solitons with each other \cite{nguyen_2014} and with potential barriers \cite{marchant_2013}.

\begin{figure}[b]
	\centering
		\includegraphics[width=0.9\textwidth]{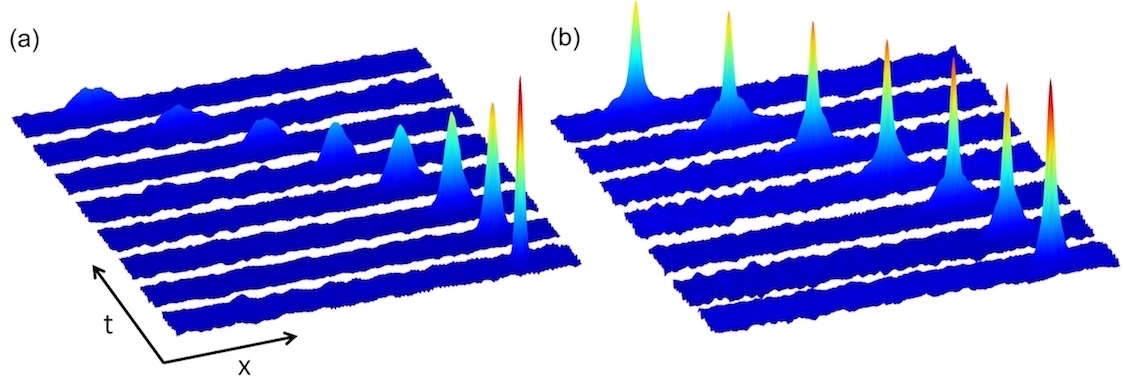}
	\caption{(a) As a repulsive BEC travels along a waveguide (with tight transverse harmonic trapping and very weak axial confinement) it spreads out. (b) For attractive interactions, the condensate is seen to maintain its shape over time, characteristic of a bright soliton.  Image courtesy of S. L. Cornish (University of Durham).}
	\label{fig:bs_exp}
\end{figure}

\section*{Problems}
\addcontentsline{toc}{section}{Problems}

\begin{prob}
\label{dark_soliton_integrals}
For a dark soliton, the integrals of motion in Eqs. 
(\ref{eqn:sol_norm},\ref{eqn:sol_mom},\ref{eqn:sol_en}) are renormalized 
so as to remove the contribution from the background and lead to finite 
values, \index{solitons!integrals of motion}
\begin{eqnarray}
N_{\rm s}&=&\int \limits_{-\infty}^{+\infty}(n_0-|\psi|^2)~{\rm d}x \nonumber 
\\
\mathcal{P}_{\rm s}&=&\frac{i\hbar}{2} \int \limits_{-\infty}^{+\infty}\left(\psi \frac{\partial \psi^*}{\partial x} -\psi^* \frac{\partial \psi}{\partial x}\right)\left(1-\frac{n_0}{|\psi|^2} \right) {\rm d}x \nonumber
\\
E_{\rm s}&=&\int \limits_{-\infty}^{+\infty}\left(\frac{\hbar^2}{2m}\left|\frac{\partial \psi}{\partial x} \right|^2 + \frac{g}{2}(|\psi|^2-n_0)^2 \right) {\rm d}x \nonumber 
\end{eqnarray}
Evaluate these integrals using the dark soliton solution 
Eq.~(\ref{eqn:grey-soliton}), leaving your answers in terms of 
$\xi, n_0, u$ and $c$.

\end{prob}

\begin{prob}
\label{dark_soliton_turning}
Consider a dark soliton in a harmonically-trapped condensate.  Approximating the background condensate with the Thomas-Fermi profile $n(x)=n_0(1-x^2/R_x^2)$ (for $x \leq R_x$, otherwise $n=0$) and treating the soliton depth $n_{\rm d}$ to be constant, obtain an expression for the soliton speed as a function of its position $x$ and depth $n_{\rm d}$.  Hence obtain an expression for the turning points of its motion.  


\end{prob}

\begin{prob}
\label{soliton_mu}
Show that the static ($u=0$) bright soliton solution, obtained from Eq. (\ref{eqn:bs}), is a solution to the 1D attractive time-independent GPE with $V(x)=0$, i.e,
\begin{equation}
\mu \psi = -\frac{\hbar^2}{2m}\frac{{\rm d}^2 \psi}{{\rm d}x^2}-|g||\psi|^2 \psi,
\end{equation}
and hence determine an expression for the chemical potential of the soliton.
\end{prob}

\begin{prob}
\label{bright_soliton_integrals}
Using the general bright soliton solution, Eq. (\ref{eqn:bs}), evaluate the soliton integrals of motion according to Eqs. (\ref{eqn:sol_norm}),(\ref{eqn:sol_mom}) and (\ref{eqn:sol_en}).  The soliton solution is already normalized to the number of atoms, $N$.  Show that the bright soliton behaves as a classical particle with positive mass.   \index{solitons!integrals of motion}


\end{prob}

\begin{prob}
\label{3d_bright_soliton}
Consider a 3D bright soliton in a cylindrically-symmetric waveguide with tight harmonic confinement (of frequency $\omega_r$) in $r$ and no trapping along $z$.  We can construct the ansatz for the soliton, 
\begin{equation}
\psi(z,r)=A {\rm sech} \left(\frac{z}{l_r \sigma_z} \right) \exp\left(-\frac{r^2}{2 l_r^2 \sigma_r^2} \right),
\end{equation}
where $l_r=\sqrt{\hbar/m \omega_r}$ is the harmonic oscillator length in the radial plane and $\sigma_r$ and $\sigma_z$ are the dimensionless variational length parameters. \index{solitons!in 3D}

\begin{itemize}
\item[(a)]~ Normalize the ansatz to $N$ atoms to show that $A=(N/2 \pi l_r^3 \sigma_r^2 \sigma_z)$.
\item[(b)] ~Show that the variational energy of this ansatz is,
\begin{equation}
E = \hbar \omega_r N \left(\frac{1}{6 \sigma_z^2} + \frac{1}{2 \sigma_r^2} + \frac{\sigma_r^2}{2} +  \frac{\gamma}{3 \sigma_r^2 \sigma_z}\right),
\end{equation}
where $\gamma = N a_{\rm s}/l_r$.
\item[(c)] ~ Make a 2D plot of the variational energy per particle, $E/N\hbar \omega_r$ (scaled by the transverse harmonic energy) as a function of the two variational length parameters, and plot this for $\gamma=-0.5$.  Locate the variational solution in this 2D ``energy landscape''.  Repeat for $\gamma=-1$; what happens to the variational solution?  By varying $\gamma$ estimate the critical value at which the solutions no longer exist (and they become prone to collapse).  
 \end{itemize} 
\end{prob}

\begin{prob}
\label{v_Landau}
Consider an object of mass $M$ moving at velocity ${\bf v}_i$ which creates an
excitation of energy $E$ and momentum ${\bf p}=\hbar {\bf k}$.
Show that Landau's critical velocity, $v_c={\rm min}(E/p)$, is equivalent
to $dE/dp=E/p$.
Compare Landau's critical velocity for the ideal gas (dispersion relation
$E(p)=p^2/2M$) against the weakly-interacting Bose gas. Finally
show that in liquid helium~II, Landau's critical velocity is
$v_c\approx 60~{\rm m/s}$. Hint:
assume that near the roton minimum the dispersion relation, shown in 
Fig.~\ref{fig:waves}(b), has the
approximate form $E(p)=\Delta_0 +(p-p_0)^2/(2 \mu_0)$ where (at very low
pressure) $\Delta_0=1.20 \times 10^{-22}~{\rm J}$ is the energy gap,
$p_0=\hbar k_0=2.02 \times 10^{-24}~{\rm kg~m/s}$ is the momentum 
at the roton minimum,
$\mu_0=0.161~m_4$ is the effective roton mass, and 
$m_4=6.65 \times 10^{-27}~{\rm kg}$ is the mass of one $^4$He atom.
\end{prob}

\chapter{Vortices and Rotation}
\label{chap:vortices} 


\abstract{As well as being free from viscosity, the Bose-Einstein condensate has another striking property - it is constrained to circulate only through the presence of whirlpools of fixed size and quantized circulation.  In contrast, in conventional fluids, the eddies can have arbitrary size and circulation.   Here we establish the form of these quantum vortices, their key properties, and how they are formed and modelled.}

\section{Phase defects}
\label{}
The condensate's wavefunction is a complex quantity. We have seen that
it can be written as $\Psi({\bf r},t)=R({\bf r},t)e^{i S({\bf r},t)}$
(Madelung transform), where $R({\bf r},t)$ and $S({\bf r},t)$
are respectively the phase and amplitude distributions at time $t$.
Consider following a closed path $C$ of arbitrary shape
through a region of the condensate.  As we go around the path, 
the integrated change in the phase is

\begin{equation}
\Delta S = \oint_C \nabla S \cdot {\rm d}\bell,
\end{equation}

\noindent
where the vector ${\rm d}\bell$ is the line element of integration.
Let the wavefunction be $\Psi_0$ and $\Psi_1$ respectively
at the starting point and at the final point of $C$. Since the two
points are the same and $\Psi$ must be single-valued, the condition
$\Psi_1=\Psi_0$ means that,

\begin{equation}
\Delta S =2\pi q,
\qquad
q=0,\pm 1, \pm 2, \cdots
\label{eqn:phase_defect}
\end{equation}

If the integer number $q \neq 0$  then, somewhere within the region
enclosed by $C$, there must be a {\em phase defect}, a point where
the phase wraps by the amount $2\pi q$. At this point the phase of the wavefunction takes on every value, and the only way that $\Psi$ can remain single-valued here is if $\Psi$ is exactly zero.  

\section{Quantized vortices}

What does the presence of a phase defect mean for the condensate as a fluid?  
Recalling that the phase distribution defines the fluid's velocity via
${\bf v}=(\hbar/m) \nabla S$, Eq.~(\ref{eqn:phase_defect}) implies that
the {\em circulation} $\Gamma$ around the path $C$ is 
either zero or a multiple
of the {\em quantum of circulation} $\kappa$,
\begin{equation}
\Gamma=\oint_C {\bf v} \cdot {\rm d}\bell= q \kappa,
\qquad
\kappa=\frac{h}{m}.
\label{eqn:kappa}
\end{equation}
This important result (the {\em quantization of the circulation})\index{circulation}
tells us that the condensate flows very differently
from ordinary fluids, where the circulation takes arbitrary values.

Assume that $q \neq 0$, and that the path $C$ is a circle of radius $r$
centred at the singularity.  Consider the simple case of two-dimensional
flow in the $xy$ plane. Using polar coordinates $(r, \theta)$,
the line element is ${\rm d}\bell=r {\rm d}\theta~\ehat_{\theta}$, where  
$\ehat_{\theta}$  is the unit vector in the azimuthal direction $\theta$.
Then the circulation becomes,
\begin{equation}
\Gamma=\oint_C {\bf v} \cdot {\rm d}\bell= 
\int^{2\pi}_0 r {\bf v} \cdot \ehat_{\theta}~ {\rm d}\theta=
2\pi r  v_{\theta}.
\end{equation}
Comparison with Eq.~(\ref{eqn:kappa}) shows that the fluid's azimuthal speed 
around the singularity is,
\begin{equation}
v_{\theta}=\frac{q\hbar}{m r}=\frac{q \kappa}{2 \pi r}.
\label{eqn:vortex-flow}
\end{equation}
Since the
condensate is a fluid without viscosity, this flow around the 
singularity should go on forever, at least in principle!

For $q\neq 0$, Eq.~(\ref{eqn:vortex-flow}) tells us that 
the velocity around the singularity decreases to zero at infinity
($v_{\theta} \to 0$ as $r \to \infty$), and that, as we approach the axis,
the flow becomes faster and faster, and diverges ($v_{\theta} \to \infty$
as $r \to 0$). 
If we increase $q$, the flow speed increases discontinuously, because 
$q$ takes only discrete values.  The sign of $q$ determines the 
direction of the flow (clockwise or anticlockwise) around the singularity. 

We now have a better picture of the nature of the singularity: it is
a {\em quantized vortex line}, a whirlpool in the fluid. The quantity
$q$ is called the {\it charge} of the vortex. \index{vortex!charge}
Figure~{\ref{fig:line} (left)
represents a straight vortex line through the
origin, parallel to the $z$ axis. Since the flow is the same on all planes
perpendicular to the $z$ axis,  the flow of the (three-dimensional)
straight vortex can be more simply described as the flow due to
a two-dimensional {\em vortex point} on the $xy$ plane, as
in Fig.~\ref{fig:line} (middle).
If these conditions are not met, such as the curved vortex line shown in Fig.~\ref{fig:line}(right), then the flow
is fully three-dimensional and cannot be represented by a vortex point. 

\begin{figure}[h]
\centering
\includegraphics[width=0.9\columnwidth,angle=0]{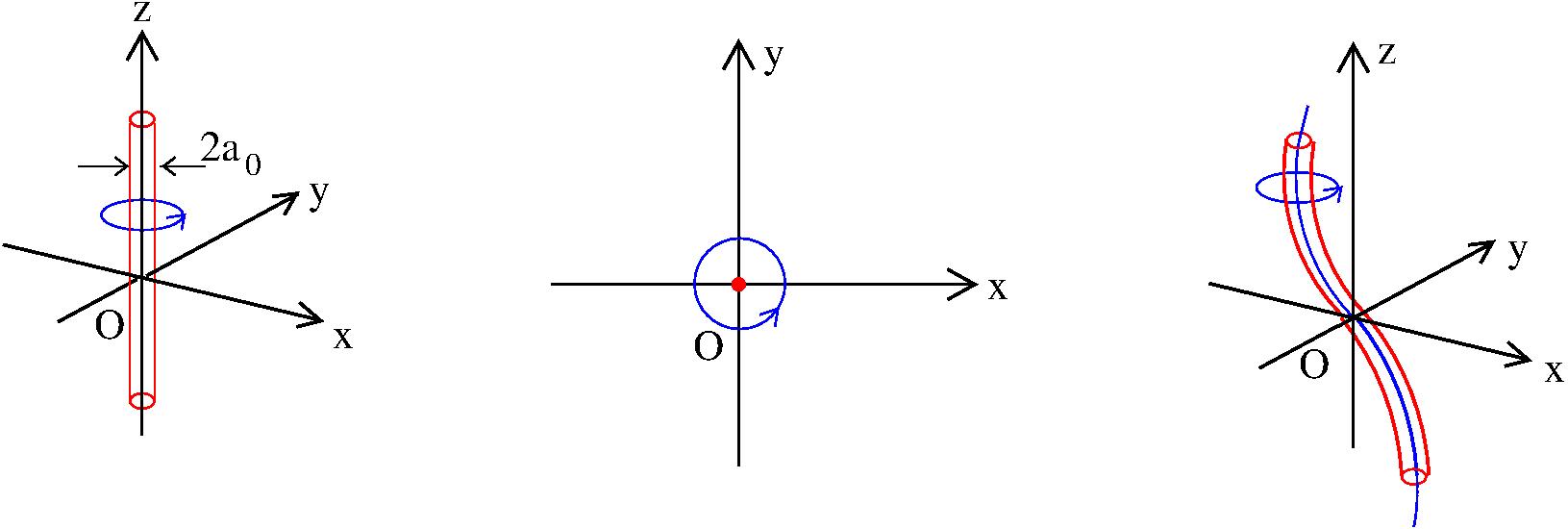}
\caption{
Left: Schematic (three-dimensional)
straight vortex line through the origin and parallel to the
$z$ axis. The red tube around the vortex axis of radius $a_0$ represents the
vortex core. 
Middle: Since the vortex line is straight, it suffices to consider the
two-dimensional flow of a vortex point
on the $xy$ plane (the flow on other planes parallel to the $xy$ plane will be the same).
Right: For a more general bent vortex line the flow is fully three-dimensional.
}
\label{fig:line}
\end{figure}

\section{Classical vs quantum vortices}
\index{vortex}
The flow of the condensate is  different from the
flow of an ordinary fluid in two respects.  Firstly, and as we showed in Section \ref{sec:fluid},  it is inviscid (there is
no viscosity to slow down the flow and bring it to a stop).  Secondly,
the circulation is quantized, as we showed above. To appreciate the second difference we recall
the {\em vorticity} field (the local rotation), defined as,
\begin{equation}
\bom=\nabla \times {\bf v}.
\end{equation}

\noindent
The following examples illustrate velocity fields with the associated
vorticity fields:

\begin{itemize}
\item[(i)]~Consider water inside a bucket rotating at 
constant angular velocity $\Omega$. We use cylindrical coordinates
$(r,\theta,z)$ where $z$ 
is the axis of rotation\footnote{We recall that 
in cylindrical coordinates, the curl of the
vector ${\bf A}=(A_r,A_{\theta},A_z)$ is
\begin{displaymath}
\nabla \times {\bf A}=\left( 
\frac{1}{r} \frac{\partial A_z}{\partial \theta}
-\frac{\partial A_{\theta}}{\partial z}, 
\frac{\partial A_r}{\partial z} - \frac{\partial A_z}{\partial r},
\frac{1}{r}\frac{\partial (r A_{\theta})}{\partial r}
-\frac{1}{r}\frac{\partial A_r}{\partial \theta}
\right).
\end{displaymath}
}.
The velocity  field is
${\bf v}=v_{\theta}\ehat_{\theta}= \Omega r\ehat_{\theta}$ and the vorticity 
is  $\bom=2 \Omega \ehat_z$
(where $\ehat_{\theta}$ and $\ehat_z$ are the unit vectors along 
 $\theta$ and $z$).  The azimuthal speed $v_{\theta}$ of this flow as a 
function of $r$
is shown by case (i) of Fig.~\ref{fig:rotations}(a). This flow is called
{\em solid body rotation}.

\item[(ii)]~
As derived above, the velocity field around a vortex line in a condensate is
$v_{\theta}=q\hbar/(m r)$, shown by case (ii) in Fig.~\ref{fig:rotations}(a). 
It is easy to verify that
its vorticity is zero:  we say that this flow is {\em irrotational}.
Physically, a parcel of fluid which goes around the vortex
axis does not `turn' (as it does in solid body rotation), but
retains its orientation 
(like a gondola of a Ferris wheel); this flow is depicted 
in case (ii) of Fig.~\ref{fig:rotations}(b).  The property of irrotationality also
follows mathematically: the condensate's
velocity is proportional to the gradient of the quantum mechanical
phase, and the curl of a gradient is always zero. However,  
the singularity itself contributes vorticity according to, \index{vorticity}
\begin{equation}
\bom= \kappa \delta^2({\bf r}) \ehat_{z},
\label{eqn:vort}
\end{equation}
where $\delta^2({\bf r})$ is the two-dimensional delta function satisfying 
$\delta^2({\bf r}=0)=1$ and $\delta^2({\bf r}\neq 0)=0$.
At first it may surprise that a
quantum vortex has zero vorticity, but the result is
expected -  the key point is that
motion in the condensate is irrotational, but isolated vortex line 
singularities are allowed. 

\item[(iii)]~
The velocity of the wind around the centre of a hurricane, case (iii) of
Fig.~\ref{fig:rotations}(a), combines solid body rotation in the
inner region ($r \ll a_0$) with irrotational motion in the outer region
($r \gg a_0$) where $a_0$ is called the {\em vortex core radius}. 

\end{itemize}

\begin{figure}[h]
\centering
\includegraphics[width=0.8\columnwidth,angle=0]{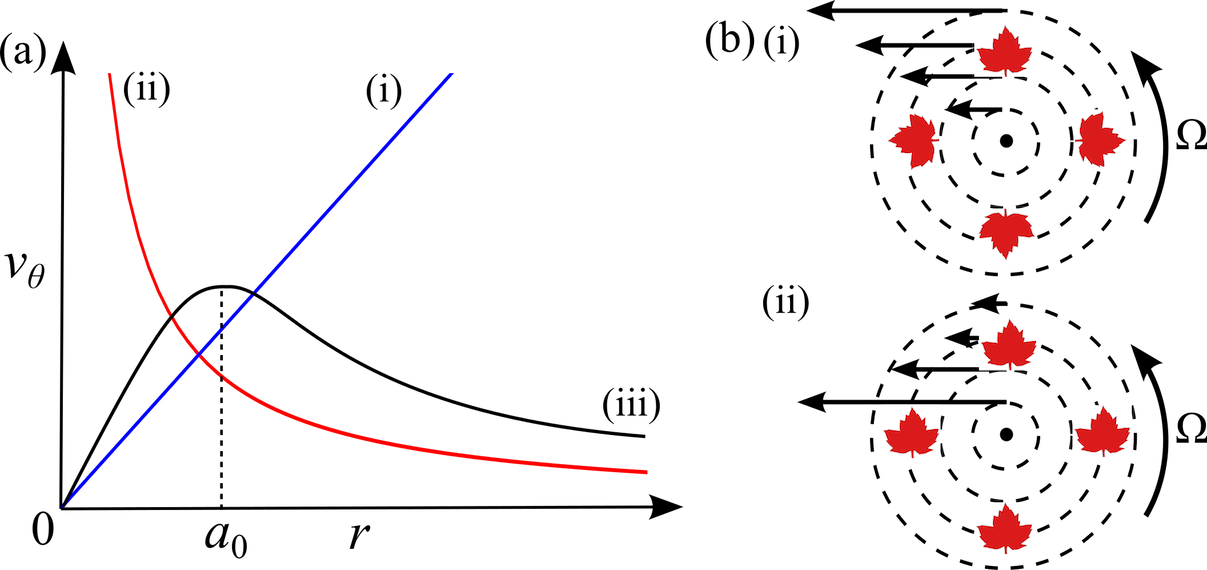}
\caption{(a) 
Examples of rotation curves.
(i) solid body rotation, (ii) vortex line in a condensate (irrotational flow), and 
(iii) flow around a hurricane or a bathtub vortex, which combines
solid body rotation in the inner region $r \ll a_0$ and irrotational
flow in the outer region $r\gg a_0$.  (b) Schematic of the two-dimensional flow for cases (i) and (ii), showing the orientation of an object, here a leaf, in the flow. 
}
\label{fig:rotations}
\end{figure}

In ordinary fluids the vorticity $\bom$ is arbitrary, and therefore
vortices can be weak or strong, big or small. In a condensate, 
Eq.~(\ref{eqn:kappa}) is a strict quantum mechanical constraint:
motion around a singularity has fixed form and intensity.

\section{The nature of the vortex core}

A natural question is: what is the structure of the vortex, particularly towards the axis of the vortex ($r \to 0$), where, according
to Eq.~(\ref{eqn:vortex-flow}), the velocity becomes
infinite?  Using cylindrical coordinates
$(r,\theta,z)$ again, we consider a straight vortex line aligned in the $z$ 
direction in a homogeneous condensate ($V=0$). 
Assuming $\Psi(r,\theta,z)=A(r) e^{i q \theta}$ and substituting
into the GPE of Eq. (\ref{eqn:tigp1}) we obtain the following differential 
equation \index{vortex}
\footnote{We have expressed the Laplacian in its cylindrically symmetric form,
\begin{equation}
\nabla^2=\frac{1}{r}\frac{\partial }{\partial r}\left(r\frac{\partial}{\partial r} \right)+\frac{1}{r^2}\frac{\partial^2}{\partial \theta^2}+\frac{\partial^2}{\partial z^2}.
\end{equation}},
for the function $A(r)$,
\begin{equation}
\mu A=-\frac{\hbar^2}{2m} \frac{1}{r} \frac{\rm d}{{\rm d}r} \left(r \frac{{\rm d} A}{{\rm d}r}\right)
+\frac{\hbar^2 q^2}{2 m r^2} A + g A^3, 
\end{equation}
The terms on the right-hand side arise from the quantum kinetic energy, the kinetic energy of the circulating flow and the interaction energy, respectively.
The boundary conditions are that
$A(r) \to 0$ for $r \to 0$ and  $A(r) \to \psi_0$ for
$r \to \infty$.  The equation has no exact solution and must be
solved numerically for $A(r)$;
the corresponding density profile
$n(r)=A^2$ is shown in Fig.~\ref{fig:vortex-profile} (a). 
It is apparent that the axis of the vortex is
surrounded by a region of depleted density, essentially a
tube of radius $a_0 \approx 5 \xi$, called the {\em vortex core radius}. For small $r$, the density scales as $r^{|q|}$.}
We see that although the velocity diverges for $r \to 0$,
the density vanishes - no atom moves at infinite speed! 
We can therefore interpret a vortex as a `hole' surrounded by
(quantized) circulation.  
Recall from Section \ref{sec:healing} that if a static and otherwise homogeneous condensate is pinned to zero density, then the density `heals` back to the background density with a characteristic profile $\tanh^2(x/\xi)$.  The vortex density profile is slightly wider than this profile and relaxes more slowly to the background density, as seen in Fig. \ref{fig:vortex-profile}(a).  This is due to the kinetic energy of the circulating flow, which gives rise to an outwards centrifugal force on the fluid.  

While there is no exact analytic form for the vortex density profile, a useful approximation for a single-charged vortex is,
\begin{equation}
n(r)=n_0\left(1-\frac{1}{1+r'^2} \right),
\label{eqn:vortex_profile}
\end{equation}
where $r'=r/\xi$.

This result (a vortex line is
a `hole' surrounded by circulating flow) has an interesting
mathematical consequence: a condensate
with vortices is a multiply-connected region, and the classical
Stokes Theorem\footnote{Stokes Theorem states that
\begin{displaymath}
\oint_C {\bf A} \cdot {\rm d}\bell
=\int_S (\nabla \times {\bf A}) \cdot {\rm d}{\bf S},
\end{displaymath}
\noindent \index{Stokes theorem}
where the surface $S$
enclosed by the oriented curve $C$ is simply-connected,
i.e. any closed curve on $S$ can be shrunk continuously to a point within $S$.}
does not apply.

In a trapped condensate the vortex creates a similar tube surrounded
by quantised circulation; the only difference is that the density
of the condensate is not uniform (as
in a homogeneous condensate).  
In typical 2D column-integrated images of the condensate, the vortex appears as a low density dot.
Since the healing length depends on the local density, in a
trapped condensate the thickness of the vortex core depends on
the position.  If the condensate is in the Thomas-Fermi regime and the vortex along the $z$ axis, then an approximation for the density profile can be constructed as the product of the static Thomas-Fermi profile, Eq. (\ref{eqn:3dtf}), and the vortex density, Eq. (\ref{eqn:vortex_profile}), i.e.,
\begin{equation}
n(x,y,z)=n_0\left(1-\frac{x^2}{R_x^2} -\frac{y^2}{R_y^2} -\frac{z^2}{R_z^2}\right)\left(1-\frac{1}{1+r'^2} \right),
\label{eqn:vortex_trap}
\end{equation}
where $r'=r/\xi$ is defined is terms of the healing length evaluated at the condensate centre. 

\begin{figure}[t]
\centering
\includegraphics[width=0.99\columnwidth,angle=0]{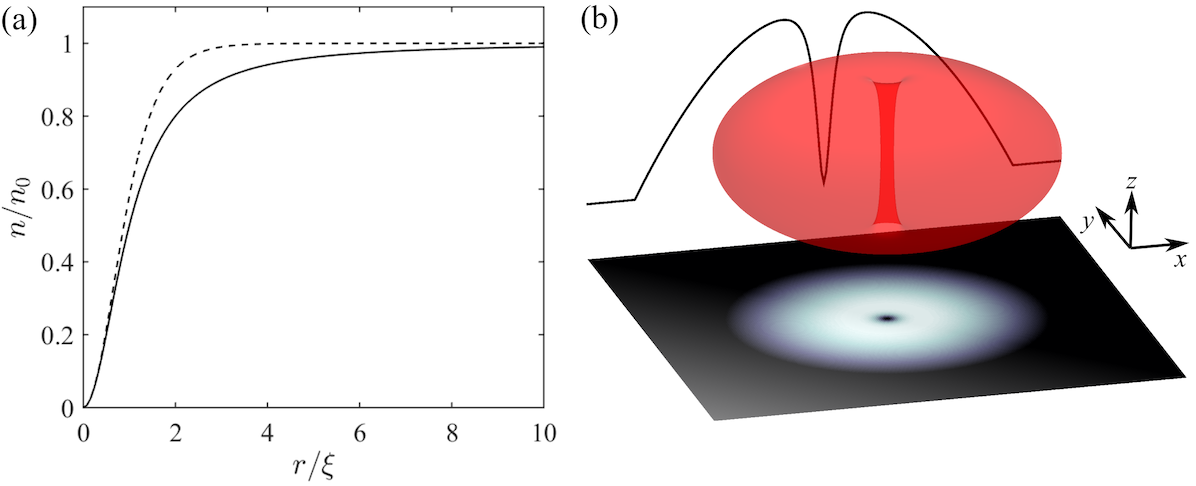}
\caption{Left:  The radial density profile $n(r)$ of a $q=1$ vortex in a 
homogeneous condensate (solid line).  Shown for comparison is the `healing' profile for a static condensate whose density is pinned to zero. Right: Appearance of a vortex lying along the axis of a trapped condensate.  Shown is an isosurface of the 3D density (with the vortex appearing as a central tube), a 2D density profile column integrated along $z$ (with the vortex appearing as a black dot), and a 1D density profile column-integrated along $y$ and $z$.
}
\label{fig:vortex-profile}
\end{figure}

\section{Vortex energy and angular momentum}
\label{sec:energy}
We now evaluate some useful properties associated with a quantum
vortex: its energy and angular momentum. For simplicity,
we still consider the case 
of a single straight vortex lying along the $z$-axis of a 
cylindrically-symmetric condensate of constant density; assuming
that the condensate's size is much larger than the healing length,
the density depletion at the axis of the vortex and near the walls
can be neglected. A cylindrical bucket of
height $H_0$ and radius $R_0$ containing superfluid liquid helium would be
a realistic example.  For trapped atomic condensates, where the vortex size is significant relative to the system size and the condensate density varies in space, these ideas can be generalized by, for example, taking the density profile to be of the form of Eq. (\ref{eqn:vortex_trap}), or by estimating the necessary integrals numerically.

The kinetic energy $E_{\rm kin}$ of  
the swirling fluid is obtained from summing the contributions of the
atoms, each carrying kinetic energy 
$m v_{\theta}^2/2$ where ${\bf v}=v_{\theta} \ehat_{\theta}
=(q \hbar/mr)~\ehat_{\theta}$ is the velocity. Summing over all
atoms we have,
\begin{eqnarray}
E_{\rm kin}=\int \frac{1}{2}  m n({\bf r}) v_{\theta}^2({\bf r})~{\rm d}^3{\bf r},
\label{eqn:kin}
\end{eqnarray}
where the integral is performed over the bucket's volume.  Using cylindrical coordinates,
\begin{equation}
E_{\rm kin}=\int_0^{H_0} {\rm d}z \int_0^{2 \pi} {\rm d}\theta \int_0^{R_0}  \frac{m n_0}{2} 
\left(\frac{q \hbar}{m r}\right)^2~r~{\rm d}r
=\pi  H_0 \frac{n_0 q^2  \hbar^2}{m} \int_0^{R_0} \frac{{\rm d}r}{r}.
\end{equation}

\bigskip
To prevent the integral from diverging at $r \to 0$ we introduce
a cutoff length $a_0$\footnote{Often this cutoff is taken instead as the 
healing length $\xi$.}, the vortex core radius; in doing so, we
recognize that the density vanishes at the axis of the vortex, but
simplify the core structure, assuming that the core is hollow up to the
distance $r=a_0$. 
Notice that without the outer limit of integration (the size of
the container $R_0$) the integral would also diverge at $r \to \infty$. 
We then obtain,

\begin{equation}
E_{\rm kin}=\pi   H_0 \frac{n_0 q^2 \hbar^2}{m} \int_{a_0}^{R_0} \frac{{\rm d}r}{r}
=\pi  H_0 \frac{n_0 q^2 \hbar^2}{m} \ln \left(\frac{R_0}{a_0} \right).
\label{eqn:vortex-energy}
\end{equation} \index{vortex!energy}

\bigskip
We conclude that the kinetic energy per unit length of the vortex,
$E_{\rm kin}/H_0=\pi n_0 (q^2 \hbar^2/m)\ln{(R_0/a_0)}$, is constant.

Each atom swirling around the axis of the vortex carries
angular momentum $L_z=m v_{\theta}r$. 
The total angular momentum of the flow is therefore,
\begin{equation}
L_z=\int m n ({\bf r}) v_{\theta} ({\bf r}) r~{\rm d}^3 {\bf r}.
\end{equation} \index{momentum!angular momentum}
Proceeding as for the kinetic energy, we find,
\begin{equation}
L_z=  
2 \pi H_0 n_0 q \hbar \left( \frac{R_0^2}{2} - \frac{a_0^2}{2} \right) 
\approx \pi H_0 n_0 q \hbar R_0^2.
\end{equation} \index{vortex!momentum} 

Consider a condensate in a state with an arbitrary high angular momentum $L_z$.
We can construct this state as either (i) one vortex with large $q$ or 
(ii) many vortices with $q=1$.   Which situation is preferred?  
Since $E_{\rm kin}$ scales as $q^2$, a state with
many singly-charged vortices has less energy than a state with a single
multi-charged vortex.
Experiments confirm that this is indeed
the case: in Ref. \cite{shin_2004} a $q=2$ vortex was seen to quickly decay into two singly-charged
vortices. Hereafter we assume that all vortices are singly-charged,
with $q=\pm 1$.

\section{Rotating condensates and vortex lattices}
\label{sec:rotating}
\subsection{Buckets}
Vortices are easily created by rotating the condensate 
\cite{Fetter-2009,Tsubota-Kasamatsu-Ueda-2002}. Consider again
a cylindrical condensate of height $H_0$, radius $R_0$ and uniform density.
A vortex appears only if the system, by creating a vortex,
lowers its energy. In a rotating
system at very low temperature, it is not the energy $E$ which must be 
minimized, but rather the
free energy $F=E-\Omega L_z$ where $\Omega$ is the angular velocity of rotation.
A state without any vortex, hence without angular momentum, has free energy \index{energy!free energy}
$F_1=E_0$ where $E_0$ is the internal energy. A state with a vortex has free
energy
$F_2=E_0+E_{\rm kin}-\Omega L_z$. The free energy difference is thus,
\begin{equation}
\Delta F=F_2-F_1=E_{\rm kin}-\Omega L_z=
2 \pi H_0 \frac{\hbar^2}{m^2} \ln \left(\frac{R_0}{a_0} \right) -\Omega \pi H_0 n_0 \hbar R_0^2.
\end{equation} \index{rotation!in a bucket}

\noindent
Therefore $\Delta F<0$ (the free energy is reduced by creating a vortex) 
provided that the rotational velocity is larger than a critical 
value $\Omega_{c1}$,
\begin{equation}
\Omega > \Omega_{c1}=\frac{\hbar}{m R_0^2} \ln \left(\frac{R_0}{a_0} \right).
\end{equation} \index{vortex!critical rotation frequency}
For superfluid helium 
($m=6.7\times10^{-27}$kg, $\kappa=9.97 \times 10^{-8}~\rm{m^2/s}$,
$a_0 \approx 10^{-10}~\rm m$) inside a 
container of radius $R_0=10^{-2}~\rm m$, the critical angular velocity
is $\Omega_{c1}=3\times 10^{-3}~\rm s^{-1}$. States with two, 
three and more vortices
onset at higher critical velocities $\Omega_{c2}$, $\Omega_{c3}$
etc, as shown in Fig.~\ref{fig:Yarmchuck-Packard-1979} for 
superfluid helium and in Fig.~\ref{fig:raman}
for atomic condensates. 
Note that the vortices are parallel to the rotation axis and arrange
themselves in a {\em vortex lattice} like atoms in a crystal with
triangular symmetry.   \index{vortex!lattice}
The vortex lattice is therefore
a steady configuration in the frame of reference rotating at 
angular velocity $\Omega$. 

Vortices are topological defects which can only be created at a boundary or spontaneously with an oppositely-charged vortex \footnote{An exception is through the technique of phase imprinting, in which the condensate phase can be directly and almost instantaneously imprinted with a desired distribution.  In this manner vortices can be suddenly formed within the condensate.}.  Where then do the vortices in a vortex lattice originate from?

For a rotating container of helium, with even a relatively small rotation frequency, the roughness of the container surface is expected to seed vortices, providing a constant source of vortices from which to develop a vortex lattice in the bulk if the critical rotation frequency is exceeded.

\begin{figure}
\centering
\includegraphics[width=0.8\columnwidth,angle=0]{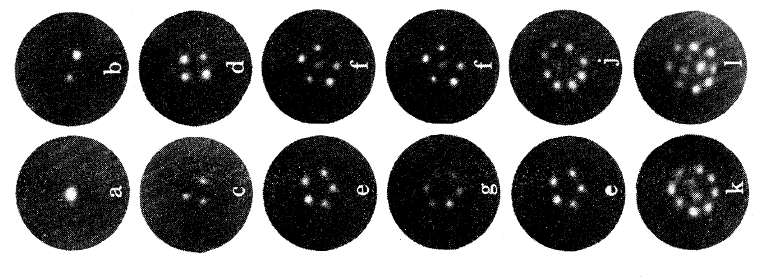}
\caption{Experimental images of vortex lattices at increasing angular 
velocities $\Omega$ in superfluid helium.  Reprinted figure with permission from \cite{Yarmchuck-Packard-1979}. Copyright 1979 by the American Physical Society.}
\label{fig:Yarmchuck-Packard-1979}
\end{figure}

According to {\em Feynman's rule}, the density of vortices
(number of vortices per unit area) is,
\begin{equation}
n_v=\frac{2 \Omega}{\kappa}.
\label{eqn:Feynman}
\end{equation} 
Since each vortex contributes vorticity according to Eq. (\ref{eqn:vort}), the average vorticity per unit area is, \index{vorticity}
\begin{equation}
\bar{\bom} = \kappa n_v \ehat_z = 2 \Omega \ehat_z.
\end{equation}
This tells us that the averaged vorticity  (averaged
over distance larger than the inter-vortex spacing) reproduces the vorticity $2 \Omega$ of an ordinary fluid in rotation.   Similarly, 
the large-scale azimuthal flow is
${\bf v} \approx \Omega r \ehat_{\theta}$.  Remarkably, the many quantized 
vortices mimic classical solid body rotational flow.  
Note that the local velocity field around 
vortices can remain rather complicated.

In the frame rotating at angular frequency $\Omega$ about the $z$-axis, the GPE of Eq. (\ref{eqn:gp2}) is,
\begin{equation}
i \hbar \frac{\partial \psi}{\partial t}=
-\frac{\hbar^2}{2m}\nabla^2 \psi
+ g \vert \psi \vert^2 \psi + V \psi + \Omega L_z \psi- \mu \psi,
\label{eqn:rotating_gp}
\end{equation}\index{Gross-Pitaevskii equation!rotating frame}
where, 
\begin{equation}
L_z=i \hbar\left(y \frac{\partial }{\partial x}  - x \frac{\partial}{\partial y}\right),
\end{equation}
is the angular momentum operator in the $z$ direction.\index{momentum!angular momentum}  The vortex lattices are the ground-state stationary solutions of this equation (providing $\Omega$ is large enough).  
Figure \ref{fig:lattice_bucket} shows such a vortex lattice solution for a condensate being rotated in a bucket.  The above bucket scenario is modelled through the {\em bucket potential},
\begin{equation}
V(r)=
\left\{
\begin{array}{lr}
0 &  {\rm ~if~} r \le R_0,\\[.5em]
\infty                     &  {\rm ~if~} r > R_0.
\end{array}
\right.
\end{equation}
The lattice features $N_{\rm v}=56$ vortices.  Note the appearance of the phase ``dislocations'' in the phase profile at each vortex position.  At the boundary there are as many $2 \pi$ phase slips as there are vortices.  The average flow speed around the edge of the bucket can then be approximated by evaluating the magnitude of ${\bf v}=(\hbar/m)\nabla S$ around the boundary, i. e.,
\begin{equation}
v_r(r=R_0)=\frac{\hbar}{m}\frac{2 \pi N_{\rm v}}{2 \pi R_0}  = \frac{\hbar}{m}\frac{56}{29 \xi} \approx 1.93 c.
\end{equation}
This is close to what one would expect for solid body rotation, $v_r(r=R_0) = \Omega R_0= 2.3 c$.  \index{rotation!in a bucket}

\begin{figure}
\centering
\includegraphics[width=0.9\columnwidth,angle=0]{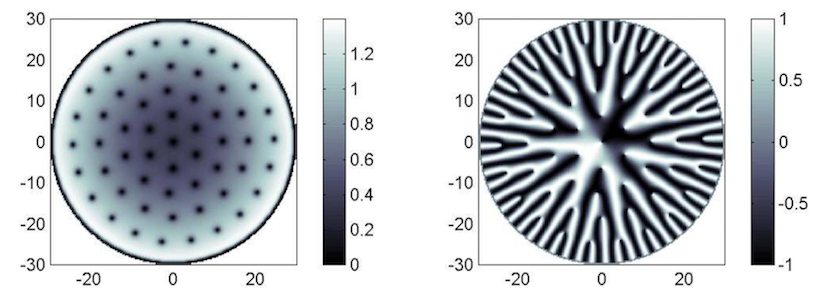}
\caption{Vortex lattice formed in a bucket potential rotating about the $z$-axis.   Shown are the (a) density (in arbitrary units) and (b) phase (in units 
of $\pi$)  in the $xy$-plane (position presented in units of the healing length $\xi$), corresponding to the stationary solution of the rotating-frame GPE of Eq. (\ref{eqn:rotating_gp}).  The bucket has radius $R=29 \xi$ and the rotation frequency is $\Omega=0.08 ~c/\xi$.  Image courtesy of Thomas Winiecki \cite{winiecki_thesis}.}
\label{fig:lattice_bucket}
\end{figure}

In a small system, at the same value of $\Omega$ one often observes 
vortex configurations
which are slightly different from each other.  This is because there is a very small energy difference between these slightly rearranged states.
For example, Fig.~\ref{fig:Yarmchuck-Packard-1979} shows two states
with six vortices each (in one case the six vortices are distributed
around a circle, in the other case there are five vortices around a circle
and one vortex in the middle).

Notice how the background density for the rotating bucket solution in Fig. \ref{fig:lattice_bucket} features a {\em meniscus}, that is, it is raised towards the edge of the bucket.  Let us determine this background density profile.  We denote the rotation vector ${\bf \Omega}=\Omega \ehat_z$.

Recall the fluid interpretation of the GPE.  Using the Madelung transformation $\psi=\sqrt{n}e^{iS}$ and the fluid velocity definition ${\bf v}=(\hbar/m)\nabla S$, the rotating-frame GPE of Eq. (\ref{eqn:rotating_gp}) is equivalent to the modified fluid equations, \index{fluid equations!rotating frame}
\begin{eqnarray}
\dfrac{\partial n}{\partial t} &=&-\nabla \cdot \left[n\left({\bf v}-{\bf \Omega}\times {\bf r}\right) \right],  \label{eqn:rotatinghydro1}
\\
\displaystyle m \frac{\partial {\bf v}}{\partial t} &=&- \nabla \cdot \left(\frac{1}{2}mv^2+ V+ gn - \frac{\hbar^2}{2m} \frac{\nabla^2 \sqrt{n}}{\sqrt{n}}- m {\bf v}\cdot \left[{\bf \Omega}\times {\bf r} \right] \right), \label{eqn:rotatinghydro2}
\end{eqnarray}
where the $\bom \times {\bf r}$ terms account for frame rotation and ${\bf v}$ is the velocity field in the laboratory frame (expressed in the coordinates of the rotating frame).
We assume the Thomas-Fermi approximation by neglecting the quantum pressure term in Eq. (\ref{eqn:rotatinghydro2}), and seek the stationary density profile. Setting $\partial {\bf v}/\partial t=0$ and integrating gives,
\begin{equation}
\frac{1}{2}mv^2+ V+ gn - m {\bf v}\cdot \left[{\bf \Omega}\times {\bf r}  \right] = \mu,
\label{eqn:rotating_profile}
\end{equation} \index{Thomas-Fermi!rotating solutions}
where the chemical potential $\mu$ is the integration constant.  

We consider a coarse-grained scale, ignoring the structure of the 
individual vortices and for which the velocity field approximates the solid body form ${\bf v}(r)=\Omega r \ehat_\theta$.  We then obtain,
\begin{equation}
gn +V - \frac{1}{2}m \Omega^2 r^2 = \mu,
\end{equation}  
where we have used $\ehat_z \times \ehat_r=\ehat_\theta$. 
Rearranging for the density,
\begin{equation}
n(r)= \frac{1}{g}\left(\mu -V+ \frac{1}{2}m \Omega^2 r^2 \right),
\label{eqn:rotating_profile1}
\end{equation}  
which is valid for $n(r)>0$; otherwise $n(r)=0$.  We conclude that rotation causes a parabolic increase in the coarse-grained density, consistent with the behaviour visible in Fig. \ref{fig:lattice_bucket}.  The is due to centrifugal effects, and is observed in rotating classical fluids. Note that $\mu$ can be determined by normalizing the profile to the required number of atoms or average density.  

\subsection{Trapped condensates}

To predict the critical rotation frequency for vortices to become favoured in a harmonically-trapped condensate, one can repeat the above approach but the inhomogeneous density profile must be accounted for (i.e. replacing $n_0$ above with $n({\bf r})$).  One way to approximate this is by the Thomas-Fermi density profile.  For a trap which is symmetric in the plane of rotation, with frequency $\omega_\perp$,  the critical rotation frequency is then,
\begin{equation}
\Omega_{\rm c1} = \frac{5}{2} \frac{\hbar}{m R_\perp^2} 
 \ln\left( \frac{0.67 R_\perp}{\xi}\right),
\end{equation}
where $R_\perp$ is the Thomas-Fermi radius in the plane of rotation.  For typical atomic condensates,  $\Omega_{\rm c1}\sim 0.3 \omega_\perp$. \index{vortex!critical rotation frequency}

Rotating an axi-symmetric harmonic trap applies no torque to the condensate, and so in practice the trap is made slightly anisotropic in the plane of rotation in order to form a vortex lattice.  Surprisingly, experiments observed vortices at rotation frequencies $\Omega \sim 0.7 \omega_\perp$, considerably higher than the frequency at which they become energetically favourable.  The traps are so smooth that vortex nucleation is very different to that of helium.

We can examine this by considering the planar potential to be weakly elliptical, with frequencies $\omega_x=\sqrt{1-\epsilon}\omega_\perp$ and $\omega_y=\sqrt{1+\epsilon}\omega_\perp$, where $\epsilon$ is the {\em trap ellipticity}.  \index{trap!ellipticity} We follow the approaches of Refs. \cite{Recati_2001,Sinha_2001}.  We seek the stationary solutions of the trapped vortex-free condensate under rotation about $z$.  Under the Thomas-Fermi approximation, 
the solutions must satisfy Eq. (\ref{eqn:rotating_profile}).  Furthermore, we look for solutions with the phase profile, and corresponding velocity profile, given by,
\begin{equation}
S(x,y)=\beta xy, \quad {\bf v}(x,y)=\frac{\hbar}{m} \nabla S=\frac{\beta \hbar}{m}(y \ehat_x + x \ehat_y).
\label{eqn:phase_vel}
\end{equation}
where $\beta$ is a parameter to be determined below.  
Inserting into Eq. (\ref{eqn:rotating_profile}), and noting that ${\bf \Omega}\times {\bf r}=\Omega(x \ehat_y-y \ehat_x)$, leads to the density profile,
\begin{eqnarray}
n=\frac{1}{g}\left(\mu -\frac{1}{2}m (\tilde{\omega}_x^2 x^2+\tilde{\omega}_y^2 y^2 + \omega_z^2 z^2)\right),
\end{eqnarray}  
where the effect of the rotation is to introduce {\em effective trap frequencies} in the $xy$-plane,
\begin{eqnarray}
\tilde{\omega}_x^2&=&(1-\epsilon)\omega_\perp^2+\beta^2-2\beta\Omega,
\\
\tilde{\omega}_y^2&=&(1+\epsilon)\omega_\perp^2+\beta^2+2\beta\Omega.
\end{eqnarray}
Plugging this density profile into the rotating-frame continuity equation, Eq. (\ref{eqn:rotatinghydro1}), and setting $\partial n/\partial t=0$, leads to an expression for $\beta$,
\begin{equation}
\beta^3+\beta(\omega_\perp^2-2\Omega^2)-\epsilon \Omega \omega_\perp^2=0.
\end{equation}
Hence the stationary solution of the condensate in the rotating frame has been completely specified.  
In the laboratory frame, this solution has an elliptical density profile which rotates about $z$.  However, the fluid remains irrotational, thanks to the special velocity field which distorts the density is such a way as to mimic rotation, as depicted in Fig. \ref{fig:irrot}(a).  \index{rotation!in a harmonic trap}

\begin{figure}
\centering
\includegraphics[width=0.8\columnwidth,angle=0]{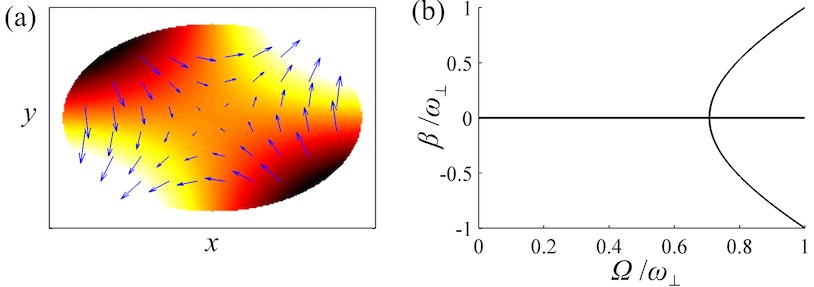}
\caption{(a) Illustration of the irrotational flow pattern of a rotating elliptically-trapped condensate, according to Eqs. (\ref{eqn:phase_vel}).  The color indicates the phase $S(x,y)$ while the velocity field is shown by arrows.  (b) The velocity field amplitude $\beta$ as a function of rotation frequency $\Omega$ for an axi-symmetric trap ($\epsilon=0$).  At $\Omega=\omega_\perp/\sqrt{2}$ the solutions trifurcate.  In this region, these solutions become unstable.}
\label{fig:irrot}

\end{figure}
Analysing the case of $\epsilon=0$ for simplicity, there exists one solution, with $\beta=0$, for $\Omega \leq \omega_\perp/\sqrt{2}$; this represents a motion-less and axi-symmetric condensate.  However, for $\Omega > \omega_\perp/\sqrt{2}$ the solutions trifurcate, with two new branches with $\beta \neq 0$ and corresponding to non-axisymmetric solutions of the form shown in Fig. \ref{fig:irrot}.  This trifurcation leads to an instability of the condensate (as can be confirmed via linearizing about these solutions \cite{Sinha_2001}) in which perturbations grow at the condensate surface and develop into vortices.  Experiments \cite{Madison_2001}  and simulations \cite{Parker_2006} of the GPE show that this instability then allows the condensate to evolve into a vortex lattice, the lowest energy state.  

\begin{figure}
\centering
\includegraphics[width=0.55\columnwidth,angle=0]{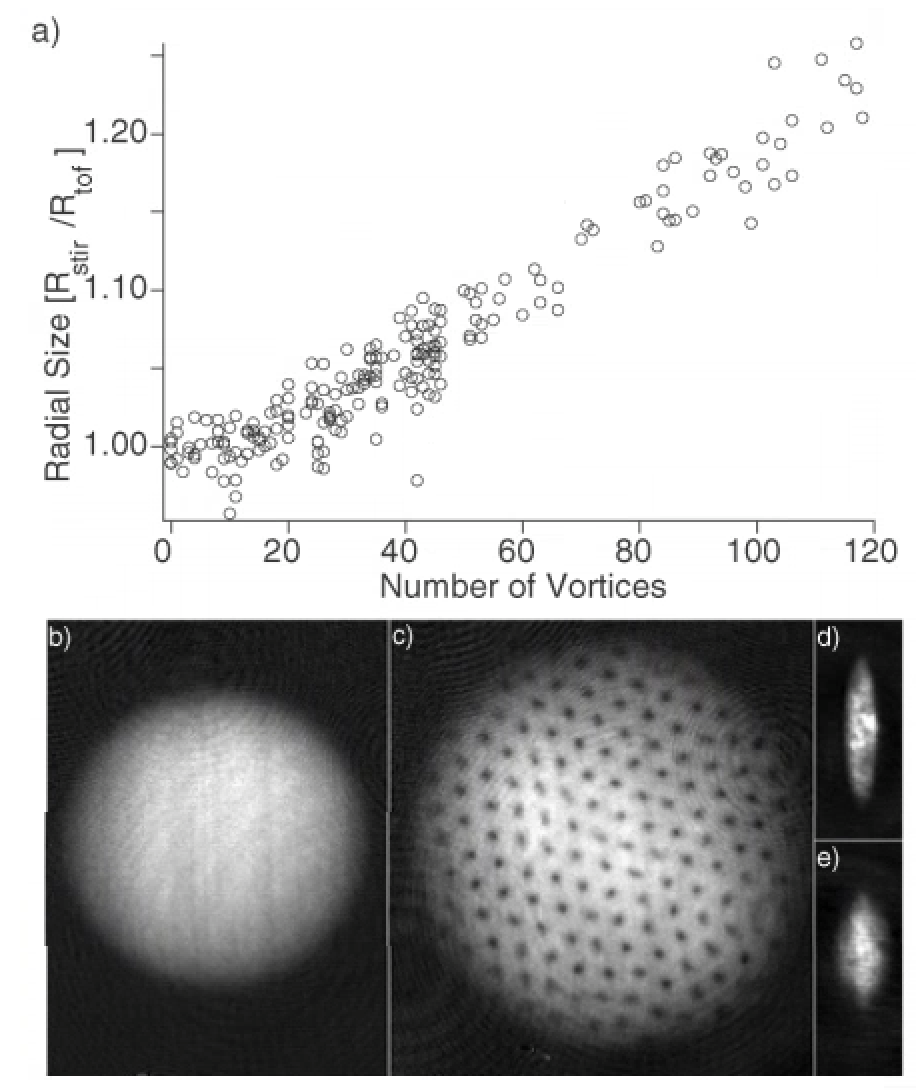}
\caption{An experimental vortex lattice (c) formed in a flattened trapped rotating atomic condensate (image represents the condensate density).  (b) shows the profile of the corresponding non-rotating condensate.  (d) and (e) show the side views of the non-rotating and rotating condensates, respectively.  The condensate grew in radius with the number of vortices, as shown in (a). Reprinted figure with permission from \cite{Raman_2001}. Copyright 2001 by the American Physical Society.}
\label{fig:raman}
\end{figure}

Figure \ref{fig:raman} shows a vortex lattice produced in a rotating trapped atomic condensate.  Note the regularity and density of the vortex lattice.  Note also that the rotating condensate is significantly broader than the non-rotating condensate.  In the presence of the vortex lattice, we can predict the coarse-grained density profile of the condensate.  Considering an axi-symmetric trap ($\omega_x=\omega_y\equiv \omega_\perp$), then the coarse-grained density profile of Eq. (\ref{eqn:rotating_profile1}) gives,
\begin{equation}
n(r)= \frac{1}{g}\left(\mu -\frac{1}{2}m(\omega_r^2 r^2-\Omega^2) r^2 \right).
\end{equation}
There is a competition between the quadratic trapping potential, which pushes atoms inwards, and the quadratic centrifugal potential, which pushes atoms outwards.  The net potential is quadratic with effective harmonic potential $\omega_r^2-\Omega^2$.  As $\Omega$ is increased, the condensate expands, and when  $\Omega \geq \omega_r$ it becomes untrapped! \index{expansion}

\section{Vortex pairs and vortex rings}
An important property of a vortex is that it moves with the local fluid velocity, and this means that two vortices in proximity induce each other to move.  We now consider some important examples.

\subsection{Vortex-antivortex pairs and corotating pairs}
Consider a pair of vortices of opposite circulation
and separation $d$, a state called a {\em vortex-antivortex pair} or {\em vortex dipole}, shown schematically in Fig.~\ref{fig:pair}.
In the figure, the flow around the vortex at the left is anticlockwise, and the
flow around the anti-vortex at the right is clockwise.  Each vortex is carried along by the flow field of the other vortex, and at each vortex the flow field has speed $v = \hbar/m d$ acting perpendicular to the line separating the vortices. Moreover, this flow acts in the {\em same direction} for both vortices, and hence they propagate together at this speed. \index{vortex!pairs}

\begin{figure}
\centering
\includegraphics[width=0.65\columnwidth,angle=0]{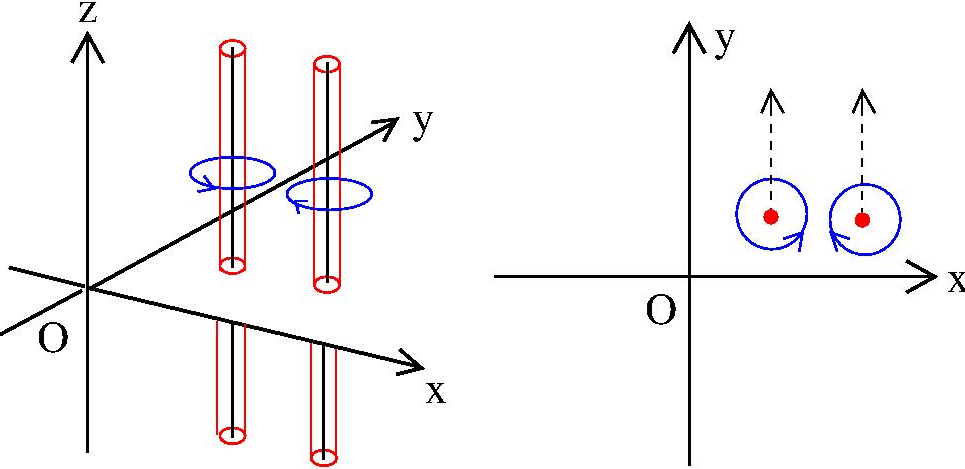}
\caption{Schematic of a vortex-antivortex pair.
}
\label{fig:pair}
\end{figure}

If instead the vortices have the same circulation, then the flow which carries each vortex now acts in {\em opposite directions} (again, perpendicular to the line separating the vortices and with the above speed).   The net effect is for the vortices to {\em co-rotate} about their mid-point.  The angular frequency of this motion is $\omega=2v/d = 2\hbar/m d^2$.  From this simple example, one can imagine how many vortices of the same circulation rotate together in a vortex lattice.  Note that the above predictions for the pair speed ignore core effects, and so are only valid for $d \gg a_0$.

We can estimate the energy of the vortex pairs in a cylindrical condensate (radius $R_0$, height $H_0$) by assuming a uniform density and integrating the kinetic energy, as we did to calculate the  energy of a single vortex line in Eq. (\ref{eqn:vortex-energy}).  The vortices have circulation $q_1$ and $q_2$, and individual velocity fields 
${\bf v}_1$ and ${\bf v}_2$, 
respectively.  The net velocity field of the two vortices is 
${\bf v}_1+{\bf v}_2$.
  Assuming $\xi \ll d \ll R_0$ then the (kinetic) energy of the pair is,
\begin{equation}
E_{\rm kin}=\int m n_0 |{\bf v}_1+{\bf v}_2|^2~{\rm d}{\bf r} =
  \frac{\pi n_0 H_0 \hbar^2}{m} \left[q_1^2 \ln \frac{R_0}{a_0}+q_2^2 \ln \frac{R_0}{a_0}+ 2 q_1 q_2 \ln \frac{R_0}{d}  \right].
\end{equation}
The first two terms are the energies of the individual vortices if they were isolated.  The second term is the {\em interaction energy}, the change in energy arising from the interaction between the vortices.  For a vortex-antivortex pair ($q_1=-q_2$) the interaction energy is negative.  This is because the flow fields tend cancel out in the bulk, reducing the total kinetic energy.  Indeed, in the limit $d \rightarrow a_0$, the flow fields completely cancel and the total energy tends to zero; in reality the vortices annihilate with each other in this limit.  For a corotating pair ($q_1=q_2$), the interaction energy is positive; in the bulk the flow fields tend to reinforce, increasing the total kinetic energy.  

In the presence of dissipation on the vortices, this result also informs us that vortex-antivortex pairs will shrink (ultimately annihilating when their cores begin to overlap) and corotating pairs will expand.  Interestingly, 
at finite temperature and in 2D condensates, vortex-antivortex pairs 
can be created spontaneously \cite{Simula_2006}.

\subsection{Vortex rings}
\index{vortex!rings}
A vortex line either terminates at a boundary (e.g. the vortex in the
cylindrical container discussed in the previous section)
or is a closed loop. A circular vortex loop is called a {\em vortex ring}.  It is the three-dimensional analog of the (two-dimensional) vortex-antivortex pair: each element of the ring moves due to the flow induced by the rest of the ring, resulting in the ring travelling in a straight line at a constant speed which is inversely
proportional to its radius. Figure~\ref{fig:Youd-ring-line} shows
a vortex ring travelling towards, and interacting with, a straight vortex line.

Both vortex rings and vortex-antivortex pairs are forms of solitary waves, since they propagate without spreading.  Moreover, like dark solitons, they are stationary (excited) solutions of the homogeneous condensate in the frame moving with the ring/pair.  

\begin{figure}
\centering
\includegraphics[width=0.8\columnwidth,angle=0]{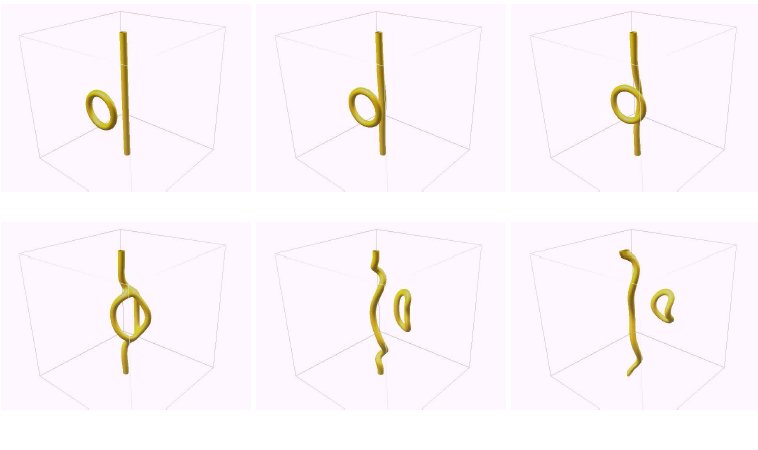}
\caption{Vortex ring travelling towards a vortex line
computed by numerically solving the GPE (courtesy of A.J. Youd) in
a periodic box (hence the vortex line appears to terminate at the
top and at the bottom). }
\label{fig:Youd-ring-line}
\end{figure}

\subsection{Vortex pair and ring generation by a moving obstacle}

Vortex rings are easily generated in ordinary fluids by pushing the fluid
through an orifice:
cigarette smokers, volcanoes and dolphins can make vortex rings.  In condensates and helium, rings and vortex-antivortex pairs can be formed by moving obstacles.  

To understand this mechanism, recall Landau's criterion for the generation of excitations in the condensate (Section \ref{sec:landau}).  
In the hydrodynamic picture, the speed of the atom/impurity is replaced by the local fluid velocity. \index{Landau criterion} \index{critical velocity}  Consider the scenario of a homogeneous condensate flowing with bulk speed $v_\infty$ past a cylindrical obstacle (this is equivalent to the cylindrical obstacle moving at speed $v_\infty$ through a static condensate but more convenient to simulate).  For low $v_\infty$, the condensate undergoes undisturbed laminar flow around the obstacle, as shown in Fig. \ref{fig:pair_generation}(left).  Note that the local flow speed is approximately twice as large, i.e. $2v_\infty$, at the poles of the obstacle than it is in the bulk (indeed, for 
an inviscid Euler fluid one would expect it to be exactly $2 v_\infty$).  When $v_\infty \approx 0.5 c$, the local flow at the poles exceeds the speed of sound, and, as per Landau's prediction, excitations are created.  These take the form of pairs of opposite circulation vortices, which periodically peal off from the poles of the obstacle and travel downstream, as seen in Fig. \ref{fig:pair_generation}(right).    

\begin{figure}
\centering
\includegraphics[width=0.49\columnwidth,angle=0]{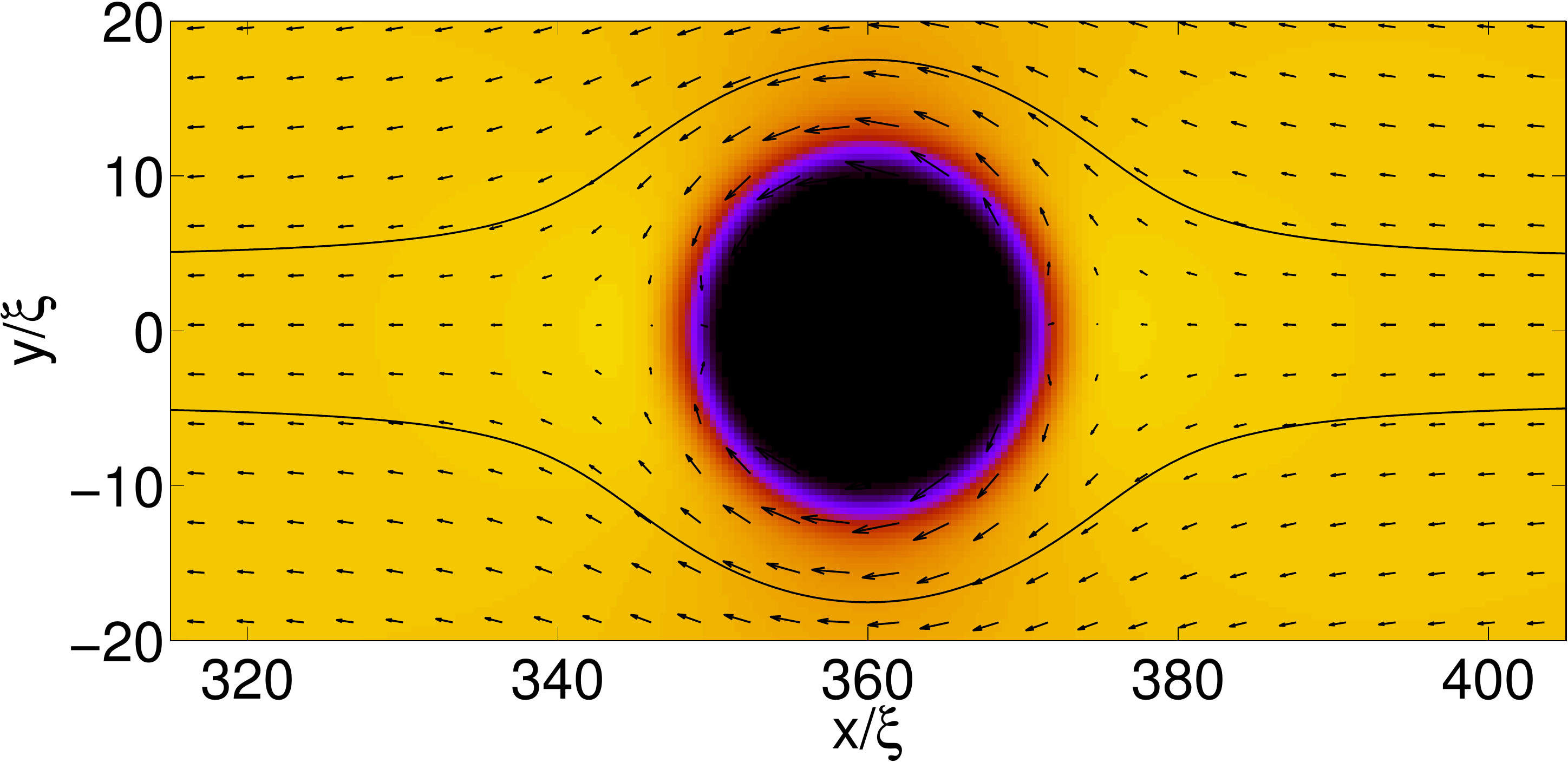}
\includegraphics[width=0.49\columnwidth,angle=0]{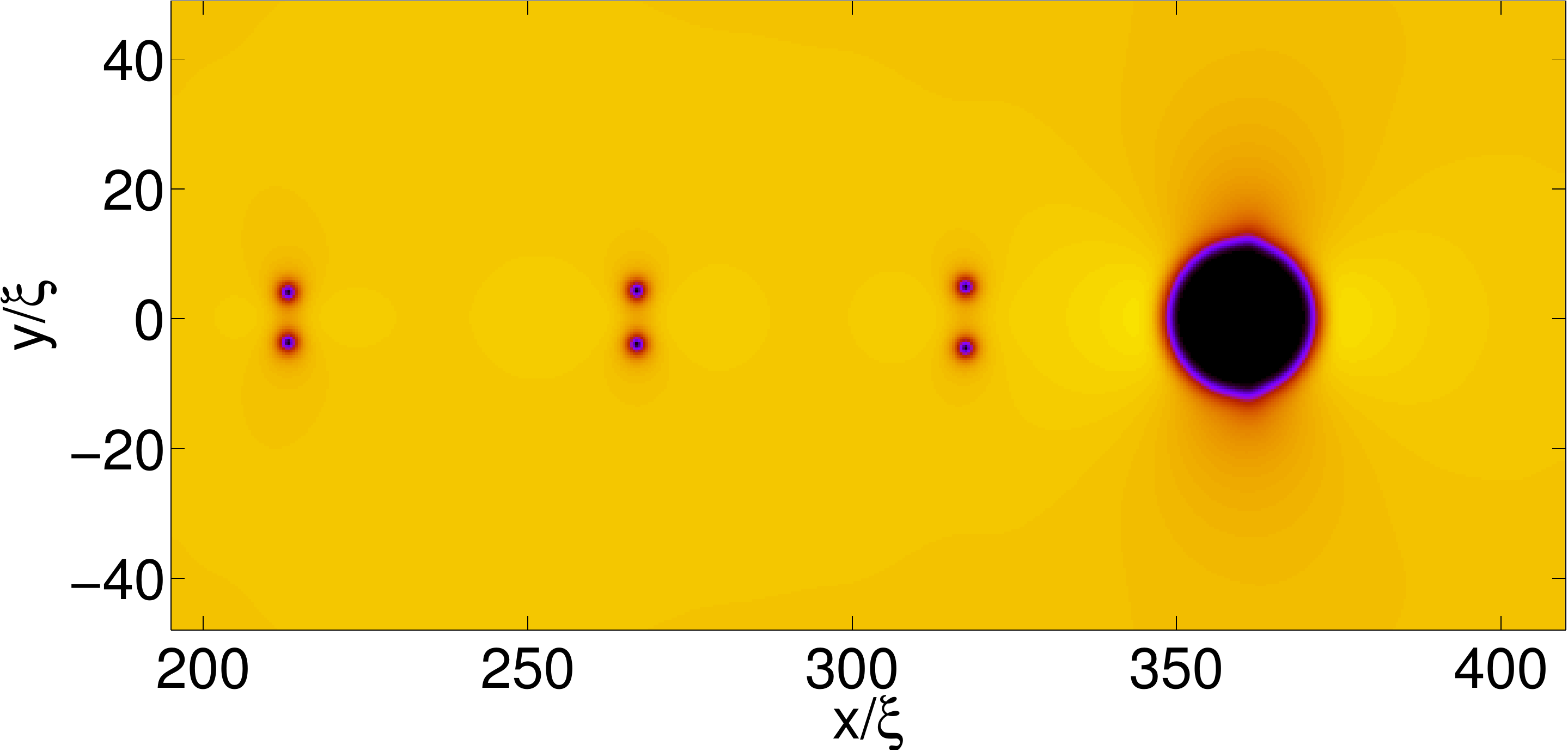}
\caption{Flow of a homogeneous condensate past a cylindrical obstacle, below (left) and above (right) the critical velocity.  Shown is the condensate density, and the arrows (left) show the velocity field.  Note that the obstacle punches a large hole in the condensate.  Results are based on simulations of the 2D GPE in the moving frame.  Figure reproduced from Ref. \cite{Stagg_2015b} under a \href{https://creativecommons.org/licenses/by/3.0/}{CC BY licence}.}
\label{fig:pair_generation}
\end{figure}

\begin{figure}
\centering
\includegraphics[width=0.77\columnwidth,angle=0]{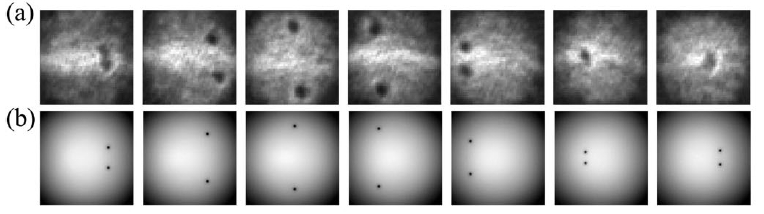}
\includegraphics[width=0.22\columnwidth,angle=0]{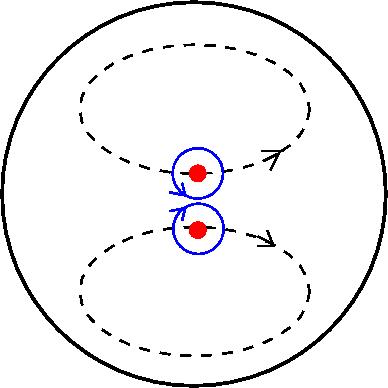}
\caption{(a) Experimental images of a vortex-antivortex pair moving within
a trapped condensate. (b) The vortex-antivortex pair's trajectory is reproduced by numerically
solving the GPE.  Note that the vortex core appears larger in the experimental images since the condensate is first expanded to aid in resolving the cores.  Figure adapted with permission from Ref. \cite{Neely_2010}. Copyrighted by the American Physical Society.  A schematic of the trajectory of a vortex-antivortex pair in a trapped condensate is shown on the right.
}
\label{fig:Neely-pair}
\end{figure}

This process has been studied experimentally in atomic condensates 
\cite{Neely_2010,Kwon_2015}.  The obstacle is engineered by a laser beam 
which  
exerts a localized repulsive potential on the condensate, and is moved relative to the condensate.  
Figure~\ref{fig:Neely-pair} shows an experimental vortex-antivortex pair which moves within
a trapped condensate (top). The dynamics  can be reproduced by simulating the GPE (bottom). Note that  whereas in an infinite condensate the vortex-antivortex pair has
constant translational velocity, within a harmonically-trapped
condensate the motion of each vortex of the pair follows a curved
trajectory. \index{vortex!pairs}

Similarly, vortex rings arise when a spherical obstacle exceeds a critical speed relative to the condensate.  They can be created in superfluid helium by injecting electrons
with a sharp high-voltage tip; the electron's zero point motion carves
a small, charged spherical bubble in the liquid of 
radius approximately $16 \times 10^{-10}~\rm m$
which can be accelerated by an applied electric field. 
Upon exceeding a critical velocity, a vortex ring peels off at the bubble's
equator; subsequently the electron falls into the vortex core, leaving
a vortex ring with an electron bubble attached; 
the last part of the sequence is shown in 
Fig.~\ref{fig:Winiecki-Adams-2000}. \index{vortex!rings}

\begin{figure}
\centering
\includegraphics[width=0.75\columnwidth,angle=0]{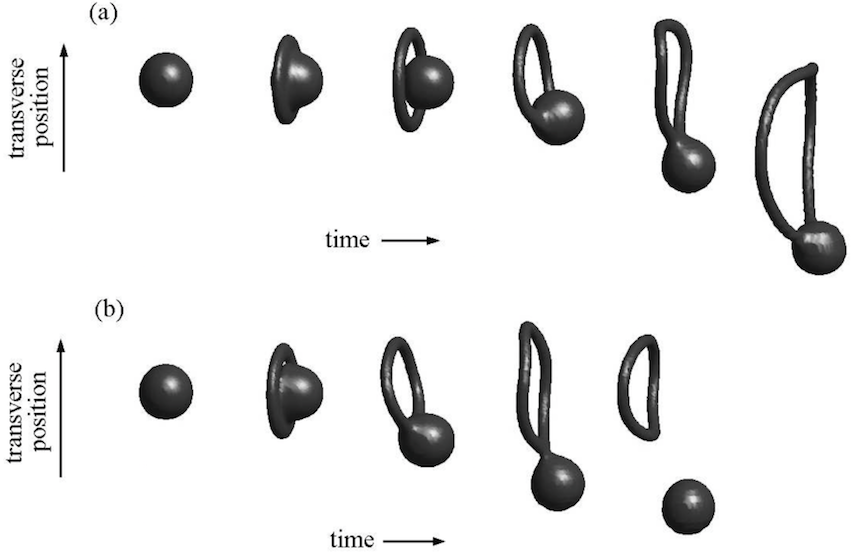}
\caption{Vortex rings nucleated by moving 
bubbles, computed by numerically solving the
GPE. Figure reproduced from \cite{Winiecki-Adams-2000} with permission from EDP Sciences.}
\label{fig:Winiecki-Adams-2000}
\end{figure}

\section{Motion of individual vortices}

We have seen how vortices move due to their interactions with other vortices.  Isolated vortices can also move under a variety of scenarios.

First imagine a condensate in a static bucket with a straight vortex line positioned close to the edge.  The fluid velocity must be zero at the boundary.  In effect, it is as if an {\em image vortex}, with opposite circulation, exists on the other side of the boundary.  As such the vortex moves around the boundary of the container as a virtual pair with its image.

In a harmonically-trapped condensate,  an off-centre vortex precesses about the trap centre.  The slow variation of the density towards the edge complicates an image interpretation.  Instead, we can interpret the precession in terms of a Magnus force.  Imagine the vortex line as a rotating cylinder, shown in Fig. \ref{fig:magnus}(left).  The vortex line feels a radial force due to its position in the condensate, and this gives rise to a motion of the vortex line which is perpendicular to the force, ${\bf v_{\rm L}}$, an effect well known in classical hydrodynamics.  This force can be deduced from the free energy of the system.  This energy decreases with the vortex position, $r_0$, as shown in Fig. \ref{fig:magnus}(right).  This radial force, which follows as $-\partial E/\partial r_0$, acts outwards and has contributions from the ``buoyancy'' of the vortex, which behaves like a bubble, as well as its kinetic energy.  
This force balances the Magnus force 
$-m n {\boldsymbol \kappa} \times {\bf v}_L$, leading to the expression,
\begin{equation}
\frac{\partial E}{\partial r_0} \widehat{\mbox{\boldmath $e$}}_r= m n {\boldsymbol \kappa} \times {\bf v_L},
\end{equation}
where ${\boldsymbol \kappa}$ is the circulation vector.  The net effect is a precession of the vortex about the trap centre.  More generally, the vortex follows a path of constant free energy; for example, it will trace out a circular path in an axi-symmetric harmonic trap and an elliptical path in a non-axi-symmetric harmonic trap.  The experiment of Ref. \cite{Freilich_2010} pioneered the 
real-time imaging of vortices in condensates and was able to directly 
monitor the precession of a vortex, finding it to agree well with 
theoretical predictions. \index{vortex!precession} \index{Magnus force}

At the trap centre, $E(r_0)$ becomes flat such that the vortex ceases to precess; in fact, the trapped condensate with a central vortex line is a stationary state.  For a non-rotating condensate, this state is energetically unstable ($E(r_0)$ is a maximum at the origin).  Under sufficiently fast rotation, however, $E(r_0)$ changes shape such that this state becomes a minimum and thus energetically stable, consistent with discussion in Section \ref{sec:rotating}.

This analysis assumes the vortex line to be straight.  This is valid is flattened, quasi-2D geometries, but in 3D geometries, the vortex line can bend and support excitations.

\begin{figure}
\centering
\includegraphics[width=0.8\columnwidth,angle=0]{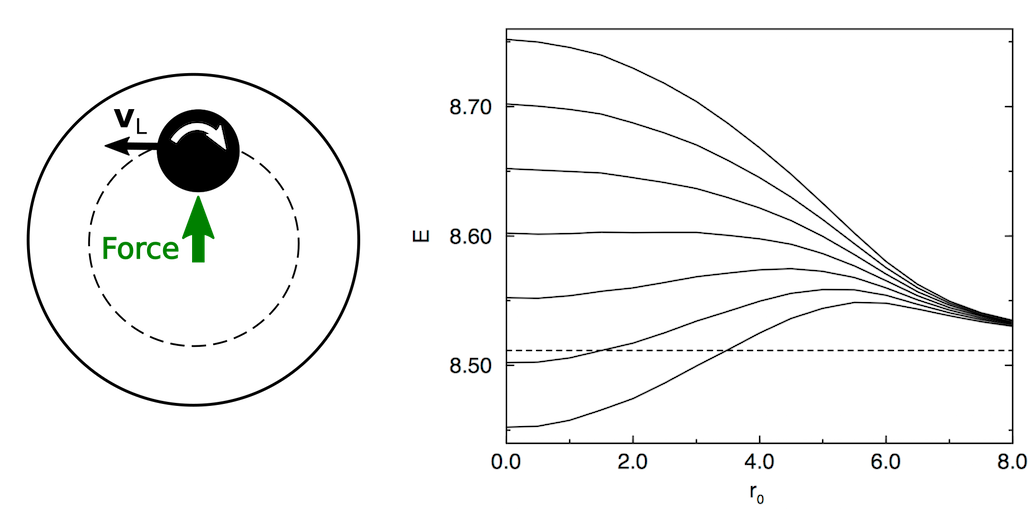}
\caption{Left: Schematic of the Magnus effect which causes an off-centre vortex to precess in a trapped condensate. Right: Free energy $E$ of a trapped condensate versus the radial position of a vortex, $r_0$.  The top line is for a non-rotating system, while the lower lines have increasing rotation frequencies.  Reprinted figure with permission from \cite{Jackson_1999}. Copyright 1999 by the American Physical Society.}
\label{fig:magnus}
\end{figure}

\section{Kelvin waves}

A sinusoidal or helical perturbation of the vortex core away from
its rest position is called a {\em Kelvin wave}. Figure~\ref{fig:kelvin} (left)
shows a Kelvin wave of amplitude $A$ and wavelength $\lambda$.
A Kelvin wave of infinitesimal amplitude $A$ and wavelength $\lambda \gg a_0$
rotates with angular velocity\index{vortex!Kelvin waves},
\begin{equation}
\omega_0 \approx \frac{\kappa k^2}{4 \pi} \left( 
\ln{\left( \frac{1}{ka_0} \right)} -0.116 
\right),
\end{equation}
\noindent
where $k=2 \pi/\lambda$ is the wavenumber; in other words, the shorter
the wave the faster it rotates. 
The time sequence shown in Fig.~\ref{fig:Youd-ring-line}
shows a vortex ring which hits a straight vortex. It is apparent that
after the collision the straight vortex is perturbed by Kelvin waves.
Vortex rings can also be perturbed by Kelvin waves, see 
Fig.~\ref{fig:kelvin} (right); the 
vortex ring with waves travels slower than the unperturbed 
circular ring.  Vortex lines also support excitations in the form of 
breathers \cite{Salman_2013}.

\begin{figure}
\centering
\includegraphics[width=0.25\columnwidth,angle=0]{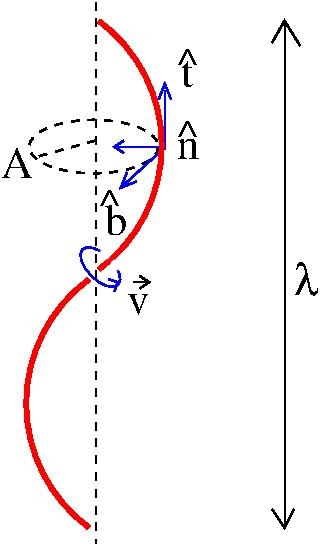}\hspace{3cm}
\includegraphics[width=0.18\columnwidth,angle=0]{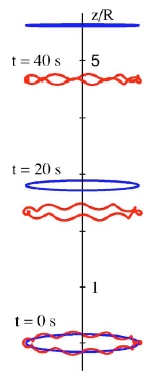}
\caption{Left: Schematic of Kelvin waves of amplitude $A$ and wavelength $\lambda$.
The three unit vectors in the tangent, normal and binormal directions
are shown. The waves rotate along the binormal direction, in the
direction opposite to the direction of the flow. Right: Comparison between motion of a
vortex ring (radius $R=0.1~\rm cm$, blue)
and vortex ring perturbed by Kelvin
waves (relative amplitude $A/R=0.05$, red). Calculation performed with the
vortex filament model \cite{Hanninen-2006}.  Figure adapted with permission from Ref. \cite{Hanninen-2006}. Copyrighted by the American Physical Society.
 }
\label{fig:kelvin}
\end{figure}

\section{Vortex reconnections}

When two quantum vortex lines approach each other, they reconnect, 
changing the topology of the flow. The effect, 
illustrated in Fig. \ref{fig:recon}, has been experimentally observed in
superfluid helium \cite{Bewley-2008} and in atomic condensates 
\cite{Serafini-2015}.
In classical inviscid fluids (governed by the Euler equation)
vortex reconnections are not
possible. Reconnections of quantum vortices thus arise
from the presence of the quantum pressure term in the Gross-Pitaevskii
equation.
In classical viscous fluids (governed by the Navier-Stokes equation)
reconnections are possible but involve dissipation of  energy, 
whereas in condensates reconnections take place while conserving the energy.
Figure~\ref{fig:Zuccher-1and2} shows the reconnection of two vortices computed
using the GPE. A vortex-antivortex pair, initially slightly bent, propagates to the right.  The curvature of the vortices quickly
increases at their midpoint, they move faster and hit each other, 
reconnecting and then moving away. \index{vortex!reconnections}

\begin{figure}[b]
\centering
\includegraphics[width=0.6\columnwidth,angle=0]{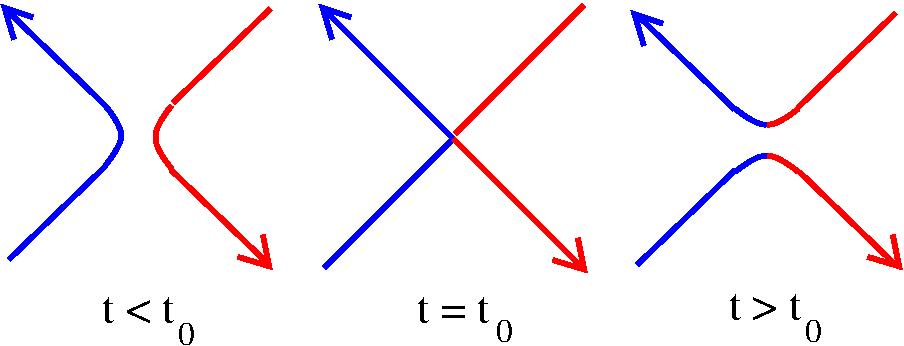}
\caption{Schematic vortex reconnection of two vortex lines.
The arrows indicate the direction of the vorticity (the rotation of
the fluid around the axis of the vortex).
Left: before the reconnection ($t<t_0)$; 
Middle: at the moment of reconnection, $t=t_0$;
Right: after the reconnection ($t< t_0$).
}
\label{fig:recon}
\end{figure}

\begin{figure}
\centering
\includegraphics[width=0.75\columnwidth,angle=0]{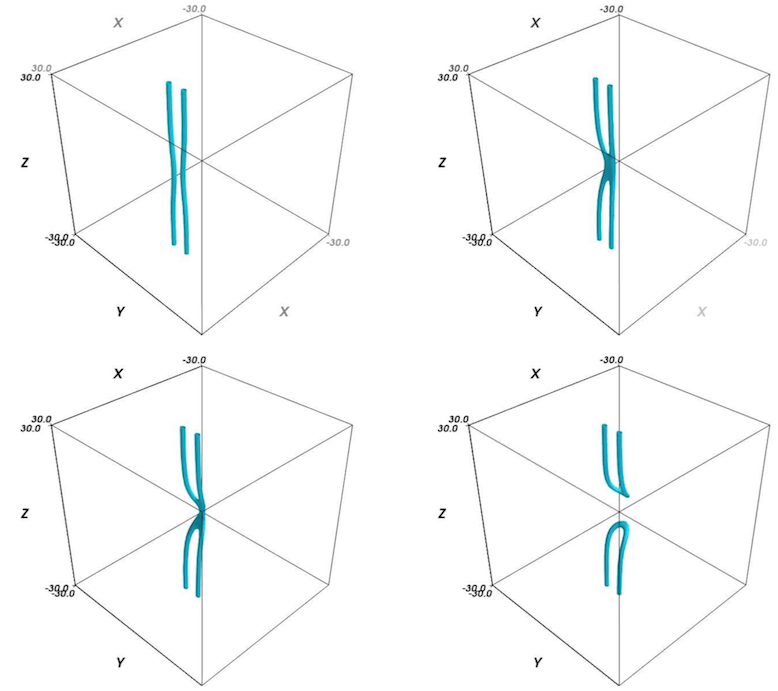}
\caption{Reconnection of antiparallel vortex lines computed by solving
the GPE in a periodic box.
Shown is the isosurface of the condensate density $\rho=0.2$, where
$\rho=1.0$ is the bulk value.  Reprinted from \cite{Zuccher-2012}. with the permission of AIP Publishing.
}
\label{fig:Zuccher-1and2}
\end{figure}

In 2D, vortex reconnections become {\it annihilation} events in which
two vortex points of opposite polarity destroy each other.  This can occur through the interaction with a third vortex, and leaves behind a soliton-like rarefaction pulse of sound \cite{Stagg-2015}.  Recently, it has been argued that a fourth vortex is required to turn
the rarefaction pulse into sound waves which then spread to infinity
\cite{Groszek-Simula-2016,Cidrim-2016}, making the annihilation a four-vortex
process.

%

%

\section{Sound emission}
Even in the absence of thermal effects, vortices can lose energy, and they do so by creating sound waves.  
This occurs when vortices and vortex elements accelerate, 
for example, Fig.~\ref{fig:Parker-sound}(left)
shows the pattern of spiral sound waves emitted outwards by a co-rotating 
pair of vortices. It also arises during vortex reconnections, which release 
a sharp pulse of sound, as seen in 
Fig.~\ref{fig:Parker-sound} (right).  
In 2D annihilation events leave behind only sound waves. \index{sound!emission} \index{vortex!sound emission}

In all of these scenarios, the pattern 
of the condensate phase changes.
The information about this change can travel outwards from the 
vortices no faster than the speed of sound.  Beyond 
this ``information horizon'', the condensate phase has the old 
pattern.  The sound waves act to smooth between the new and old 
patterns, and prevent discontinuities in the phase at this horizon.

The time evolution of a condensate described by the GPE (that is, a condensate at very small temperatures) conserves the total energy,
although the relative proportion of kinetic energy (due to vortices)
and sound energy (due to waves) may change.  In general, a collection of freely-evolving vortices will decay into sound waves, with the energy being transferred into the ``sound field'', although this decay is typically very slow.  The decay can be prohibited, or even reversed, by suitable driving of the system, and under certain conditions, intense sound waves can create 
vortices \cite{Berloff-2004}.


\begin{figure}[t]
\centering
\includegraphics[width=0.42\columnwidth,angle=0]{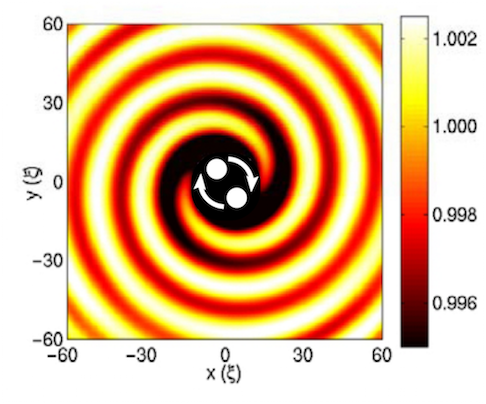}\hspace{0.6cm}
\includegraphics[width=0.38\columnwidth,angle=0]{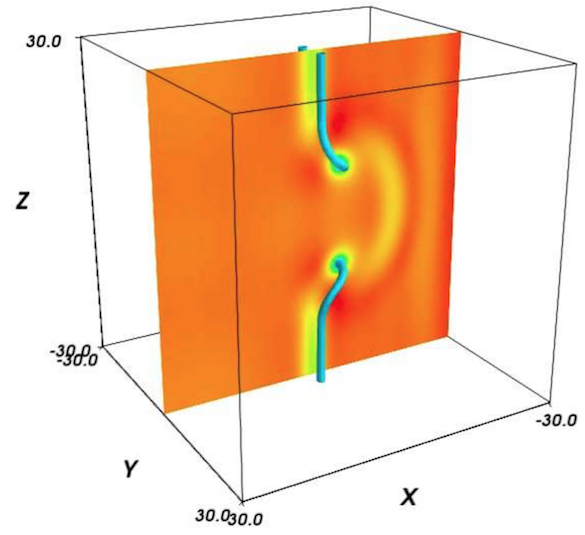}
\caption{Left: Pattern of sound waves (density variations)
on the $xy$ plane generated by a rotating pair of vortices (shown by the white dots).  Image adapted from \cite{Parker-thesis}.  Note the small amplitude of the sound waves, relative to the background density of one.
Right: Rarefaction sound pulse generated by the vortex reconnection of Fig.~\ref{fig:Zuccher-1and2}, shown as density
variations on the central plane.  Reprinted from \cite{Zuccher-2012}. with the permission of AIP Publishing.
}
\label{fig:Parker-sound}
\end{figure}

\section{Quantum turbulence}
\label{sec:QT}

Besides lattices, Kelvin waves and vortex rings, other complex
vortex states have been studied recently, e.g., U- and
S-shaped vortices \cite{Aftalion-Danaila-2003} and vortex knots
\cite{Proment-2012}, see Fig.~\ref{fig:knots}. 
But the most challenging vortex state is turbulence. \index{quantum turbulence}

\begin{figure}[b]
\centering
\includegraphics[width=0.47\columnwidth,angle=0]{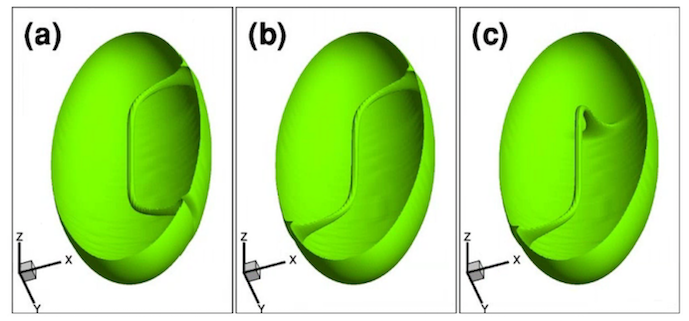}\quad \quad
\includegraphics[width=0.4\columnwidth,angle=0]{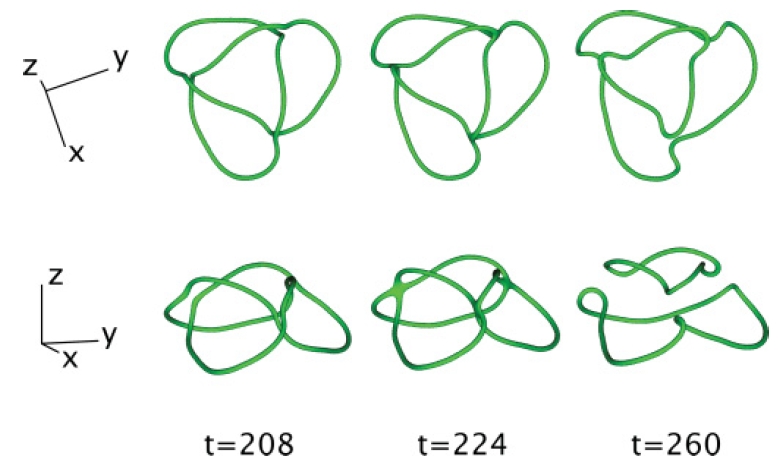}
\caption{Left: U and S-shaped vortices in a spheroidal condensate.  Reprinted figure with permission from \cite{Aftalion-Danaila-2003}. Copyright 2003 by the American Physical Society.  Right: The break-up of a $T_{2,3}$ vortex knot
into two vortex rings.  Reprinted figure with permission from \cite{Proment-2012}. Copyright 2012 by the American Physical Society.
}
\label{fig:knots}
\end{figure}

A disordered vortex configuration of many vortices
is called a {\em vortex tangle}; \index{vortex!tangle}
it represents a state of {\em quantum turbulence}.  
Vortex reconnections and the resulting generation of smaller and smaller
vortex loops in a cascade process were first
conjectured by Richard Feynman in his pioneering 1955 article
on the applications of quantum mechanics to liquid helium \cite{Feynman-1955}.  
Figure~\ref{fig:Feynman} illustrates this cascade.  
Vortices move in an irregular way around each other, undergoing
reconnections which trigger Kelvin waves \index{vortex!Kelvin waves}and generate small
vortex loops. In a statistical steady state, the intensity of the
turbulence is usually measured (experimentally and numerically) by
the {\em vortex line density} $L$, defined as the length of vortex
lines per unit volume. \index{vortex!line density} From the vortex line density $L$ one estimates
that the typical distance between vortices is $\ell \approx L^{-1/2}$.  As well as vortices, quantum turbulence also features sound waves.

\begin{figure}[t]
\centering
\includegraphics[width=0.45\columnwidth,angle=0]{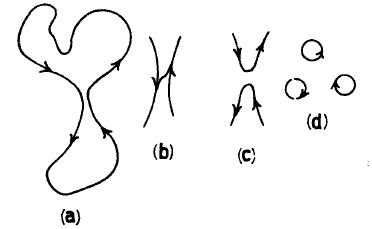}
\caption{Schematic of vortex reconnections and generation of small vortex
loops, as envisaged by Richard Feynman \cite{Feynman-1955}.
}
\label{fig:Feynman}
\end{figure}

Current work \cite{Barenghi-PNAS-2014,Tsatsos-2016} studies properties
of quantum turbulence such as velocity and acceleration statistics
\cite{White-2010},
the emergence of coherent structures out of disorder, and the
energy spectrum $E_k$ (representing the distribution of
the kinetic energy over the length scales); in particular, 
the energy spectrum
is defined from, \index{quantum turbulence!energy spectrum}
\begin{equation}
E'=\frac{1}{\cal V} \int_{\cal V} \frac{{\bf v}^2}{2} {\rm d}^3{\bf r}
=\int_0^{\infty} E_k ~{\rm d}k,
\label{eqn:specrum}
\end{equation}
where $E'$ is energy per unit mass, $\cal V$ is the volume and $k$ the
wavenumber.

The two main tools to study quantum turbulence are the GPE and the
{\it vortex filament model}, which we describe in 
Section~\ref{sec:thin-core}; the latter is directly relevant to
superfluid helium, but is important in general, as it isolates vortex
interactions, neglecting finite core-size effects and sound waves.
In the next subsections we describe recent results for 3D and 2D turbulence.

\subsection{Three-dimensional quantum turbulence}
\label{subsec:3D-QT}

Quantum turbulence at very low temperatures is generated in superfluid
helium by stirring with grids,  wires or propellers, 
or by injecting vortex rings.  \index{quantum turbulence!in 3D}
Observations of the decay of the vortex line density and the energy spectrum
reveal two turbulent regimes \cite{Walmsley-Golov-2008}. 
In the first regime \cite{Tabeling-1998}, called 
{\it quasi-classical turbulence} and illustrated in Fig.~\ref{fig:turbo1},
the energy spectrum obeys the same Kolmogorov scaling
\index{quantum turbulence!energy spectrum}
of ordinary turbulence ($E_k \sim k^{-5/3}$) over the hydrodynamic range
$k_D \ll k \ll k_{\ell}$ (where $k_{\ell}=2 \pi/\ell$, $k_D=2 \pi/D$ and
$D$ is the system size). This result is confirmed by numerical simulations
based on the GPE \cite{Nore-1997} and the vortex filament model
\cite{Araki-Tsubota-Nemir-2002,Baggaley-2012,Baggaley-2012-ultra}. 
Kolmogorov scaling suggests the existence of a classical {\it cascade}, which,
step-by-step, transfers energy from large eddies to smaller  
eddies.  The concentration of energy at the
largest length scales (near $k_D$) arises from the emergence of
transient bundles of
vortices of the same polarity \cite{Laurie-2012} 
which induce large scale flows.
Without forcing, quasi-classical turbulence decays as $L \sim t^{-3/2}$.  \index{quantum turbulence!decay regimes}

However, under other conditions, $E_k$ peaks at the intermediate scales 
followed at large wavenumbers by the $k^{-1}$ dependence 
typical of isolated vortices, suggesting a random 
vortex configuration without cascade \cite{Baggaley-2012-ultra}. 
In the absence of forcing, this regime,
called {\it ultra-quantum turbulence} \cite{Walmsley-Golov-2008}, 
decays as $L \sim t^{-1}$.

\begin{figure}[t]
\centering
\includegraphics[width=0.4\columnwidth,angle=0]{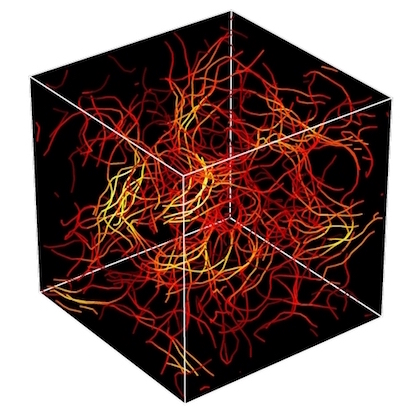}
\includegraphics[width=0.5\columnwidth,angle=0]{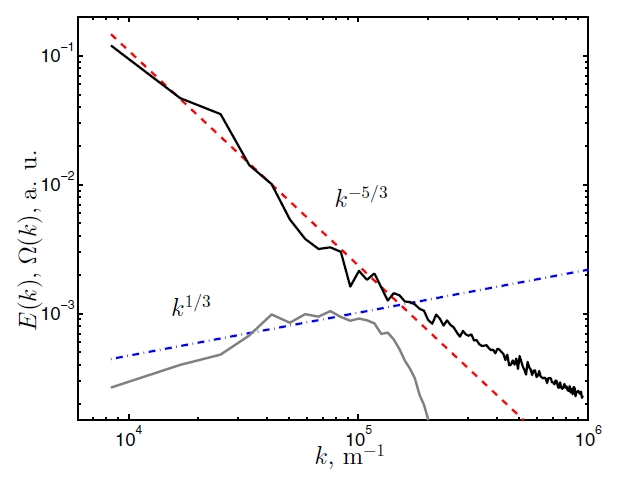}
\caption{
Left: Quantum turbulence in superfluid helium
computed in a periodic box using the
vortex filament method \cite{Laurie-2012}. Lighter colour denotes
bundles of vortex lines with the same orientation: they are
responsible for the emergence of the classical $k^{-5/3}$ Kolmogorov
spectrum.  Figure adapted with permission from Ref. \cite{Laurie-2012}. Copyrighted by the American Physical Society. \index{quantum turbulence!energy spectrum}
Right: Energy spectrum of the kinetic energy $E_k$ vs $k$,
 computed using the vortex
filament method \cite{Baggaley-2012}: note the $k^{-5/3}$ Kolmogorov
scaling for $k< k_{\ell} \approx 1.8 \times 10^5~{\rm m^{-1}}$. 
The curve at the bottom shows that the spectrum of the 
coarse-grained vorticity is consistent with the $k^{1/3}$ scaling of 
Kolmogorov theory.
}
\label{fig:turbo1}
\end{figure}

\begin{figure}
\centering
\includegraphics[width=0.26\columnwidth,angle=0]{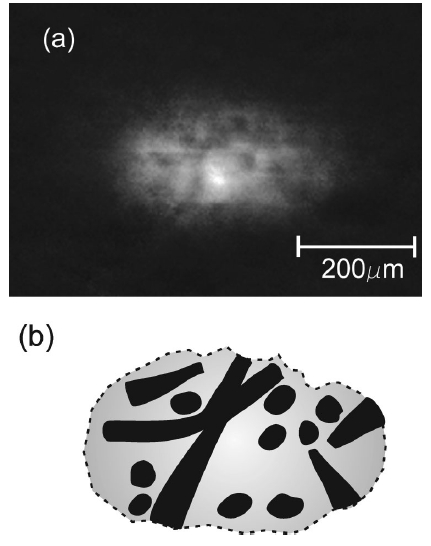}
\includegraphics[width=0.37\columnwidth,angle=0]{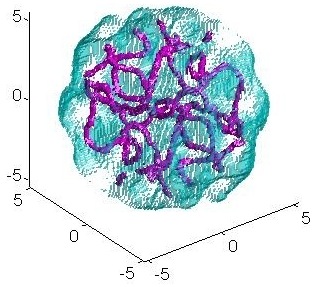}
\includegraphics[width=0.32\columnwidth,angle=0]{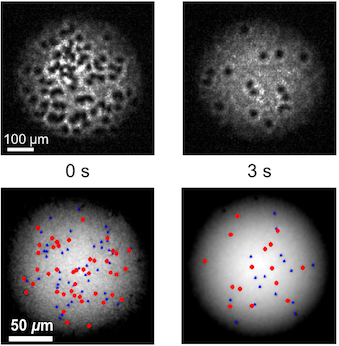}
\caption{
Left: Absorption images of turbulent 3D atomic condensate (top)
and schematic diagram of the inferred distribution of
vortices (bottom) \cite{Henn-2009}.  Reprinted figure with permission from \cite{Henn-2009}. Copyright 2009 by the American Physical Society.
Middle: Quantum turbulence in a harmonically confined atomic
condensate computed using the GPE. The surface of the
condensate is pale blue, the surface of the vortex cores is purple. Figure adapted with permission from Ref. \cite{White-2010}. Copyrighted by the American Physical Society. Right:  Experimental absorption (top) images of an condensate in a 
state of 2D turbulence \cite{kwon_2014}.  Images courtesy of Y. I. Shin. 
Corresponding images of (unexpanded) condensate density from GPE 
simulations (bottom) \cite{Stagg-2015}.  Images courtesy of G. W. Stagg.  Vortices with positive (negative) 
circulation are highlighted by red circles (blue triangles).  
The vortices appear much smaller since the condensate has not been expanded.
}
\label{fig:turbo2}
\end{figure}

Turbulence in atomic condensates has been generated by stirring the gas with a 
laser beam or by shaking the confining trap \cite{Henn-2009,Tsatsos-2016}.  
Current 3D condensates
created in the laboratory are relatively small, see 
Fig.~\ref{fig:turbo2}. The limited 
separation of length scales (unlike helium,
$D$ is not much bigger than $\ell$, which is
not much bigger than $a_0$) and the difficulty in directly
measuring the velocity
have so far prevented measurements of the energy spectrum, although
the Kolmogorov regime has been predicted \cite{Kobayashi-Tsubota-2007}.

\subsection{Two-dimensional quantum turbulence}
\label{sec:2d_turbo}

Due to the ability to engineer the effective dimensionality, atomic condensates
also allow the study of 2D turbulence, which consists of a disordered 
arrangement of vortex points and waves. This is a remarkable feature
of quantum fluids, because (with the possible exception of soap films)
ordinary flows are never really 2D (for example, only by considering large-scale
patterns the atmosphere can be approximated by a 2D flow).  
Figure 5.21 (right) shows experimental and simulated images 
of 2D turbulence in a trapped condensate.  The turbulence is not being 
driven and so the number of vortices decays over time.
\index{quantum turbulence!in 2D}

In fluid dynamics, 2D turbulence is expected to shown 
unique features such as an {\it inverse cascade} where increasingly \index{quantum turbulence!inverse cascade}
large vortical structures form over time (an example is
Jupiter's great Red Spot). The inverse cascade
involves the clustering of vortices
with the same sign, predicted by Onsager, and represents a phase 
transition associated with a state of negative effective
temperature (defined in terms of the entropy of the vortex
configuration).
In the opposite limit the
vortices tend to form dipoles \cite{Simula-2014,Billam_2014}.


\section{Vortices of infinitesimal thickness}
\label{sec:thin-core}

In this section we derive mathematical tools to model
quantized vortex lines as {\em vortex filaments} (in 3D)
or {\em vortex points} (in 2D). Both methods are based
on the classical Euler equation. They assume that the fluid
is incompressible, thus neglecting sound waves, and treat the vortex
cores as line (in 3D) or point (in 2D) singularities. This approximation
is realistic for helium turbulence experiments, 
where there is a wide separation of length scales between the
system size ($D \approx 10^{-2}$ to $10^{-1}~\rm m$), 
the inter-vortex distance ($\ell \approx 10^{-6}$ to $10^{-4}~\rm m$)
and the vortex core radius ($a_0 \approx 10^{-10}~\rm m$).} The
approximation is less good for atomic condensates,
but the model is useful to isolate pure vortex dynamics from
sound and healing length effects.

We have seen that, at length scales larger than the healing length $\xi$,
the Gross-Pitaevskii equation reduces to classical continuity equation
and the compressible Euler equation. In the further limit of velocities much
less than the speed of sound (i.e. small Mach numbers), density variations 
can be neglected; in this limit,
the compressible Euler equation reduces to the incompressible Euler
equation,

\begin{equation}
\frac{\partial {\bf v}}{\partial t}+ ({\bf v} \cdot \nabla) {\bf v} 
=-\frac{1}{\rho} \nabla p,
\label{eqn:incompressible-Euler}
\end{equation}

\noindent
where $\rho$ is constant, and the continiuty equation becomes the
solenoidal condition $\nabla \cdot {\bf v}=0$.

\subsection{Three-dimensional vortex filaments}

\noindent
We introduce the vector potential ${\bf A}$ defined such that,
${\bf v}=\nabla \times {\bf A}$.
Since the divergence of a curl is always zero, we have
$\nabla \cdot {\bf A}=0$,
and ${\bf A} \to {\rm constant}$ for ${\bf x} \to \infty$.
The vorticity $\bom$ can be written as,
\begin{equation}
\bom=\nabla \times {\bf v}=\nabla \times (\nabla \times {\bf A})
=\nabla (\nabla \cdot {\bf A}) -\nabla^2 {\bf A}=-\nabla^2 {\bf A},
\end{equation}

Given the vorticity distribution $\bom({\bf r},t)$ at the time $t$,
the vector potential ${\bf A}({\bf r},t)$
is obtained by solving Poisson's equation,
\begin{equation}
\nabla^2 {\bf A}=-\bom.
\label{eqn:Poisson}
\end{equation}\index{vortex!local induction approximation} \index{vortex!filament method}

\noindent
The solution of Eq.~(\ref{eqn:Poisson}) at the point ${\bf s}$ is,
\begin{equation}
{\bf A}({\bf s},t)=
\frac{1}{4 \pi} \int_\mathcal{V} 
\frac{\bom({\bf r},t)}{\vert {\bf s}-{\bf r} \vert} {\rm d}^3 {\bf r},
\end{equation}

\noindent
where ${\bf r}$ is the variable of integration and $\mathcal{V}$ is  volume. 
Taking the curl (with respect
to ${\bf s}$), we obtain the {\em Biot-Savart law}, \index{vortex!Biot-Savart law}

\begin{equation}
{\bf v}({\bf s},t)
=\frac{1}{4 \pi} \int_V \frac{\bom({\bf r},t) \times ({\bf s}-{\bf r})}
{\vert {\bf s}-{\bf r} \vert^3} {\rm d}^3{\bf r}.
\label{eqn:BS0}
\end{equation}

In electromagnetism, the Biot-Savart law 
determines the magnetic field as a function of
the distribution of currents. In vortex dynamics,
the Biot-Savart law
determines the velocity as a function of the distribution of
vorticity.
If we assume that the vorticity  $\bom$ is
concentrated on filaments of infinitesimal thickness with circulation
$\kappa$, we can formally
replace $\bom({\bf r},t) {\rm d}^3{\bf r}$ with
$\kappa {\rm d}{\bf r}$.  The volume integral, Eq.~(\ref{eqn:BS0}),
becomes a line integral over the vortex line configuration $\cal L$,
and the Biot-Savart law reduces to,

\begin{equation}
{\bf v}({\bf s},t)
=-\frac{\kappa}{4 \pi} \oint_{\cal L} \frac{({\bf s}-{\bf r})}
{\vert {\bf s}-{\bf r} \vert^3} \times {\rm d}{\bf r}.
\label{eqn:BS1}
\end{equation}

Equation~(\ref{eqn:BS1}) is the cornerstone of the vortex filament method,
in which we model quantized vortices
as three dimensional oriented space curves ${\bf s}(\xi_0,t)$ 
of circulation $\kappa$,
where the parameter $\xi_0$ is arc length. Since, according to Helmholtz's
Theorem, a vortex line moves with the flow, the time evolution of
the vortex configuration is given by,
\begin{equation}
\frac{{\rm d}{\bf s}}{{\rm d}t}={\bf v}^{\rm self}({\bf s}),
\end{equation}

\noindent
where,

\begin{equation}
{\bf v}^{\rm self}({\bf s})
=-\frac{\kappa}{4 \pi} \oint_{\cal L} 
\frac{ ({\bf s}-{\bf r})}
{\vert {\bf s}-{\bf r} \vert^3} \times {\rm d}{\bf r}.
\label{eqn:BS2}
\end{equation}

\noindent
(the {\em self-induced velocity})
is the velocity which {\em all} vortex lines present in the flow induce
at the point ${\bf s}$.

To implement the vortex filament method, vortex lines are discretized
into a large number of points ${\bf s}_j$ ($j=1,2, \cdots$), each point
evolving in time according to Eq.~({\ref{eqn:BS2}). Vortex reconnections
are performed algorithmically.  Since the integrand
of Eq.~(\ref{eqn:BS2}) diverges as ${\bf r} \to {\bf s}$,  it must be
desingularized; a physically sensible cutoff length scale is
the vortex core radius $a_0$. This cutoff idea is also behind the 
following {\em Local Induction Approximation} (LIA) to the Biot-Savart law,

\begin{equation}
{\bf v}^{\rm self}({\bf s})
=\beta {\bf s}' \times {\bf s}'',
\qquad
\beta=\frac{\kappa}{4 \pi} \ln \left(\frac{R}{a_0}\right),
\label{eqn:LIA}
\end{equation}

\noindent
where ${\bf s}'={\rm d}{\bf s}/{\rm d}\xi_0$ is the unit {\em tangent} 
vector at the point ${\bf s}$, 
${\bf s}''={\rm d}^2{\bf s}/{\rm d}\xi_0^2$ is in the {\em normal} direction, 
and $R=1/\vert {\bf s}'' \vert$ is the local
radius of curvature. The physical interpretation of the LIA is simple:
at the point ${\bf s}$, a vortex moves in the {\em binormal} direction
with speed which is inversely proportional to the local radius of
curvature. Note that a straight vortex line does not move, as its radius of
curvature is infinite. 

To illustrate the LIA, we compute the velocity of
a vortex ring of radius $R$ located on the $z=0$ plane at $t=0$.
The ring is described by the space curve 
${\bf s}=(R \cos{(\theta)},R \sin{(\theta)},0)$, where $\theta$ is the
angle and $\xi_0=R \theta$ is the arc length. Taking derivatives with
respect to $\xi_0$ we have
${\bf s}'= (-\sin{(\xi_0/R)},\cos{(\xi_0/R)},0)$ and
${\bf s}''=(-1/R)(\cos{(\xi_0/R)},\sin{(\xi_0/R)},0)$. Using
Eq.~(\ref{eqn:LIA}), we conclude that the vortex ring moves in the
$z$ direction with velocity,
\begin{equation}
{\bf v}^{\rm self}=\frac{\kappa}{4 \pi R} 
\ln{(R/a_0)} \ehat_z.
\label{eqn:ring-LIA}
\end{equation}

The result is in good agreement with a more precise solution of the
Euler equation based
on a hollow core at constant volume, which is,

\begin{equation}
{\bf v}^{\rm self}=\frac{\kappa}{4 \pi R} 
\left( \ln\left( \frac{8 R}{a_0}\right) -\frac{1}{2} \right) \ehat_z.
\label{eqn:ring-BS}
\end{equation}

Using the GPE, Roberts and Grant \cite{Roberts-Grant-1971} found that
a vortex ring of radius much larger than the healing length moves with
velocity,

\begin{equation}
{\bf v}^{\rm self}=\frac{\kappa}{4 \pi R} 
\left( \ln \left( \frac{8 R}{a_0}\right) -0.615 \right) \ehat_z.
\label{eqn:ring-BS}
\end{equation}

\subsection{Two-dimensional vortex points}

As in the previous section, we consider inviscid, 
incompressible ($\nabla \cdot {\bf v}=0$),  irrotational
($\nabla \times {\bf v}={\bf 0}$) flow, and allow singularities.
We also assume that the flow is two-dimensional on the
$xy$ plane, with velocity field,
\begin{equation}
{\bf v}(x,y)=(v_x(x,y),v_y(x,y)),
\end{equation}  \index{vortex!points}

\noindent
The introduction of the {\em stream function}
$\psi$ (not to be confused with the wavefunction), defined by, \index{stream function}
\begin{equation}
v_x=\frac{\partial \psi}{\partial y},
\qquad
v_y=-\frac{\partial \psi}{\partial x},
\end{equation}

\noindent
guarantees that $\nabla \cdot {\bf v}=0$.
The irrotationality of the flow 
implies the existence of a velocity potential $\phi$ such that
${\bf v}=\nabla \phi$, \index{velocity potential}

\begin{equation}
v_x=\frac{\partial \phi}{\partial x},
\qquad
v_y=\frac{\partial \phi}{\partial y}.
\end{equation}

\noindent
It follows that both stream function and velocity potential satisfy the
two-dimensional Laplace's equation ($\nabla^2 \psi=0$, $\nabla^2 \phi=0$), 
and well-known techniques of complex variables can be applied. 
For this purpose, let $z=x+iy$
be a point of the complex plane (rather than the vertical coordinates). 
We introduce the {\em complex potential},
\begin{equation}
\Omega(z)=\phi+i\psi.
\end{equation}
\noindent
It can be shown that the velocity components $v_x$ and $v_y$ are obtained from,
\begin{equation}
v_x-iv_y=\frac{{\rm d}\Omega}{{\rm d}z},
\end{equation}

Any complex potential $\Omega(z)$ can be interpreted as a two-dimensional
inviscid, incompressible, irrotational flow. Since Laplace's equation 
is linear, the sum of solutions is another solution, and we 
can add the
complex potential of simple flows to obtain
the complex potential of more complicated flows.
In particular,

\begin{equation}
\Omega(z)=U_0 e^{-i\eta}z,
\end{equation}
\noindent
represents a uniform flow of speed $U_0$ at angle $\eta$ with
the $x$ axis, and,

\begin{equation}
\Omega(z)=-\frac{i \kappa}{2 \pi}\log{(z-z_0)},
\end{equation}
\noindent
represents a positive (anticlockwise) vortex point of circulation
$\kappa$ at position $z=z_0$.

\section*{Problems}
\addcontentsline{toc}{section}{Problems}

\begin{prob}
\label{critical-rotation}
Consider the bucket of Sections \ref{sec:energy} and \ref{sec:rotating} 
to now feature a harmonic potential $V(r)=\frac{1}{2}m \omega_r^2 r^2$ 
perpendicular to the axis of the cylinder.  Take the condensate to adopt 
the Thomas-Fermi profile.
\begin{itemize}
\item[(a)]~Show that the energy of the vortex-free condensate is $E_0=\pi m n_0 \omega_r^2 H_0 R_r^4 / 6$, where $R_r$ is the radial Thomas-Fermi radius and $n_0$ is the density along the axis.
\item[(b)]~Now estimate the kinetic energy $E_{\rm kin}$ due to a vortex along the axis via Eq. (\ref{eqn:kin}).  Use the fact that $a_0 \ll R_r$ to simplify your final result. 
\item[(c)]~Estimate the angular momentum of the vortex state, and hence estimate the critical rotation frequency at which the presence of a vortex becomes energetically favourable.  
\end{itemize}

\end{prob}

\begin{prob}
\label{LIA-Kelvinwave}
Use the LIA 
(Eq.~\ref{eqn:LIA})
to determine the angular frequency of rotation of
a Kelvin wave of wave length $\lambda=2\pi/k$ (where $k$ is the
wavenumber) on a vortex with circulation $\kappa$.
\end{prob}

\begin{prob}
\label{vortex-antivortex}
Using the vortex point method and the complex potential, determine
the translational speed of a vortex-antivortex pair 
(each of circulation $\kappa$)
separated by the distance $2D$.
\end{prob}

\begin{prob}
\label{vortex-vortex}
Using the vortex point method and the complex potential, determine
the period of rotation of a vortex-vortex pair 
(each of circulation $\kappa$)
separated by the distance $2D$.
\end{prob}

\begin{prob}
\label{tangle-energy}
Consider a homogeneous, isotropic, random vortex tangle 
(ultra-quantum turbulence) of vortex line
density $L$, contained in a cubic box of size $D$.
Show that the kinetic energy is approximately
\begin{displaymath}
E \approx \frac{\rho \kappa^2 L D^3}{4 \pi} \ln \left( \frac{\ell}{a_0} \right),
\end{displaymath}
where $\rho$ is the density, $\kappa$ the quantum of circulation,
$\ell \approx L^{-1/2}$ is the inter-vortex distance and $a_0$ is
the vortex core radius.

\end{prob}

\begin{prob}
\label{ultra-quantum-decay}
In an ordinary fluid of kinematic viscosity $\nu$, the decay of the 
kinetic energy per unit mass, $E'$, obeys the equation
\begin{displaymath}
\frac{dE'}{dt}=-\nu \omega^2,
\end{displaymath}
\noindent
where $\omega$ is the rms vorticity. Consider ultra-quantum turbulence
of vortex line density $L$. Define the rms superfluid vorticity as
$\omega=\kappa L$, and show thet the vortex line density obeys
the equation,

\begin{displaymath}
\frac{dL}{dt}= -\frac{\nu}{c} L^2,
\end{displaymath}
\noindent
where the constant $c$ is,
\begin{displaymath}
c=\frac{1}{4 \pi} \ln \left( \frac{\ell}{a_0}\right),
\end{displaymath}
\noindent
hence show that, for large times, the turbulence decays as

\begin{displaymath}
L \sim \frac{c}{\nu} t^{-1}.
\end{displaymath}
\end{prob}

\appendix
\chapter{Simulating the 1D GPE}
\label{intro} 

The GPE is a nonlinear partial differential equation, and its
solution must, in general, be obtained numerically.  A variety of 
numerical methods exist to solve the GPE, including those based on 
Runge-Kutta methods, the Crank-Nicolson method and the split-step Fourier 
method  \footnote{A. Minguzzi, S. Succi, F. Toschi, M. P. Tosi, P. Vignolo, Phys. Rep. {\bf 395}, 223 (2004)}.  The latter (also known as the 
time-splitting spectral method) is particularly compact and efficient,
and here we apply it to the 1D GPE. Furthermore, we introduce
the imaginary time method for obtaining ground state solutions.  
Basic Matlab code is provided.

\section{Split-Step Fourier Method}
The split-step fourier method is well-established for numerically solving the time-dependent Schrodinger equation, written here in one-dimension,
\begin{equation}
i \hbar \frac{\partial \psi(x,t)}{\partial t}=\hat{H} \psi(x,t).
\end{equation}
The Hamiltonian $\hat{H}$ can be expressed as $\hat{H}=\hat{T}+\hat{V}$, where $\hat{T} \equiv -\frac{\hbar^2}{2m}\frac{\partial^2}{ \partial x^2}$ and $\hat{V}\equiv V(x)$ are the kinetic and potential energy operators. Integrating from $t$ to $t+\Delta t$ (and noting the time-independence of the Hamiltonian) leads to the time-evolution equation,
\begin{equation}
\psi(x,t+\Delta t)=e^{-i \Delta t \hat{H} / \hbar}\psi(x,t).
\end{equation}
The operators $T$ and $V$ do not commute, hence
$e^{- i \Delta t\hat{H}/\hbar} \neq e^{-i \Delta t \hat{T} / \hbar} e^{-i \Delta t \hat{V} /\hbar}$.  Nonetheless, the following approximation,
\begin{equation}
e^{-i \Delta t \hat{H} /\hbar} \psi \approx e^{-i \Delta t \hat{V}/2\hbar} e^{-i \Delta t \hat{T} / \hbar} e^{-i \Delta t \hat{V} / 2\hbar} \psi,
\label{eqn:split1}
\end{equation}
holds
with error $\mathcal{O}(\Delta t^3)$.  In position space $\hat{V}$ is 
diagonal, and so the operation $e^{-i \Delta t \hat{V} /2 \hbar} \psi$ 
simply corresponds to multiplication of $\psi(x,t)$ by 
$e^{-i \Delta t  V(x) / 2 \hbar}$.  Although
$\hat{T}$ is not diagonal in position space, it becomes diagonal in 
reciprocal space.  Conversion to reciprocal space is achieved by taking the Fourier transform $\mathcal{F}$ of the wavefunction $\tilde{\psi}(k,t) = \mathcal{F} [\psi(x,t)]$, where $k$ denotes the 1D wavevector.  Then the kinetic energy operation corresponds to multiplication of $\tilde{\psi}(k,t)$ by $e^{- i \hbar \Delta t  k^2/2m}$.  Thus Eq. (\ref{eqn:split1}) can be written as,
\begin{equation}
\psi(x,t+\Delta t) \approx e^{-\frac{i}{2\hbar} V(x) \Delta t} \cdot \mathcal{F}^{-1} \left[e^{-\frac{i \hbar k^2}{2m}\Delta t} \cdot \mathcal{F}[e^{-\frac{i}{2\hbar} V(x) \Delta t} \cdot \psi(x,t)] \right].
\label{eqn:split2}
\end{equation}
In practice, the computational expense of performing forward and backward Fourier transforms to evaluate Eq. (\ref{eqn:split1}) is small (particularly when using numerical fast Fourier transform techniques) compared to the significant expense of evaluating the kinetic energy term directly in position space.  Note that the split-step method naturally incorporates periodic boundary conditions.

The above method was developed for the linear Schrodinger equation 
with time-independent Hamiltonian.  Remarkably, it holds for the GPE 
(despite its nonlinearity and time-dependent Hamiltonian)
under the replacement $V(x) \mapsto V(x) + g |\psi|^2$.  
Errors of $\mathcal{O}(\Delta t^3)$ are maintained, providing the most up-to-date $\psi$ is always employed during the sequential operations in Eq. (\ref{eqn:split2}) \cite{javanainen_2006}.

\section{1D GPE Solver}
We now outline the approach to solve the 1D GPE using the split-step method, with reference to the Matlab code included below.  To make the numbers more convenient, the GPE is divided through by $\hbar$ (equivalent to considering energy in units of $\hbar$). 
We consider a 1D box, discretized into grid points with spacing $\Delta x$ (\texttt{dx}), and extending over the spatial range $x=[-M \Delta x, M \Delta x]$, where $M$ ({\texttt{M}) is a positive integer.  Position is described by a vector $x_i$ (\texttt{x}), defined as $x_i=- M \Delta x + (i-1)\Delta x$, with $i=1,...,2M+1$.  The potential $V(x)$ is defined as the vector $V_i=V(x_i)$.  
Starting from the initial time, the wavefunction $\psi(x)$, 
represented by the vector $\psi_i=\psi(x_i)$ (\texttt{psi}), 
is evolved over the time interval $\Delta t$ (\texttt{dt}) 
by evaluating Equation (\ref{eqn:split2}) numerically by 
replacing the Fourier transform $\mathcal{F}$ 
(and its inverse $\mathcal{F}^{-1}$) 
by the discrete fast Fourier transform.  Here, wavenumber is 
discretized into a vector $k_i$ (\texttt{k}), defined as 
$k_i=-M \Delta k+(i-1)\Delta k$, with 
$\Delta k = \pi/M \Delta x$  (\texttt{dk}).  
This time iteration step is repeated $N_t$ (\texttt{Nt}) times to 
find the solution at the desired final time.

The Matlab code below simulates a BEC of 5000 $^{87}$Rb atoms with $a_s=5.8$ nm and trapping frequencies
$\omega_\perp = 2 \pi \times 100$ Hz and $\omega_x = 2 \pi \times 40$ Hz.  
Starting from the narrow non-interacting ground state (Gaussisan) profile, 
the condensate undergoes oscillating expansions and contractions, 
due to the competition between repulsive interactions and  
confining potential.  Note - under different scenarios, reduced time and grid spacings may be required to ensure numerical convergence.

\small
\begin{verbatim}
% SOLVES THE 1D GPE VIA THE SPLIT-STEP FOURIER METHOD
clear all;clf; %Clear workspace and figure

hbar=1.054e-34;amu=1.660538921e-27; %Physical constants
m=87*amu;as=5.8e-9; %Atomic mass; scattering length
N=1000;wr=100*2*pi;wx=40*2*pi; %Atom number; trap frequencies

M=200; Nx=2*M+1;
dx=double(2e-7); x=(-M:1:M)*dx; %Define spatial grid
dk=pi/(M*dx); k=(-M:1:M)*dk; %Define k-space grid
dt=double(10e-8); Nt=200000; %Define time step and number

lr=sqrt(hbar/(m*wr)); lx=sqrt(hbar/(m*wx)); %HO lengths
g1d=2*hbar*hbar*as/(m*lr^2); %1D interaction coefficient

V=0.5*m*wx^2*x.^2/hbar; %Define potential
psi_0=sqrt(N/lx)*(1/pi)^(1/4)*exp(-x.^2/(2*lx^2)); %Initial wavefunction

%[psi_0,mu] = get_ground_state(psi_0,dt,g1d,x,k,m,V); %Imaginary time

Nframe=100; %Data saved every Nframe steps
t=0; i=1; psi=psi_0; spacetime=[]; %Initialization

for itime=1:Nt %Time-stepping with split-step Fourier method
     psi=psi.*exp(-0.5*1i*dt*(V+(g1d/hbar)*abs(psi).^2));    
     psi_k=fftshift(fft(psi)/Nx);
     psi_k=psi_k.*exp(-0.5*dt*1i*(hbar/m)*k.^2);
     psi=ifft(ifftshift(psi_k))*Nx; 
     psi=psi.*exp(-0.5*1i*dt*(V+(g1d/hbar)*abs(psi).^2)); 
     if mod(itime,Nt/Nframe) == 0  %Save wavefunction every Nframe steps
       spacetime=vertcat(spacetime,abs(psi.^2)); t
     end   
     t=t+dt;
end

subplot(1,3,1); %Plot potential
plot(x,V,'k'); xlabel('x (m)'); ylabel('V (J/hbar)');

subplot(1,3,2); %Plot initial and final density
plot(x,abs(psi_0).^2,'k',x,abs(psi).^2,'b'); 
legend('\psi(x,0)','\psi(x,T)');xlabel('x (m)');ylabel('|\psi|^2 (m^{-1})');

subplot(1,3,3); % Plot spacetime evolution as pcolor plot
dt_large=dt*double(Nt/Nframe);
pcolor(x,dt_large*(1:1:Nframe),spacetime); shading interp;
xlabel('x (m)'); ylabel('t (s)');

\end{verbatim}
\normalsize

\section{Imaginary time method}
A convenient numerical method for obtaining ground state solutions of the Schrodinger equation/GPE is through imaginary time propagation.  The wavefunction $\psi(x,t)$ can be expressed as a superposition of eigenstates $\phi_m(x)$ with time-dependent amplitudes $a_m(t)$ and energies $E_m(t)$, i.e. $\psi(x,t)= \sum_m a_m(t) \phi_m(x)$, for which, after the substitution
 $t \rightarrow -i\Delta t$, the evolution equation (\ref{eqn:split1}) becomes,
\begin{equation}
\psi(t+\Delta t) = e^{-\Delta t \hat{H}/\hbar} \psi(x,t) =  \sum_m a_m(t) \phi_m(x) e^{-\Delta t E_m/\hbar}.
\end{equation}
The amplitude of each eigenstate contribution decays over time, with the ground state (with lowest $E_m$) decaying the slowest.
Thus, by renormalizing $\psi$ after each iteration (to ensure the conservation of the desired norm/number of particles), 
$\psi$ will evolve towards the ground state.  

Convergence may be assessed by monitoring the chemical potential.  This is conveniently evaluated using the relation $\mu = (\hbar/\Delta t) \ln |\psi(x,t)/\psi(x,t+\Delta t)|$ at some coordinate within the condensate; this relation is obtained by introducing the eigenvalue $\mu$ and imaginary time into Equation (\ref{eqn:split1}).  

The Matlab function \verb|get_ground_state| below obtains the GPE ground state via imaginary time propagation.  Uncommenting line 19 in the above GPE solver calls this function prior to real time propagation; as one  expects, 
the profile remains static in time.

\small
\begin{verbatim}
% SOLVES THE 1D GPE IN IMAGINARY TIME USING THE SPLIT-STEP METHOD
function [psi,mu] = get_ground_state(psi,dt,g1d,x,k,m,V)

hbar=1.054e-34; dx=x(2)-x(1); dk=2*pi/(x(end)-x(1));
N=dx*norm(psi).^2; Nx=length(x);
psi_mid_old=psi((Nx-1)/2); mu_old=1; j=1; mu_error=1;
 while mu_error > 1e-8
     psi=psi.*exp(-0.5*dt*(V+(g1d/hbar)*abs(psi).^2));    
     psi_k=fftshift(fft(psi))/Nx;
     psi_k=psi_k.*exp(-0.5*dt*(hbar/m)*k.^2);
     psi=ifft(ifftshift(psi_k))*Nx;
     psi=psi.*exp(-0.5*dt*(V+(g1d/hbar)*abs(psi).^2)); 
     
     psi_mid=psi((Nx-1)/2);
     mu=log(psi_mid_old/psi_mid)/dt; mu_error=abs(mu-mu_old)/mu;

     psi=psi*sqrt(N)/sqrt((dx*norm(psi).^2));
    if mod(j,5000) == 0    
         mu_error
     end
     if j > 1e8
         'no solution found'
         break
     end
     psi_mid_old=psi((Nx-1)/2); mu_old=mu; j=j+1;
 end
end

\end{verbatim}

\section*{Problems}
\addcontentsline{toc}{section}{Problems}

\begin{prob}
\label{Density profiles}
Obtain the ground-state density profiles for a 1D condensate under harmonic confinement with i) no interactions, ii) repulsive interactions and iii) attractive interactions.  Compare ii) with the corresponding Thomas-Fermi profile.  
\end{prob}

\begin{prob}
\label{Quantum carpets}
Starting from the Gaussian harmonic oscillator ground state, release the non-interacting condensate into an infinite square well (achieve by setting the potential to a high value towards the edge of the box, and zero elsewhere).  Repeat for repulsive and attractive interactions.  How does the initial expansion (before reflection from the box walls) depend on the interactions?

Now simulate the longer-term behaviour.  The wavefunction undergoes revivals, known as the Talbot effect, and forms a ``quantum carpet" \cite{marzoli_1998}.
\end{prob}

\begin{prob}
\label{Collective modes}
Form the ground state solution for a repulsively-interacting condensate in a harmonic trap.  Excite a centre-of-mass (``sloshing") oscillation by shifting the trap by some distance at $t=0$.  Similarly, excite a monopole mode by slightly weakening the trap at $t=0$.  Extract the frequencies of these modes.  Do the frequencies depend on the number of particles and the interaction sign/strength?
\end{prob}

%

%

\backmatter

\Extrachap{References}

\addtocontents{toc}{\setcounter{tocdepth}{0}}

\printindex


\end{document}